\documentclass[12pt,a4paper,openany]{book}
\usepackage{anysize}
\usepackage[latin1]{inputenc}
\usepackage[OT2, T1]{fontenc}
\usepackage[portuges,russian,english]{babel}
\usepackage[ruled,chapter]{algorithm}
\usepackage{makeidx,subfig}
\usepackage{amsmath,amssymb,amsthm,amsfonts,mathrsfs,braket}
\usepackage{graphicx}
\usepackage{fancyhdr,array,geometry}
\usepackage{simplewick}
\usepackage{latexsym}
\usepackage[all]{xy}
\usepackage{euscript}
\usepackage{enumerate}
\usepackage{dsfont,slashed}
\usepackage[plainpages=false,pagebackref=true,pdftex]{hyperref}
\usepackage{cite}

\makeindex

\marginsize{25mm}{25mm}{25mm}{25mm}

\hypersetup{pdfpagelabels,hyperindex,colorlinks=true,breaklinks=true,bookmarks=true,linkcolor=black,citecolor=black,
urlcolor=black}

\begin{document}

\setlength{\abovecaptionskip}{0.0cm}
\setlength{\belowcaptionskip}{0.0cm}
\setlength{\baselineskip}{24pt}

\pagestyle{fancy}
\lhead{}
\chead{}
\rhead{\thepage}
\lfoot{}
\cfoot{}
\rfoot{}

\fancypagestyle{plain}
{
	\fancyhf{}
	\lhead{}
	\chead{}
	\rhead{\thepage}
	\lfoot{}
	\cfoot{}
	\rfoot{}
}

\renewcommand{\headrulewidth}{0pt}
\newcommand{\Tr}{{\rm Tr}}
\newcommand{\med}{{\rm med}}
\makeatletter
\renewcommand\mainmatter{%
    \clearpage
  \@mainmattertrue
  \pagenumbering{arabic}}
\makeatother

\hyphenation{Levitin}


\frontmatter 

\thispagestyle{empty}

\begin{figure}[h]
	\includegraphics[scale=.8]{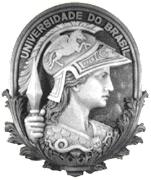}
\end{figure}

\vspace{15pt}

\begin{center}

\textbf{UNIVERSIDADE FEDERAL DO RIO DE JANEIRO}

\textbf{INSTITUTO DE FÍSICA}

\vspace{30pt}

{\Large \bf Quantum Speed Limits for General Physical Processes}

\vspace{25pt}

{\large \bf Márcio Mendes Taddei}

\vspace{35pt}

\begin{flushright}
\parbox{10.3cm}{Ph.D. Thesis presented to the Graduate Program in Physics of the Institute of Physics of the Federal University of Rio de Janeiro - UFRJ, as part of the requirements to the obtention of the title of Doctor in Sciences (Physics).}

\vspace{18pt}

{\large \bf Advisor: Dr. Ruynet Lima de Matos Filho}

\vspace{12pt}

\end{flushright}

\vspace{90pt}

\textbf{Rio de Janeiro}

\textbf{February, 2014}

\end{center}



\newpage
\mbox{}\vspace{5cm}

\begin{center}
\begin{tabular}{|c|}
\hline\\
\begin{minipage}{15cm}
\rule{15cm}{0pt} \flushright

\begin{minipage}{12.5cm}
\hspace{-2cm} T121 \hspace{0,68cm} Taddei, Márcio Mendes

\hspace{2cm} Quantum Speed Limits for General Physical Processes / Márcio Mendes Taddei. - Rio de Janeiro, 2014.

\hspace{2cm} \pageref{fimdopre}, \pageref{ofim}f. : il. ; 30 cm.\\

\hspace{2cm} Tese (Doutorado em Física) - Programa de Pós-graduação em Física, Instituto de Física , Universidade Federal do Rio de Janeiro.

\hspace{2cm} Orientador: Dr. Ruynet Lima de Matos Filho

\hspace{2cm} Bibliografia: f. \pageref{bibcomeco}-\pageref{bibfim}.\\

\hspace{2cm}
1. Física Quântica. 2. Limite Quântico de Velocidade. 3. Informação Quântica de Fisher. 4. Geometria de Estados Quânticos. 
 I. Matos Filho, Ruynet Lima de. II. Universidade Federal do Rio de Janeiro. Instituto de Física. III. Quantum Speed Limits for General Physical Processes.
\flushright{CDD 530.12}
\bigskip
\end{minipage}
\end{minipage}
\\
\hline
\end{tabular}
\end{center}



\newpage

\thispagestyle{empty}

\noindent

\begin{figure}[ht]
  \begin{center}
	\includegraphics[width=\linewidth]{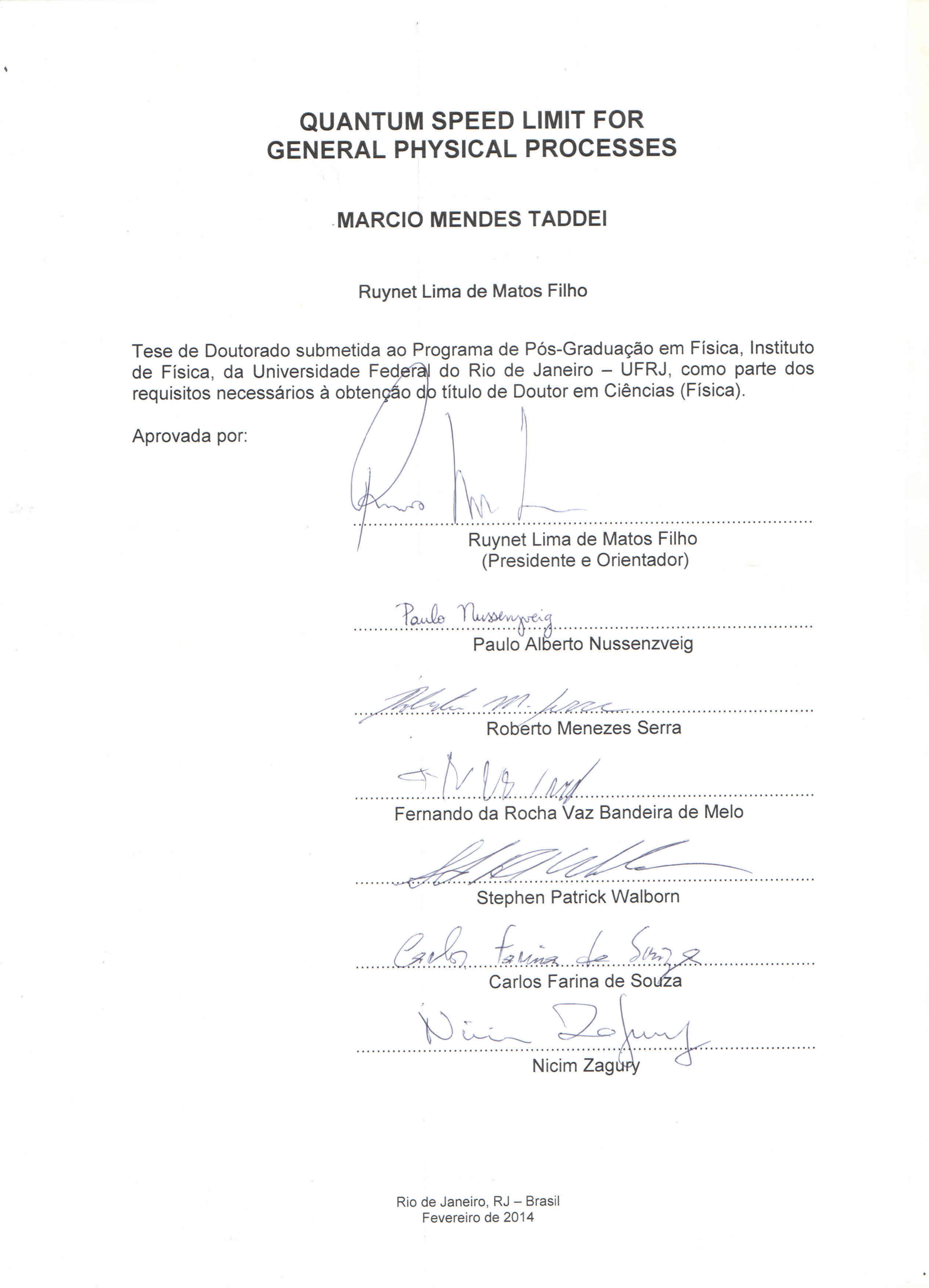}
  \end{center}
\end{figure}

\clearpage



\newpage

\noindent

\vspace*{20pt}
\begin{center}
{\LARGE\bf Resumo}\\
\vspace{15pt}
{\Large\bf Limites Quânticos para a Velocidade\\ de Processos Físicos Gerais}\\
\vspace{6pt}
{\bf Márcio Mendes Taddei}\\
\vspace{12pt}
{\bf Orientador:  Ruynet Lima de Matos Filho}\\
\vspace{20pt}
\parbox{14cm}{Resumo da Tese de Doutorado apresentada ao Programa de Pós-Graduação em Física do Instituto de Física da Universidade Federal do Rio de Janeiro - UFRJ, como parte dos requisitos necessários à obtenção do título de Doutor em Ciências (Física).}
\end{center}
\vspace*{35pt}

Limites quânticos de velocidade são relações que fornecem limites inferiores ao tempo de evolução de sistemas quânticos. Este tipo de resultado começou com Mandelstam e Tamm, que associaram o tempo mínimo necessário para um estado ficar ortogonal com a variância de energia de tal estado. Outro limite inferior, obtido mais tarde, foi o de Margolus e Levitin, que relacionaram este tempo mínimo com a energia média do sistema quântico. Tais limites são comumente usados para discutir o papel do emaranhamento na velocidade de evolução de sistemas quânticos, e têm potencial de aplicação a sistemas de evolução especialmente rápida, como os de computação (clássica ou quântica) ou mesmo nos esquemas explicativos do surgimento do comportamento clássico a partir de um substrato quântico, que dependem de uma descoerência extremamente veloz.

Tais resultados, especialmente o de Mandelstam-Tamm, já foram generalizados de algumas formas, em particular a incluir evoluções a estados não-ortogonais. Entretanto, havia uma lacuna na literatura da área, pois apenas evoluções unitárias -- sistemas quânticos fechados -- eram consideradas. Nesta tese, esta limitação é superada: nosso principal resultado é um limite para evoluções gerais, unitárias ou não, de sistemas quânticos, que corretamente recupera os limites de Mandelstam-Tamm no caso unitário. Aplicações desse limite para certos casos concretos interessantes são apresentadas. Esse limite também é usado para estender para o caso não-unitário a discussão do papel do emaranhamento em evoluções rápidas, fornecendo resultados não-triviais.

Para a dedução dos resultados, emprega-se uma abordagem geométrica que já havia se mostrado extremamente útil quando desenvolvida no caso unitário por Anandan e Aharonov, em especial por permitir uma clara interpretação dos limites e a discussão dos critérios para sua saturação. Faz-se mister ressaltar que não se pressupõe neste trabalho conhecimento prévio do leitor sobre geometria de estados quânticos. Nota-se ainda que o problema de limites ao tempo de evolução é intimamente ligado à metrologia quântica, em particular, ao problema de estimação quântica de parâmetros. Talvez o sinal mais manifesto desta proximidade seja a utilização, em nosso limite geométrico, da informação quântica de Fisher, grandeza de larga aplicação em metrologia quântica.

Por fim, apresentam-se resultados adicionais na direção de obter uma interpretação geométrica de limites do tipo de Margolus-Levitin.

\vspace{15pt}

\textbf{Palavras-chave:} 1. Limite Quântico de Velocidade, 2. Informação Quântica de Fisher, 3. Processos não-unitários, 4. Geometria de Estados Quânticos.



\newpage

\noindent

\vspace*{20pt}
\begin{center}
{\LARGE\bf Abstract}\\
\vspace{15pt}
{\Large\bf Quantum Speed Limits for General Physical Processes}\\
\vspace{6pt}
{\bf Márcio Mendes Taddei}\\
\vspace{12pt}
{\bf Advisor:  Ruynet Lima de Matos Filho}\\
\vspace{20pt}
\parbox{14cm}{\emph{Abstract} of the Ph.D. Thesis presented to the Graduate Program in Physics of the Institute of Physics of the Federal University of Rio de Janeiro - UFRJ, as part of the requirements to the obtention of the title of Doctor in Sciences (Physics).} 
\end{center}
\vspace*{35pt}

Quantum speed limits are relations yielding lower bounds on the evolution time of quantum systems. This kind of result was started by Mandelstam and Tamm, who associated the minimal time needed for a state to turn orthogonal with the \textit{energy variance} of said state. Another lower bound, of much later obtention, was that of Margolus and Levitin, who related this minimal time to the \textit{average energy} of the quantum system. These bounds are commonly used to discuss the role of entanglement in the evolution speed of quantum systems, and have the potential to be applied to extremely fast-evolving systems, such as those in both classical and quantum computation or in the explanatory schemes for how classical behavior arises from a quantum substrate, which depend on an exceedingly fast decoherence.

These results, especially that of Mandelstam and Tamm, have been generalized in some ways, in particular by including evolutions to non-orthogonal states. However, there was a gap in the literature on this area, for only unitary evolutions -- closed quantum systems -- had been considered. On this thesis, such limitation is overcome: our main result is a bound for quantum-system evolutions in general, whether unitary or not, and correctly recovers the Mandelstam-Tamm bounds in the unitary case. Applications of this bound to several concrete cases of interest are herein presented. This bound is also used to extend to the non-unitary case the discussion of the role of entanglement in fast evolutions, leading to nontrivial results.

For the derivation of the results, a geometric approach has been employed, which had already shown to be extremely useful for unitary evolutions when developed by Anandan and Aharonov, especially for allowing a clear interpretation of the bounds and the discussion of the criteria for their saturation. It should be noted that in this work no previous knowledge of quantum-state geometry by the reader is assumed. An important remark is that the problem of lower bounds on evolution time is closely related to quantum metrology, in particular, to quantum parameter estimation. Perhaps the most manifest sign of such proximity is the utilization, in our geometric bound, of the quantum Fisher information, a quantity largely applied in quantum metrology.

Finally, additional results towards obtaining a geometric interpretation of Margolus-Levitin bounds are presented.

\vspace{15pt}




\newpage

\noindent

\vspace*{20pt}

\begin{center}

{\LARGE\bf Acknowledgments}

\end{center}

\vspace*{40pt}

First and foremost, I acknowledge the contribution of my research group not only to this thesis, but, more importantly, to science as practiced in Brazil. As part of the Quantum Optics and Quantum Information Group, I am able to do science at an international level and be a living witness --- and even proof! --- against the notion that quality science can only be done in fully developed countries. There sure still are difficulties and limitations, especially regarding funding, but research in Brazil has come a great way and my group is part of that effort. It is particularly satisfying to watch conferences and talk to colleagues from all over the world on an equal footing, be they from Toronto, Hannover, Turku or Singapore. 

I must thank my advisor, Prof. Ruynet L. de Matos Filho, who accepted me as a student as I joined the group mid-Ph.D. During this time, Ruynet has always been interested in working with me, proposing new ideas, listening to me when I had results to show. A great part of the work on this thesis was done during the weekly meetings at his office. He has a very well defined notion of the responsibilities of advisor and advisee, and does not shirk his share of duties. He has always been understanding of my needs during this time, and helped me professionally whenever necessary. I have always had a good measure of freedom when working with him --- with a matching demand in return! He has also seen to it that I participate in conferences and meetings and had contact with scientists of related areas from the whole world. I also thank my direct collaborators Bruno M. Escher, who has often taken part in those meetings at Ruynet's office, and Prof. Luiz Davidovich, who has contributed in ways I could not have understood before his participation. The work here presented couldn't have come to fruition without them. 

I thank my former advisor, Prof. Carlos Farina, with whom I had my scientific initiation in the Casimir Effect Group. The three formative years I spent working with the group were decisive for the scientist and the person I have become. Farina is a human being with a deep sense of the qualms of those around him, and an extremely rare ability to put himself in the place of others. I have always felt the liberty to come to him with my worries, including my wish to leave his group mid-Ph.D. I thank him for not only being completely understanding of my move, but also helpful towards it (all the while without giving up of me as a student until I had completely made up my mind). Even though I have left the group, I still feel attached to the people in it, and it is with great joy that I see their ongoing success. I also thank Tarciro Mendes who, along with Farina, was my first scientific collaborator, and Marcus Venícius Cougo-Pinto, for teaching me a more formal approach to mathematics --- which has been directly helpful to this thesis --- as well as being an inspiration in terms of didactics, clarity and mathematical rigor.

I thank my parents, Jayme and Angela, for not only accepting my choices of studying first Chemistry, then Physics, but encouraging me to pursue an academic career to the point of believing that anything else would be nonsensical. I also thank my father for the corrections to the English in this thesis. And I thank my elder brother, Paulo, with whom I have had contact with interesting different perspectives on academic life.

I acknowledge the Physics Institute of the Federal University of Rio de Janeiro (IF-UFRJ) for the very professional atmosphere that can be found here. The commitment to doing science at a high level can be distinctly seen in many of the research groups I have come in contact with. It is an environment with an abnormal affluence of bright people, generation after generation. This professional environment is also reflected in the good work of Casé, Pedro and the administrative staff of the graduate courses of IF-UFRJ as a whole. 

I thank the funding agencies that have directly supported me, CNPq (Brazilian National Council for Scientific and Technological Development), Faperj (Foundation for Research Support of the State of Rio de Janeiro) and INCT-IQ (Brazilian National Institute for Science and Technology of Quantum Information), as well as sister agency CAPES, which has also supported the group. E agradeço ao povo brasileiro, que financia as instituições de apoio à pesquisa e permite que elas existam\footnote{Translation: ``And I thank the Brazilian people, which finances the research support institutions and allows them to exist.''}.

I must nevertheless play a down note. It is saddening that the academic career (worldwide) is so structured that the work one puts the most effort into during their graduation ends up being read by a small amount of people. In the writing of this thesis I have arduously striven to reach out to more people but, as with anything in life, all the dedication one can summon in an endeavor can only change the world by the tiniest bit. If they get to be so lucky.

I now take the liberty to write the remainder of the acknowledgments in Portuguese. 

Agora em bom português, agradeço aos colegas do grupo de ótica e informação quânticas. Ao Eduardo Paul e Alvaro Pimentel pelas conversas que iam do jogo da velha quântico às utilizações da palavra ``whose'' em inglês, e pelas várias vezes que entrei na sala de vocês para colocar alguma ideia no quadro (de física ou não). Ao Wellison Bastos, logo na sala em frente, pela paciência de me ouvir em muitos de meus anseios. À Mariana Barros, minha companheira de viagens. E ao grupo como um todo, mesmo quando vocês comem todo o churrasco antes de eu chegar.

Agradeço aos amigos que fiz na Física desde a época da graduação. Ao compadre Renato Aranha, ao Guilherme Bastos --- nosso querido Quantum Boy, de quem fui colega da graduação ao doutorado e em ambos os grupos de pesquisa ---, ao Wilton Kort-Kamp, Andreson Rego, Anderson Kendi, Rafael Bezerra, Gustavo Quintas de Medeiros, às pessoas do LAPE, de quem eu sempre estive próximo, Daniel Vieira, Oscar Augusto, Danielle Tostes, Daniela Szilard, Vinícius Franco.

Agradeço aos amigos que tenho e tive ao longo do doutorado, com os quais eu chorei e sorri, e sem os quais a trajetória seria muito mais penosa: Alexandre Sardinha, Edu\-ar\-do Am\-bro\-sio, Franco de Castro, Gabriella Cruz, Gustavo Rocha, Sarah Ozorio, Li\-di\-a\-ne Ca\-val\-can\-te, Thaís Alves, Tatiana Hessab, Wando Fortes, Warny Marçano, \foreignlanguage{russian}{Yuliya Samulenok}. Isso sem falar, é claro, nos amigos da 23C, Alan Albuquerque, Antonio Henrique Campello, Fabiane Almeida, Gustavo Coelho, Larissa Alves, Larissa Barbosa, Leonardo Marques, Mariana Perfeito, Mario Gesteira, Ricardo Barboza, Rodrigo de Oliveira, Tatiana Teitelroit, Ulysses Vilela (como é estranho não usar alguns dos apelidos).

Por fim, agradeço aos loucos companheiros do teatro, essa arte tão oposta à física em tantos sentidos, e por isso mesmo tão estimulante de fazer: Nathalia Colón, Ana Beatriz Machado, Beatriz Schuwartz, Daniel dos Anjos, Jailson Junior, João Pedro Marques, José Henrique Calazans, Luis Zacharias, Michelle André, Rafael de Andrade, Bárbara Cruz, Tayná Germano, Vinícius Ferreira, Wagner dos Santos. Sem vocês não seria fácil seguir dia após dia.



\newpage
\phantomsection
\addcontentsline{toc}{chapter}{Contents}
\tableofcontents

\newpage
\phantomsection
\addcontentsline{toc}{chapter}{List of Figures}
\listoffigures

\newpage
\phantomsection
\addcontentsline{toc}{chapter}{List of Tables}
\listoftables
\label{fimdopre}
\mainmatter

\begin{chapter}{Introduction}
\label{intro}

\hspace{5 mm} How fast can a quantum system evolve to an orthogonal state?
This question serves as a starting point in the search for the quantum speed limit, the maximal evolution speed of a quantum system. The goal of this search is to obtain general bounds,
valid for any particular system one may wish to apply it to, limiting (from below) the time it takes for a system to become distinguishable from its initial state. A paradigmatic first answer to this problem, appearing in the seminal work of Mandelstam and Tamm~\cite{MandelstamTamm}, was that, given an evolution dictated by a time-independent Hamiltonian $H$ and a pure initial state, the time $\tau$ necessary for the final state to be orthogonal to the given initial state is bounded by\footnote{We shall later see in Section~\ref{boundbyMT} that Mandelstam and Tamm's result in~\cite{MandelstamTamm} is more general than this.}
\begin{equation}
\tau \geq \dfrac{\pi\hbar}{2\Delta E} \ ,
\label{MTbound0}
\end{equation}
where $\Delta E$ is the standard deviation of $H$ and $\hbar$ is the reduced Planck constant. A key feature of the above bound is the dependence on the inverse of the standard deviation of the system energy.

\subsubsection{Why study such bounds?}

\hspace{5mm}The motivation for working on the quantum speed limit is fourfold. The topic originally arose from discussions on the structure of quantum mechanics, especially in attempts to define a suitable time operator and derive an uncertainty relation for time and energy analogous to that for position and momentum. The obtention of an uncertainty relation would often guide and serve as basic test for posited definitions. We remark that there has, in fact, been quite a lot of confusion regarding phrases such as ``time-energy uncertainty relation'' and that we find the term ``uncertainty relation'' unsuitable to refer to quantum speed limits such as the above (see page~\pageref{distinguishing} below for details).

The advances of computer science in its incessant search for faster computation times brought along a practical applicability for the quantum speed limit. While the evolution in computer processing power --- roughly exponential increase on clock rates observed since the 1960s, although with restrictions~\cite{tipoMoore} --- has been dictated mostly by advances in materials science, electronics (in particular, integrated circuitry and the ubiquitous transistor) and computer architecture, there are limits imposed solely by quantum mechanics~\cite{Lloyd}, irrespective of any technological hurdle. Such limits are relevant even for classical computation if it relies on quantum systems for storage or transfer of information: in the (oversimplified) case of a spin-based memory which only uses the pointer states up and down of a single spin to represent bits 0 and 1 without implementing or taking advantage of any superposition, computation times still depend on the time taken to flip a spin, which is an evolution between orthogonal states of a quantum system. (Dominant technology for hard-disk drives uses memory blocks comprising many spins each as bits~\cite{magnetoelectronics,spintronics}, as well as promising candidates for novel random-access memories~\cite{spintronics,STT}).   Naturally, the quantum-mechanically imposed limits are particularly important for quantum computing and quantum communication, which count on explicit quantum properties (superposition, entanglement).

A third problem related to exceedingly short-timed evolutions is that of the quantum-classical transition, i.e., understanding how, from a quantum-mechanical substrate --- which is experimentally demonstrated to be a more general theory ---, our human classical experience emerges. In other words, the problem consists in finding out why our everyday experience inhabits only a small fraction of quantum state space. The most promising explanatory models~\cite{Zurek,Gogolin} focus on the inevitable interaction with the environment and the consequent selection of few states and their classical mixture in exceedingly short time (decoherence). Evolution times are therefore key to this explanation, as quantum phenomena are not perceived on a macroscopic system due to the short-livedness of its quantum properties.

Lastly, the control of the state and dynamics of a quantum system, which has for quite some time received a great deal of interest for its practical usefulness in fields as diverse as bond-selective chemistry and quantum computation~\cite{WRD,GordonRice,Whither,NJPControl,Khaneja}, has always had as a concern the search for fast evolutions. A subset of the quantum control program called quantum brachistochrone problem, which consists~\cite{LuoBr} in finding the fastest evolution given initial and final states and some restriction on the resources used and on the form of evolution allowed, can be greatly aided by results on the quantum speed limit. We nonetheless remark that the quantum speed limit and the brachistochrone, albeit interconnected, are answers to different questions. The former inquires how long it must take for a state in a given process to change by some amount; the latter seeks to tailor a process (given some restrictions) so that the evolution between chosen states is as fast as possible. Their relatedness is depicted by a simple example, though: if the only restriction to the evolution of a closed system is on its energy, the brachistochrone (fastest path) is the one that saturates the quantum speed limit.

\subsubsection{Another relevant bound}

\hspace{5mm}Much work has taken the original bound further,  
but the most impacting later result on the topic was the bound obtained by Margolus and Levitin~\cite{MargolusLevitin}, 
\begin{equation}
\tau \geq \dfrac{\pi\hbar}{2\braket E} \ ,
\label{MLbound0}
\end{equation}
where the time $\tau$ necessary for a state to become orthogonal is bounded by the inverse of $\braket E$, the average energy with respect to the ground state of the system. When discussing generalized forms of these bounds, it is useful to characterize them as Mandelstam-Tamm-like bounds when they depend on the energy variance and/or reduce to Eq.~\eqref{MTbound0} or as Margolus-Levitin-like bounds, when they depend on average energies and/or reduce to Eq.~\eqref{MLbound0}. We note that Mandelstam-Tamm-like bounds are founded on a more elaborate framework, with a clear physical interpretation and with stances in which they are saturated along the course of entire evolutions, whereas for Margolus-Levitin-bounds these two features are absent (see Section~\ref{MTML} for more on the subject).

\subsubsection{Distinguishing between bounds and uncertainty relations}
\label{distinguishing}

\hspace{5mm} There has been in fact much discussion on relations involving time and energy since the dawn of quantum mechanics, and any informed reader would be readily reminded by Eq.\eqref{MTbound0} of the ``time-energy uncertainty relation''. We feel the need to clarify some distinctions among these similar-looking equations to better precise the goal of our work.

Heisenberg's uncertainty relation for canonical conjugate observables --- best known by the ubiquitous position-momentum case
\begin{equation}
\Delta X \Delta P \geq \hbar/2 \ ,
\label{relincXP}
\end{equation}
but valid for any canonical pair --- has had two main interpretations in the literature. The first claims that the relation bounds the precision of {\it sequential measurements} of position and momentum (or any canonical pair) on a given particle. This was the spirit of the famous Gedankenexperiment consisting of ``looking at an electron through a microscope'' mentioned in the original article of 1927~\cite{Heisenberg}. However, the most recurrent interpretation states that a particle cannot have, {\it at the same time}, position- and momentum-distributions whose standard deviations are below values allowed by the relation. In terms of experiments, this view would translate into comparing the standard deviations in position and momentum from measurements made on identical, but different, ensembles. (As before, the argument goes for any canonical pair.) Kennard~\cite{Kennard} and Weyl~\cite{Weyl} showed, shortly after, that which later became a standard passage on textbooks on quantum mechanics (e.g.~\cite{Cohen}): the commutation relation between canonical pairs implies the uncertainty relation in the latter view. 

If the topic seems controversial so far, it only turns worse when one attempts to write time-energy relations such as
\begin{equation}
\Delta E \Delta t \geq \hbar/2 \ .
\label{relincET}
\end{equation}
This happens for a few reasons, such as the lack of a satisfactory definition of a time observable and the conceptual infeasibility of an independent subsequent time measurement. One interpretation of this kind of relation, posited by Bohr~\cite{Bohr} and Landau and Peierls~\cite{LandauPeierls}, was that an energy measurement made during a time $\Delta t$ would induce a disturbance $\Delta E$ in the energy of the system according to Eq.\eqref{relincET}. This disturbance would be relevant for subsequent measurements of the energy, but not restrictive of the precision of the original measurement. 
Fock and Krylov~\cite{FockKrylov}, on the other hand, viewed Eq.\eqref{relincET} as a relation between the precision of an energy measurement $\Delta E$ and the time $\Delta t$ taken to perform it. These discrepant viewpoints triggered a long, heated discussion in the literature, especially between, but not restricted to, Fock on one side and Aharonov and Bohm on the other~\cite{FockKrylov,AharonovBohm,Fock,AharonovBohm2,Fock2}. The latter defended that the relation did not hold in this sense, i.e., precise energy measurements could be done arbitrarily fast, and further concluded that not even the disturbance for later measuring the energy of the system occurred by necessity: it was, in principle, possible to construct an energy measurement on a system in which the only perturbations created were in the energy of the interacting field/apparatus\footnote{This basically amounts to stating the existence of quantum non-demolition (QND) measurements.}. Aharonov and Bohm's point of view eventually gained widespread acceptance~\cite{Vorontsov,AharonovSafko,EberlySingh}; a review of this debate can be found in~\cite{Vorontsov}. The present work nevertheless does {\it not} undertake these issues of relating time duration of an energy measurement to precision of and/or disturbance caused by said measurement.

There were, however, efforts to obtain a time-energy inequality without mention of measurements and describable directly by the state of the system, akin to the more usual interpretation of Eq.\eqref{relincXP}, by developing a definition of a time observable in some sense canonically conjugate to the Hamiltonian. This search for a time observable was unfruitful and eventually abandoned, but this was the spirit in which the seminal Mandelstam-Tamm bound arose: Eq.\eqref{MTbound0} relates quantities of a given state in a given evolution, without mention of any measurement.

Because the quantum speed limit, despite the notational resemblance, is to be interpreted in ways radically different from the well-established canonical-conjugate uncertainty relation (e.g. Eq.~\ref{relincXP}), given that neither is Eq.~\eqref{MTbound0} a consequence of a commutation relation (no accepted definition of time operator) nor does it refer directly to measurements, we refrain from calling it ``uncertainty relation'' and favor terms such as ``bound on evolution time'' or simply ``quantum speed limit''.

\subsubsection{Our contribution}

\hspace{5mm}Despite the interest in quantum speed limits for decades, very little had been done~\cite{CarliniBr8,Beretta,Obada,BrodyN} for open systems, described by non-unitary evolutions, and even so, always dealing with particular cases, not seeking general expressions. Jones and Kok~\cite{JonesKok} had even argued the case of impossibility of a general bound for time evolutions for the non-unitary case. Many of the applications of the bounds would  benefit from results for non-unitary evolutions, though. They would allow one to assess noise effects on a system, which is crucial for realistic descriptions of fast computation and communication channels; the decoherence used to explain the quantum-classical transition is an intrinsically non-unitary process; quantum control in general could be improved if more evolution forms were available to choose from.

Our main original contribution to the topic comes in the form of a very general bound for the time evolution of a state, valid for any physical process, be it unitary or not, applicable for any time during the evolution. It is derived geometrically and has a straightforward geometrical interpretation; it is a Mandelstam-Tamm-like bound, saturated under a simple and clear criterion. 
An additional result is a second bound, dependent on the median of the energy distribution and applicable for unitary cases, which could lead to a more general Margolus-Levitin-like bound in the future.

This thesis is structured as follows: in Chapter~\ref{cap2}, we review previous results found in the literature and define some of the notation. Chapter~\ref{cap3} is dedicated to present our main bound, preceded by the concepts and definitions necessary to discuss it. We also show our additional results. In Chapter~\ref{cap4} we apply the bound to some examples, including some instances of the brachistochrone problem, and final remarks and perspectives are left to Chapter~\ref{final}.

In an effort to broaden the prospective readership of this thesis past examination, we have striven to make it accessible to the reader not knowledgeable on Quantum Information Theory, although Quantum Mechanics is a prerequisite. The aim at a larger reach is also the motivation for choosing to compose the present work in English.

\end{chapter}

\begin{chapter}{Main concepts and existing literature}
\label{cap2}

\hspace{5mm} In this Chapter, we introduce the main concepts of the topic of bounds on quantum state evolution by reviewing the preceding works in the literature. Although chronological order is largely respected, we are ultimately guided by the presentation and development of the conceptual framework necessary for our work, and by situating our work in the literature\footnote{Comments on works subsequent to our own are left to Appendix~\ref{later}.}. More elaborate constructs, such as the geometry of quantum states and the (quantum) Fisher information, are left for the following Chapter.

We note that the question of how long a system must evolve for so that initial and final states are orthogonal (or distinguishable to some degree) must be undertaken in the Schrödinger picture, in which evolution manifests itself fully in the state of the system. The question as expressed above would be void --- simply put, moot --- in the Heisenberg picture, since its evolution acts not on the state of the system, but on its observables. The interaction picture is also unsuitable for this discussion, for in this case the effect of evolution is partly imparted to the state and partly to the observables. Because we are comparing states at different times, orthogonality (and, later on, fidelity) in each of these pictures is not equivalent, and should be assessed in the Schrödinger picture in order for all the changes caused by evolution to be reflected in the state. One could, in principle, restate the problem in order to work in different frameworks --- e.g., phrase it in terms of the orthogonality of supports of observables in the Heisenberg picture ---, but this will not concern us here. We work in the Schrödinger picture throughout this thesis. \label{Schrpicture}

\section{The Mandelstam-Tamm bound}
\label{boundbyMT}

\hspace{5 mm} The first fundamental bound on evolution time was achieved by Mandelstam and Tamm in their seminal work of 1945~\cite{MandelstamTamm}. In its most cited form, it states that for an evolution governed by a time-independent Hamiltonian $H$ to turn a state orthogonal, it must take time $\tau$ obeying
\begin{equation}
\tau \geq \dfrac{\pi\hbar}{2\Delta E} \ , 
\label{MTbound}
\end{equation}
where $\Delta E$ is the standard deviation of $H$. We remark that the numerical value of the bound depends on the Hamiltonian at hand as well as on the initial state. This is a general feature of the quantum speed limit: for each process (in this case defined by $H$ and the initial state), a different value of $\tau$ may be obtained. That this must be the case can be seen by two extreme scenarios: firstly, if the system is initially in an energy eigenstate ($\Delta E=0$), it will never reach an orthogonal state, an aspect grasped by Eq.\eqref{MTbound} as it predicts $\tau\geq\infty$. On the other hand, with no constraint on the energy spread of a system, its evolution can be arbitrarily fast. Such is the case of a spin-flip under a field of arbitrarily high magnitude and, since the bound is still valid, it must yield a correspondingly small time to reach an orthogonal state. Because $\Delta E$ in a spin-flip is proportional to the magnitude of the field, Eq.~\eqref{MTbound} correctly indicates a vanishing time. Hence, only a process-dependent bound can produce relevant results. \label{appliedfield} We seize the opportunity to note that $\Delta E$ in the bound must be the standard deviation of the full Hamiltonian governing the evolution, which, in the case of a spin-flip, includes the applied field as well as the spin Hamiltonian\footnote{A common misconception is that the bound can do without the applied field.}.

Mandelstam and Tamm's result, however, encompasses more than just Eq.~\eqref{MTbound}, for it assesses not only orthogonality, but also how distinguishable an evolved state can be from its initial state on a gradual scale. The fidelity $F$ is a function that indicates how much two states can be distinguished by measurements. $F$ is a real-valued, symmetric function of two states which is null when they are orthogonal (ideally distinguishable) and maximal (unity) when they coincide. For two pure states, the fidelity $F$ between them is simply the modulus squared of their overlap, so that \mbox{$F(\ket{\psi_0},\ket{\psi_\tau})=|\braket{\psi_0|\psi_\tau}|^2$}. The Mandelstam-Tamm bound asserts that, given an evolution generated by a time-independent Hamiltonian $H$ and an initial state $\ket{\psi_0}$, the fidelity between the latter and the evolved state at time $\tau$, $\ket{\psi_\tau}$, is bounded by (from\footnote{In the original Russian version, it corresponds to Eq.~(9).} Eq.(10) of~\cite{MandelstamTamm})
\begin{equation}
F(\ket{\psi_0},\ket{\psi_\tau})=|\braket{\psi_0|\psi_\tau}|^2 \geq \cos^2\left(\Delta E \tau/\hbar \right) \ 
\label{MToriginal}
\end{equation}
for $0\leq\Delta E \tau/\hbar\leq\pi/2$.
The reader should note that, because of the time independence of the Hamiltonian, $\Delta E$ is constant and can be calculated on the initial or any other state along the evolution. Relation \eqref{MToriginal} can be easily inverted to yield an explicit bound on the time $\tau$ necessary for the system to reach some state $\ket{\psi_\tau}$ such that the fidelity relative to the initial state goes below a given value $F$:
\begin{equation}
\tau \geq \frac{\hbar}{\Delta E} \arccos \sqrt F \ ,
\label{MTfidelidade}
\end{equation}
where $\arccos$ is defined to have $[0,\pi]$ as image throughout this thesis. To illustrate the improvement on the previous result (Eq.~\ref{MTbound}), we notice that, for an energy eigenstate ($\Delta E =0$), Eq.~\eqref{MTfidelidade} not only grants that it fails to reach an orthogonal state, but exactly predicts that the state does not evolve at all, since $\tau\geq\infty$ for any fidelity $F\neq1$. The term ``Mandelstam-Tamm bound'' nevertheless most commonly refers to the special case of orthogonal states, Eq.~\eqref{MTbound}, recovered by Eq.~\eqref{MTfidelidade} with $F=0$.

Perhaps the best way to interpret the fidelity-dependent bound is by plotting the fidelity between initial and final states allowed by Eq.~\eqref{MTfidelidade} against time. This is done in Fig.\ref{grafbasico}, where a simple application to a qubit (two-level system) is presented, with a Hamiltonian proportional to the Pauli operator $Z$, $H=\hbar\omega Z/2$, and an initial state $\ket{\psi_0}=\left( \ket 0 + \sqrt 2\ket 1 \right)/\sqrt 3$ ($\ket 0$ and $\ket 1$ being the $\pm1$ eigenstates of $Z$, resp.).
\begin{figure}[ht]
	\centering
	\includegraphics[width=.5\columnwidth]{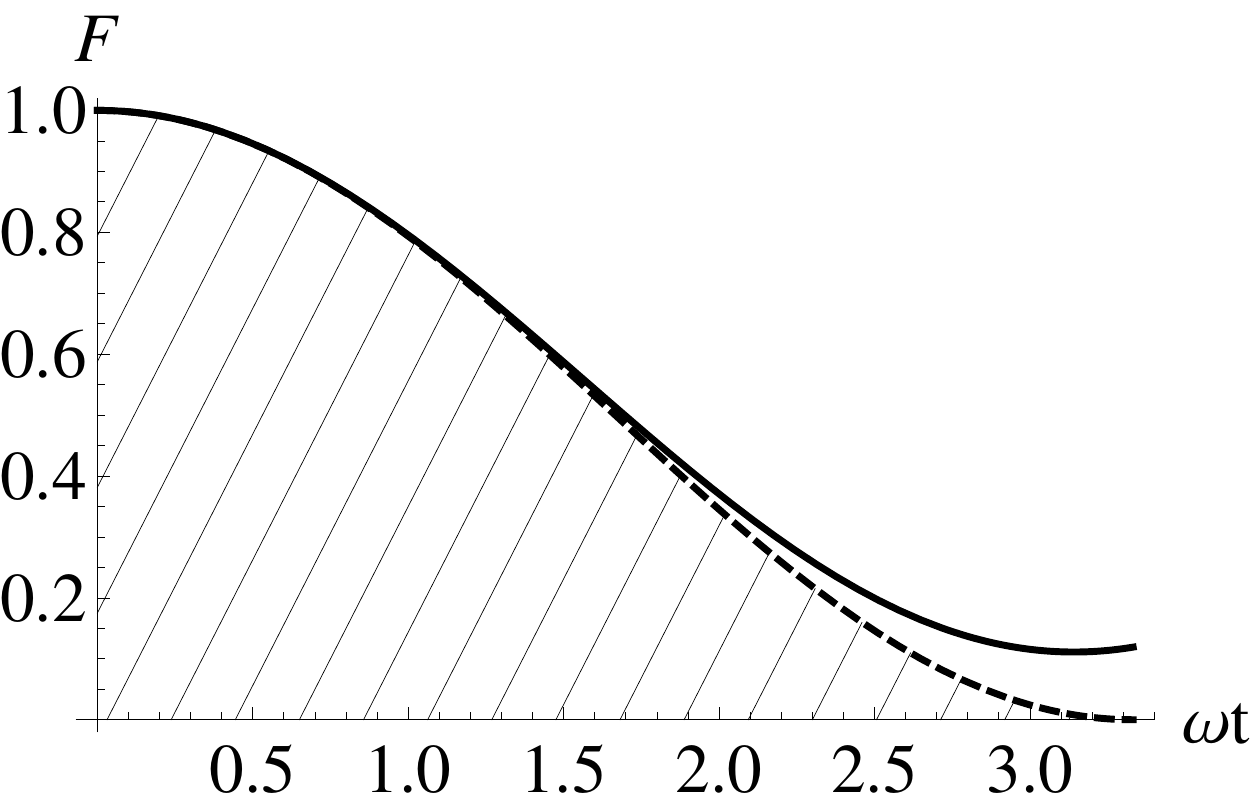}
	\caption{Bound on fidelity $F$ between initial and final states (dashed line) of the simple unitary evolution of initial state $\ket{\psi_0}=\left( \ket 0 + \sqrt 2\ket 1 \right)/\sqrt 3$ under a Hamiltonian proportional to Pauli operator $Z$, compared with actual fidelity along the evolution (solid line). The bound grants that the hatched area cannot be reached by evolution.}
	\label{grafbasico}
\end{figure}
The bound asserts that the fidelity between the initial state and the state at time $t$ must always be above the dashed line. The actual fidelity during evolution can be easily calculated in this example (solid line), and it does obey the bound, as it must. In other words, the bound affirms that the hatched area is beyond reach for the evolution.

The relations above are valid for simple evolutions, i.e., pure states evolving unitarily under a time-independent Hamiltonian. A proof of Eq.~\eqref{MTfidelidade} can be found on the very readable original article~\cite{MandelstamTamm}, which we here reproduce; for the reader interested in an alternative demonstration also founded on basic quantum mechanics, we recommend~\cite{Vaidman}. 

The original demonstration depends on two standard relations of quantum mechanics. Firstly, for any pair of observables, the product of their standard deviations is bounded by their commutator. We are interested in the case in which one of the observables is the Hamiltonian $H$ dictating the evolution,
\begin{equation}
\Delta E \Delta A \geq \frac{1}{2} |\braket{[H,A]}| \ ,
\label{incertcom}
\end{equation}
where $A$ is any observable of the system and $\Delta A$, its standard deviation. On the other hand, the type of evolution at play allows one to write, for time-independent\footnote{Time independence in the Schrödinger picture is meant, as mentioned on page~\pageref{Schrpicture}.} $A$,
\begin{equation}
\hbar \frac{d\braket{A}}{dt} = i \braket{[H,A]} \ ,
\label{dAdt}
\end{equation}
where the averages are taken in the state $\ket{\psi_t}$ at time $t$. By taking the modulus of the latter and comparing with the former, one obtains
\begin{equation}
\Delta E \Delta A \geq \frac\hbar2 \left|\frac{d\braket{A}}{dt} \right| \ .
\label{incertAE}
\end{equation}

Let us now choose $A$ to be the projection operator onto the initial state, $A=P_0=\ket{\psi_0}\bra{\psi_0}$, $P_0^2=P_0$. Then $\Delta P_0=\sqrt{\braket{P_0^2}-\braket{P_0}^2}=\sqrt{\braket{P_0}-\braket{P_0}^2}\geq0$ and from \eqref{incertAE} one finds
\begin{equation}
\Delta E \geq \frac{\hbar}{2} \left| \frac{d\braket{P_0}/dt} {\sqrt{\braket{P_0}-\braket{P_0}^2}} \right| \ .
\label{formaP0}
\end{equation}
Upon integration with respect to time from $0$ to $\tau$, one uses the fact that $\int_a^b|f(t)|dt\geq\left|\int_a^b f(t)dt\right|$ for any $f(t)$ 
 and that the whole expression depends on $P_0$ only via $\braket{P_0}$ to arrive at
\begin{equation}
\Delta E \cdot \tau \geq \hbar \arccos \sqrt{\braket{P_0}_\tau} \ ,
\label{MTquase}
\end{equation}
where $\braket{P_0}_\tau$ is the expectation value of the projector $P_0$ at time $\tau$, i.e., the modulus squared of the overlap, or fidelity, $\braket{\psi_\tau|\psi_0}\braket{\psi_0|\psi_\tau}=F(\ket{\psi_0},\ket{\psi_\tau})$ (it should be clear that $\braket{P_0}_0=1$). From this we recover \eqref{MTfidelidade}.

Interestingly, Mandelstam and Tamm present two different bounds in~\cite{MandelstamTamm}. The one mentioned above, on a par with modern usages, is presented only after a relation based on the possibility of inferring a change on some given observable of the system. The authors define $\Delta T$ as the minimum time necessary for the average value of an observable $A$ to change by an amount equal to its standard deviation. Valid, as before, for a pure-state evolution dictated by a time-independent Hamiltonian $H$, the bound (Eq.(5b) of~\cite{MandelstamTamm}) reads
\begin{equation}
\Delta E \Delta T \geq \hbar/2 .
\label{MTprimordial}
\end{equation}
Since this relation is valid for any observable $A$ of the system, $\Delta T$ must be interpreted as the time necessary for the state of the system to change, but this form does not lend itself easily to a quantitative description of state evolution.

This relation can be demonstrated by integrating Eq.~\eqref{incertAE} over time from $t$ to $t+\Delta t$ and applying again the relation $\int_a^b|f(t)|dt\geq\left|\int_a^b f(t)dt\right|$ 
, yielding
\begin{equation}
\Delta E \Delta t \geq \frac{\hbar}{2} \left( \frac{\left|\braket{A}_{t+\Delta t}-\braket{A}_t\right|}{\bar {\Delta A}} \right) \ ,
\label{incertETprimordial}
\end{equation}
where $\bar {\Delta A}:=1/\Delta t\int_t^{t+\Delta t}\Delta A dt$ is the time average of $\Delta A$ over the integration region. By definition, $\Delta T$ is the first value of $\Delta t$ for which the term in parentheses in Eq.\eqref{incertETprimordial} equals one (``shortest time during which the average value of a certain quantity is changed by an amount equal to the standard [deviation] of this quantity''\footnote{Quote from~\cite{MandelstamTamm}}). We thus have derived Eq.~\eqref{MTprimordial}.\footnote{Gray and Vogt have, for some reason, recently shown formally which of the properties of the mathematical structure of quantum mechanics are necessary and sufficient for this inequality to be valid~\cite{GrayVogt}. We remark that the standard assumption of observables $H$ and $A$ being defined over the whole Hilbert space is sufficient.}

We note that the definition of time $\Delta T$ has also been taken as a definition of a time operator~\cite{LeubnerKiener,GislasonSabelli}
\begin{equation}
T:= \frac{A}{|d\braket A/dt|} \ , \ \Delta T= \frac{\Delta A}{|d\braket A/dt|} \ ,
\label{MTdeftempo}
\end{equation}
since, when applied to Eq.~\eqref{incertAE}, it clearly leads to Eq.~\eqref{MTprimordial}.

After Mandelstam and Tamm's original paper, Bhattacharyya~\cite{Bhattacharyya} rederived and applied the bound of Eq.~\eqref{MTbound}, whereas Fleming~\cite{Fleming}, Bauer and Mello~\cite{BauerMello}, Gislason et al~\cite{GislasonSabelli}, Uffink and Hilgevoord~\cite{UffHilge,Uffink} recognized that usual decays, not being unitary evolutions, do not obey Eq.~\eqref{MTbound} and attempted definitions of time (inverse energy widths, average lifetime, etc.) to tackle such cases. Eberly and Singh~\cite{EberlySingh} presented another time definition, being followed by Leubner and Kiener~\cite{LeubnerKiener}. All of these definitions, although occasionally valid for more general cases than Eq.~\eqref{MTbound}, had a strong heuristic character, which undermined the relevance of their application: none has achieved usage by the community.

More interesting are the results that take Eq.~\eqref{MTbound} further: Pfeifer and Fröhlich~\cite{Pfeifer,PfFrohlich} have generalized it for time-dependent Hamiltonians. In the presented framework, this can be done straightforwardly: Eqs.\eqref{incertcom}-\eqref{formaP0} carry over for time-dependent Hamiltonians $H(t)$ (with their standard deviation being $\Delta E(t)$), and integration of Eq.~\eqref{formaP0} over time from $0$ to $\tau$ then yields
\begin{equation}
\int_0^\tau \Delta E (t) dt \geq \hbar \arccos \sqrt F \ ,
\label{MTtimedep}
\end{equation}
an implicit bound on the time $\tau$ necessary for the fidelity $|\braket{\psi_0|\psi_\tau}|^2$ between initial and final states to reach an amount $F$. The bound is particularly easy to apply if the Hamiltonian is self-commuting (i.e. $[H(t),H(t')]=0 \ \forall \ t,t'$), because $\Delta E(t)$ can then be calculated as an average on the initial state of the evolution (in fact, the result has been generalized for non-self-commuting Hamiltonians only later, on geometrical grounds, by Anandan and Aharonov~\cite{AA} and then Uhlmann~\cite{Uhlmann}; later still by Deffner and Lutz~\cite{Deffner}).

\section{The Margolus-Levitin bound} 
\label{boundbyML}

\hspace{5mm}A different bound was obtained by Margolus and Levitin~\cite{MargolusLevitin} fifty-three years after Mandelstam and Tamm's seminal paper. It is independent --- in the sense that it does not recover Eq.~\eqref{MTbound} in any way --- and relates the minimal time $\tau$ to reach an orthogonal state to $\braket E$, the average energy of the system relative to the ground state:
\begin{equation}
\tau \geq \dfrac{\pi\hbar}{2\braket E} \ . 
\label{MLbound}
\end{equation}
This bound is valid for pure states evolving under a time-independent Hamiltonian $H$. To demonstrate it, let us start by denoting by $\ket{\psi_0}$ the initial state, which can be decomposed in energy eigenstates $\ket{E_n}$ as $\ket{\psi_0}=\sum_nc_n\ket{E_n}$ (with $H\ket{E_n}=E_n\ket{E_n}$), and assume, without loss of generality, a zero-energy ground state, such that $E_n\geq0$. We are interested in finding the first root of the overlap $S(t)$,
\begin{equation}
S(t)=\braket{\psi_0|\psi_t}=\sum_n |c_n|^2 e^{-iE_nt/\hbar} \ ,
\label{overlap}
\end{equation}
where the summations run through all energy eigenstates. By taking the real part of the above,
\begin{equation}
{\rm Re} \ S(t) = \sum_n |c_n|^2 \cos \frac{E_nt}{\hbar} \ .
\label{Reoverlap}
\end{equation}
The derivation now makes use of a trigonometric inequality
\begin{equation}
\cos x \geq 1 - \frac{2}{\pi} \left(x+\sin x\right) \ \ \forall \ x\geq0 \ ,
\label{cosaleat}
\end{equation}
where equality only holds for $x=0$ and $x=\pi$. This choice of inequality entails a certain degree of arbitrariness that prevents a clear interpretation of the following result and is ground for criticism, especially since no derivation so far has been able to do without such an arbitrary choice. Substituting Eq.~\eqref{cosaleat} with $x=E_nt/\hbar\geq0$ in each term of~\eqref{Reoverlap}, one finds
\begin{equation}\begin{split}
{\rm Re} \ S(t) \geq &\sum_n |c_n|^2 \left[ 1 - \frac 2 \pi \left(\frac{E_nt}{\hbar}+\sin\frac{E_nt}{\hbar} \right)\right] \ ,\\
				 & =  1 - \frac{2\braket{E}}{\pi\hbar}t + \frac 2 \pi {\rm Im} \ S(t) \ .
\end{split}\label{ReS}
\end{equation}
If $\tau$ is a root of the overlap, or $\braket{\psi_0|\psi_\tau}=0$, both real and imaginary parts of $S(\tau)$ are null, leading to Eq.~\eqref{MLbound}.

Unlike the Mandelstam-Tamm relations, the Margolus-Levitin bound in its closed, analytic form only applies to orthogonal initial and final states, and is unable to describe the system throughout its evolution. \label{boundGiovannetti} Giovannetti et al~\cite{GiovannettiPRA} seeked to overcome this limitation and extend the Margolus-Levitin bound to any fidelity between initial and final states, i.e., to find a bound valid along the system evolution akin to Eq.~\eqref{MTfidelidade}. They pursued a relation of the form
\begin{equation}
\tau \geq \frac{\pi\hbar}{2\braket E}\alpha(F) 
\label{Giovgeral}
\end{equation}
bounding the time $\tau$ needed to reach a fidelity $F$ between initial and final states, where $\alpha(F)$, a function depending only on $F$, was to be found. The rather cumbersome following formula for $\alpha(F)$ was obtained
\begin{equation}
\alpha(F) = \min_\theta \left[ \max_q \left( \left[ 1 - \sqrt F(\cos\theta - q\sin\theta) \right] \frac{2}{\pi a} \right) \right] \ ,
\label{Giovalpha}
\end{equation}
where $q\in[0,\infty)$ and $\theta\in[0,2\pi]$ are parameters over which the expression has to be optimized, and $a$ is an implicit function of $q$ defined by
\begin{equation}\begin{split}
		a  &=\frac{y\sqrt{y^2(1+q^2)+q^2}}{1+y^2} \ , \\
\sin y &= \frac{a(1-qy)+q}{1+q^2} \ ,
\end{split}\label{implicit}
\end{equation}
with $y\in[\pi-\arctan1/q,\pi+\arctan q]$. The authors resorted to numerical calculation for estimation of $\alpha(F)$. Readers interested in the demonstration are referred to Appendix 1 of~\cite{GiovannettiPRA}.

\section{Main features of the bounds}
\label{MTML}

\hspace{5mm} A clear interpretation of the Mandelstam-Tamm bound has been enabled by Anandan and Aharonov~\cite{AA,Anandan}, who developed a geometric approach to the quantum speed limit, i.e., they rederived the Mandelstam-Tamm bound on geometrical foundations. Succinctly stated, they have shown that Eq.~\eqref{MTtimedep} can be interpreted as a comparison between the path followed in state space in a certain evolution with the geodesic between the endpoints of that evolution. Their approach will be discussed in greater detail in Section~\ref{GeoApp} due to its importance to our main result, also derived geometrically.

An additional sign of the relevance of the geometric approach is that it presents a straightforward criterion for saturating the speed limit (equality in Eq.~\ref{MTtimedep}), namely, that the evolution be along a geodesic. Such finding paved the way for the first works, by Horesh and Mann~\cite{HoreshMann} and Pati~\cite{Pati}, on finding the states that reach the limit. The necessary and sufficient condition is that the states be of the form
\begin{equation}
\ket{\psi_0}=\frac{1}{\sqrt 2} \left(\ket{E_n} + e^{i\phi}\ket{E_{n'}}\right) \ ,
\label{intelligent}
\end{equation}
i.e., equiprobable, coherent superpositions of two energy eigenstates, with $H\ket{E_n}=E_n\ket{E_n}$. Saturation of the Mandelstam-Tamm bound is achieved along the entire evolution for these states. Since geometric arguments are to be developed only in Section~\ref{GeoApp}, we here present a simplified proof of the saturation, limited to orthogonal states and constant Hamiltonians, due to Levitin and Toffoli~\cite{LevitinToffoli}.

This demonstration stems from a rederivation of the bound with the mentioned restrictions, bearing some resemblance to the derivation in Section~\ref{boundbyML}. Being $\ket{\psi_0}=\sum_nc_n\ket{E_n}$ the decomposition of the initial state into energy eigenstates, the fidelity can be written as 
\begin{equation}
F(\ket{\psi_0},\ket{\psi_t})=|\braket{\psi_0|\psi_t}|^2=\sum_{n,n'}|c_n|^2|c_{n'}|^2e^{-i(E_n-E_{n'})\tfrac{t}{\hbar}}=\sum_{n,n'}|c_n|^2|c_{n'}|^2\cos\left(\tfrac{E_n-E_{n'}}{\hbar}t\right) \ ,
\label{satcos}
\end{equation}
where the fact that the fidelity is real was used and summations are over all energy eigenstates. Once again a (quite arbitrary) trigonometric inequality is used,
\begin{equation}
\cos x \geq 1 - \frac{4}{\pi^2}x\sin x - \frac{2}{\pi^2}x^2 \ , 
\label{aleat2}
\end{equation}
valid for any real $x$; equality occurs if and only if $x=0$ or $x=\pm\pi$. By substituting this inequality in Eq.~\eqref{satcos} for every term,
\begin{equation}\begin{split}
F(\ket{\psi_0},\ket{\psi_t}) \geq & \sum_{n,n'}|c_n|^2|c_{n'}|^2 \left[ 1 - \frac{4}{\pi^2}\left(\tfrac{E_n-E_{n'}}{\hbar}t\right) \sin\left(\tfrac{E_n-E_{n'}}{\hbar}t\right) - \frac{2}{\pi^2}\left(\tfrac{E_n-E_{n'}}{\hbar}t\right)^2  \right] \ , \\
& = 1 + \frac{4}{\pi^2} \frac{d F}{dt}t - \frac{4}{\pi^2}\frac{(\Delta E)^2t^2}{\hbar^2} \ .
\end{split}\label{satgeq}
\end{equation}
Because $F$ is non-negative (and smooth), whenever $F=0$, $dF/dt=0$ as well, and orthogonality is only reached for a time $\tau$ if
\begin{equation}
0\geq1-\frac{4}{\pi^2}\frac{(\Delta E)^2\tau^2}{\hbar^2} \ ,
\label{falsetaMTbound}
\end{equation}
which rederives the Mandelstam-Tamm bound. The condition for saturation at $\tau$ is that of the trigonometric inequality for all $n$. This requires either $(E_n-E_{n'})\tau=0$ or $(E_n-E_{n'})\tau=\pm\pi\hbar$ for every pair $(n,n')$ with nonzero initial-state-expansion coefficients $(c_n,c_{n'})$, which can only be accomplished if $\ket{\psi_0}$ is a superposition of no more than two eigenstates. Simple calculations show that the only two-eigenstate superpositions that reach an orthogonal state are the equiprobable ones from Eq.~\eqref{intelligent}, and it can be verified that their orthogonality time is indeed $\tau=\pi\hbar/2\Delta E$. A more general demonstration based on geometrical arguments will be found in Section~\ref{GeoApp}.

No clear interpretation such as that provided by geometrical arguments has been bestowed on the Margolus-Levitin bound. Attempts to derive it geometrically have been made~\cite{JonesKok,Zwierz}, but they have been unable to recover Eq.~\eqref{MLbound} exactly. Furthermore, its saturation has only been proven for reaching orthogonality (a single instant during evolution) and occurs in evolutions along which the Mandelstam-Tamm bound saturates at all times, with the same states given by Eq.~\eqref{intelligent}. This has been proven for the first time by Söderholm et al~\cite{Soderholm}; we here reproduce Levitin and Toffoli's proof~\cite{LevitinToffoli}, based on the derivation of the Margolus-Levitin bound presented above. For equality to occur in Eq.~\eqref{ReS} at time $\tau$, each term of the sum must correspond to equality on the trigonometric relation Eq.~\eqref{cosaleat}. This implies, for every $n$ such that $c_n\neq0$, either $E_n\tau=0$ or $E_n\tau=\pi\hbar$. The state that saturates the bound must then only be composed of two energy eigenstates --- one of them being the ground state ---, and, as before, only equiprobable superpositions become orthogonal. It is easily verified that these actually saturate the Margolus-Levitin bound. We remark that, although the two demonstrations parallel one another in many senses, this is the most comprehensive, definitive proof of saturation for the Margolus-Levitin bound, whereas the above can be considered a quite restricted proof of the Mandelstam-Tamm bound given the bulk of results on the subject.

As for Giovannetti et al's $\alpha(F)$-based bound from page~\pageref{boundGiovannetti}, saturation of Eqs.~\eqref{Giovgeral}-\eqref{implicit} has only been characterized numerically, happening at a single instant of time (at most) for each evolution. Furthermore, no average-energy-based bound in the form of Eq.~\eqref{Giovgeral} can be saturable along an evolution in the neighborhood of the initial time (except for the trivial case of a non-evolving state). This is due to the fact, granted by geometrical considerations (see Section~\ref{thebound} for details), 
  that the Mandelstam-Tamm bound is valid as an equality up to second order in time, i.e., always saturates for sufficiently short times (as illustrated in Fig.\ref{grafbasico}). For an average-energy-based bound to saturate in such short times, it would have to equal the Mandelstam-Tamm bound. Since the dependence of the two bounds Eqs.~\eqref{MTfidelidade} and \eqref{Giovgeral} on $F$ is necessarily different (it is straightforward to test $\arccos\sqrt F$ as $\alpha(F)$ in Eq.~\eqref{Giovgeral} to verify it is unfit), they can only be equal when $\Delta E$ and $\braket E$ equal zero, i.e., the trivial case.

\section{Further results} 
\hspace{5mm} The subject has amassed an impressive number of publications and the above is by no means an exhaustive list of the previous works. A frequently approached issue is how entanglement of a multipartite system affects the speed of evolution, and it has been tackled with both bounds~\cite{GiovannettiEPL,GiovannettiJOB,Batle,Borras2006,Zander,Kupferman,Frowis}, with special interest on how evolution speed scales with the amount $N$ of subsystems. For pure states, it has been established that entanglement is a resource that is able to speed up evolution: the time $\tau$ necessary for pure, separable states to reach a given distinguishability scales no faster than $\tau\sim1/\sqrt N$, whereas fully entangled states can reach a $\tau\sim1/N$ scaling. This will be shown in detail in Section~\ref{Nqubit}, but we can advance the following argument: if a pure $N$-partite state begins and remains separable throughout evolution, its Hamiltonian can (effectively) be written as a sum of local Hamiltonians, $H=\sum_iH_i$, each $H_i$ acting only on subsystem $i$
. In this case, the energy variance is additive, $[\Delta E]^2 = \sum_i[\Delta E_i]^2$, which implies a scaling $[\Delta E]^2\sim N$. Applying to the Mandelstam-Tamm result, $\tau$ is bounded by a $\sim1/\sqrt N$ scaling for separable states. On the other hand, a fully entangled initial state $(\ket{000...0}+\ket{111...1})/\sqrt 2$ evolves under a specific choice of local Hamiltonian with the fidelity relative to the initial state being a function of the product $(N\tau)$, so that it reaches a $\tau\sim1/N$ scaling and breaks the bound valid to separable states. The actual demonstration is found in Section~\ref{Nqubit}.

The application of the bound to mixed states has received quite some attention~\cite{Uhlmann,GiovannettiPRA,GiovannettiEPL,Borras2006,Kupferman,Deffner,Frowis}. These works analyze how a mixed initial state evolves under a possibly time-dependent Hamiltonian, tackling 
 unitary evolutions, unable to reduce the purity of a state in any way --- i.e., evolutions describing closed systems. This approach can be useful for systems with a given constant degree of mixture, but cannot describe processes actually responsible for mixing a state. At first, the extension of the entanglement/speed-up relation to mixed states were example-based approaches (chiefly on qubits) by Giovannetti et al~\cite{GiovannettiJOB}, Borrás et al~\cite{Borras2006}, and Kupferman et al~\cite{Kupferman} that verified entanglement speeding up evolution~\cite{GiovannettiJOB,Borras2006}, but not in every case~\cite{Kupferman}. It was Fröwis~\cite{Frowis} who demonstrated that entanglement is a necessary condition to the speed-up of the unitary evolution of an $N$-qubit system: fully separable states are bounded by a $\tau\sim1/\sqrt N$ scaling, while entangled states can be faster, possibly reaching $\tau\sim1/N$. More details will be given in Section~\ref{Nqubit} after discussing the quantum Fisher information, a quantity first brought to the topic of quantum speed limits by Uhlmann~\cite{Uhlmann} and instrumental to Fröwis's result, as well as ours.

Moreover, there have been different inequalities based on higher moments of the energy distribution~\cite{LuoZhang,Andrews,ZielinskiZych,Yurtsever,FuLiLuo,Brody}, as well as other definitions of energy dispersion~\cite{Chau}. We also mention relevant works~\cite{LuoBr,BrodyHookBr,YungBr,CarliniBr6,CarliniBr7,CarliniBr8,OliveiraBr,BorrasBr,MostBr,ZhaoBr} on the related brachistochrone problem: given initial and final states, what Hamiltonian drives the system the fastest? This is particularly problematic on the so-called adiabatic paths, for which the usual route requires slow driving~\cite{BornFock,adiabatico}. The quantum speed limit has been applied under that restriction in~\cite{Andrecut}.

\section{What had not been done}
\hspace{5mm} All of the aforementioned papers have in common that they are applicable solely to unitary evolutions. In spite of the comprehensive literature on the subject, very few contributions treat the more realistic, and more general, non-unitary evolutions. As exceptions we mention Beretta~\cite{Beretta}, who derives a bound based on a nonequilibrium Massieu function, the application of which to describe evolution is admittedly ad hoc; Obada et al~\cite{Obada}, who obtain a speed limit for the specific case of a Cooper-pair box interacting with a cavity field; and Brody et al~\cite{BrodyN}, who briefly discuss speed of evolution in the context of a system with gain or loss. Carlini et al in 2008~\cite{CarliniBr8} interestingly treat the brachistochrone problem searching for paths among unitary and non-unitary evolutions.

The single publication in the literature preceding our work that does discuss a general bound for non-unitary evolution is~\cite{JonesKok} by Jones and Kok, who state that ``there is no quantum speed limit for non-unitary evolution''. However, their proof, based on a counterexample, is flawed and cannot be used to ascertain such a conclusion. The counterexample consists of a qubit, initially in a pure state, on which an interaction is turned on, driving an evolution through mixed states until another pure state, orthogonal to the initial one, is attained, point at which the interaction is turned off. This is accomplished by applying gates that entangle the initial, central qubit with $N$ other, ``satellite'', qubits. Before and after the interaction, the energy of the central qubit is given by a Hamiltonian proportional to a Pauli operator, such that initial and final states have the same energy average $\braket E_0$ and standard deviation $\Delta E_0$. Disregarding the fact that the evolution in between is not governed by this Hamiltonian, the authors expect orthogonalization time to obey $\tau\geq\pi\hbar/2\Delta E_0$ or $\tau\geq\pi\hbar/2\braket E_0$, which can clearly be violated if the interaction energies --- not accounted for in $\Delta E_0$ and $\braket E_0$ --- are high\footnote{This is particularly blatant considering the authors of~\cite{JonesKok} disregard a coupling Hamiltonian, applied to the central qubit $N$ times, whose coupling constant $g$ must be $g\geq2\Delta E_0$ for there to be a violation.}. Although the evolution of the central qubit is indeed non-unitary, the fallacy is reminiscent of a misconception mentioned in the beginning of this Chapter, page~\pageref{appliedfield}, when discussing unitary evolutions: even when initial and final energies only depend on a simpler Hamiltonian, if during evolution some field is applied, this field must be taken into account for the speed limit, since it can, naturally, speed up evolution. The dependence on $\Delta E$ \textit{along} evolution in Eq.~\eqref{MTtimedep} is also due to that factor. The bound we present, valid for non-unitary evolutions, also indirectly refutes their finding.

After our work had been developed, there have been other publications on bounds for non-unitary evolutions. These are discussed in Appendix~\ref{later}.

In the next Chapter, we present our main bound (Section~\ref{thebound}) preceded by important notions leading to it: a short review of necessary quantum information results in Section~\ref{open}, the geometry of quantum state space in Section~\ref{GeoApp}, the (quantum) Fisher information in Section~\ref{Fisher}. Additional bounds obtained are introduced in Section~\ref{additional}. Application of the bounds is left to Chapter~\ref{cap4}.

\end{chapter}

\begin{chapter}{Quantum Speed Limits for General Physical Processes}
\label{cap3}

\hspace{5 mm} This is the Chapter in which we present our main results. We begin by considerations on relevant theoretical foundations: some aspects of quantum information theory are shown in Section~\ref{open}, the geometry of quantum state space in Section~\ref{GeoApp}, and the (quantum) Fisher information in Section~\ref{Fisher}. Our most important bound is presented in Section~\ref{thebound}, product of a work done in collaboration with B.M. Escher, L. Davidovich, and R.L. de Matos Filho and has led to the publication~\cite{MMT}. Additional results are presented in Section~\ref{additional}. Applications of the results are left to the next Chapter.

\section{Useful quantum information tools}
\label{open}
\hspace{5mm} We begin by briefly reviewing some quantum-informational concepts that will be used throughout this Chapter. We will present the Bures fidelity of mixed states; discuss the purification, an important tool for dealing with open-system evolution; and make a comment on POVMs and quantum measurement theory. The contents of this Section can be found in most textbooks on quantum information theory, such as~\cite{NielsenChuang}, and the reader knowledgeable in quantum information may feel free to skip ahead to the next Section.

\subsection{The Bures Fidelity}\label{sectBuresF}
\hspace{5mm}Evolutions of open systems are (in general) non-unitary, and the applicability of our main result to non-unitary evolutions is instrumental to its relevance. Since non-unitary evolutions turn pure states into mixtures, a first prerequisite is a definition of fidelity for mixed states. In the previous Chapter, the fidelity was only defined for pure states by their overlap, so that \mbox{$F(\ket{\psi},\ket{\phi})=|\braket{\psi|\phi}|^2$}. For (possibly) mixed states, described by density operators $\rho$ and $\sigma$, we use the Bures fidelity, defined\footnote{In the literature $F_B$ is sometimes defined as the trace, instead of the trace squared as we use. This is a matter of convention; we prefer the form used in Eq.~\eqref{FBures}, which allows $F_B$ to be  interpreted as a probability.} as
\begin{equation}
F_B(\rho,\sigma) = \left( \Tr\sqrt{\sqrt{\rho} \ \sigma\sqrt{\rho}}\right)^2 \ .
\label{FBures}
\end{equation}
The square root of a positive operator is an operator with the same eigenvectors, but taking the square root of the corresponding eigenvalues. The Bures fidelity obeys the desired properties for a fidelity: i)$F_B$ is symmetric in $\rho$ and $\sigma$, although this is not trivial to show; ii) if $\rho$ and $\sigma$ are orthogonal (i.e., have orthogonal supports, or $\sigma\rho=0=\rho\sigma$), then $F_B(\rho,\sigma)=0$; iii) if $\rho$ and $\sigma$ coincide, $F_B(\rho,\sigma)=1$; iv) for any states $\rho$ and $\sigma$, $0\leq F_B(\rho,\sigma) \leq1$. Moreover, Eq.~\eqref{FBures} recovers, for pure states $\rho=\ket\psi\bra\psi$, $\sigma=\ket\phi\bra\phi$, the previous definition of fidelity, \mbox{$F_B(\rho,\sigma)=|\braket{\psi|\phi}|^2$}. We will see further motivation for this definition later in this Section.

\subsection{Purifications}\label{purif}

\hspace{5mm}A very useful tool in quantum information to describe mixed states is the concept of purification~\cite{NielsenChuang,Schumacher}. Given a state $\rho$ of a system $S$, it is always possible to find a pure state $\ket{\psi_{SE}}$ pertaining to a larger system composed of $S$ and an auxiliary system $E$ such that $\Tr_E(\ket{\psi_{SE}}\bra{\psi_{SE}})=\rho$. In this case, $\ket{\psi_{SE}}$ is said to be a \textit{purification} of $\rho$. The auxiliary system $E$ is called an environment of $S$. All the information relative to the state $\rho$ of $S$ is included in its purification, i.e., $\ket{\psi_{SE}}$ allows for a full description of $\rho$.

Although inspired by physical environments, we emphasize that purifications are, first and foremost, a theoretical construct. In other words, $E$ is a fictitious system that need not have any direct physical significance: we will not be interested in an accurate description of a physical environment. For a given state $\rho$ of $S$, there are infinitely many valid ways to purify it, and our choice of purification is guided by how to describe the system $S$ to the best of our interest, disregarding the actual physical state of $E$ (often favoring simplicity and low dimensionality).

An important reason for the adoption of the Bures fidelity can be understood by purifications. Given two states $\rho$ and $\sigma$, the fidelity $F_B$ is the maximal fidelity between their respective purifications $\ket{\psi_{SE}}$ and $\ket{\phi_{SE}}$, where the maximum is taken over the different purifications possible for each of the states. This result is comprised in Uhlmann's theorem~\cite{UhlmannTh}, which further guarantees the existence of purifications reaching $F_B$ and also that maximization only needs to be performed over the purifications of one of the states, using a fixed arbitrary purification for the other. In mathematical form, Uhlmann's theorem reads
\begin{equation}
F_B(\rho,\sigma) = \max_{\ket{\phi_{SE}}} |\braket{\psi_{SE}|\phi_{SE}}|^2 \ , 
\label{UhlmTheor}
\end{equation}
where $\ket{\psi_{SE}}$ is an arbitrary purification of $\rho$, and the maximum is taken over purifications $\ket{\phi_{SE}}$ of $\sigma$. In fact, the maximization does not even need to be performed over all purifications of $\sigma$, but can be restricted to auxiliary systems $E$ with the same dimension as $S$. Besides in the original paper~\cite{UhlmannTh}, demonstrations of this theorem can be found in~\cite{NielsenChuang,Jozsa}. 

Purifications can be used to describe non-unitary evolutions of quantum states. Given a state $\rho(t)$ of system $S$ which evolves non-unitarily, there are purifications $\ket{\psi_{SE}(t)}$ for every time $t$. The evolution of $\ket{\psi_{SE}(t)}$ is unitary (for the state is always pure), and can be described by a unitary operator $U_{SE}$. Since a purification in $S+E$ conveys a complete description of the state of system $S$, the unitary operator $U_{SE}$ fully characterizes the non-unitary evolution of $S$. There are various other ways to describe non-unitary evolutions, such as Kraus operators~\cite{Schumacher}, Lindblad equations~\cite{NielsenChuang,Schumacher} or master equations~\cite{CohenPhotonsAtoms}, but because of the further importance of purification to our results, we explicitly make use of such unitary operators. We note that this approach of a unitary evolution of a larger system can be used to derive all three aforementioned non-unitary-evolution frameworks.

\subsection{POVMs}\label{sectPOVMs}

\hspace{5mm}Quantum theory not only entails the dynamics of a system, but also its measurement possibilities. It is a standard postulate of the Copenhagen interpretation of quantum mechanics (e.g.~\cite{Cohen}) that, whereas physical observables are represented by operators $O$ on the Hilbert space, measurements on a quantum state $\rho$ are represented by projection operators $\Pi_m=\ket{\phi_m}\bra{\phi_m}$. This is to be understood in the sense that i) any measured quantity will be one of the eigenvalues $\lambda_m$ of the observable $O$ with corresponding eigenstate given by $\ket{\phi_m}$, ii) the state is projected onto $\ket{\phi_m}\bra{\phi_m}$ by the measurement and iii) the probability of finding this measurement outcome is given by $\Tr(\rho\Pi_m)$. Each such measurement corresponds to a set $\{\Pi_m\}$ of projectors onto the eigenstates of $O$; the Hermiticity of observable $O$ implies that this set is composed of \textit{orthogonal} projection operators.
 
This type of measurement, called projective or von Neumann measurement, is nevertheless not the only possible kind. In quantum information, it is useful to deal with a more general set, that of the so-called positive operator-valued measures (POVMs)~\cite{NielsenChuang}. 
 A POVM is composed of elements $E_k$, each corresponding to a measurement outcome $k$ whose occurrence probability is given by $P_k=\Tr(\rho E_k)$. A POVM is a set $\{E_k\}$ meeting two defining properties: i) each $E_k$ is a positive operator and ii) the set obeys a completeness relation of the form $\sum_kE_k=\mathbb{I}$. The simplest examples of POVMs are, in fact, complete sets of orthogonal projectors, but the fact that the $E_k$ need not be orthogonal can be used to one's advantage. Suppose~\cite{NielsenChuang} 
 one is sent a system in one of two non-orthogonal pure states $\ket0$ and $(\ket0+\ket1)/\sqrt2$ (with $\ket0$ orthogonal to $\ket1$) and has to find out which of the two was received. No measurement can perfectly distinguish the two, but the following POVM has an interesting property:
\begin{subequations}\label{exPOVM}\begin{align}
	  E_a & = (2-\sqrt2)  \ket1\bra1 ; \label{POVM1}\\
	  E_b & = (2-\sqrt2) \frac{(\ket0-\ket1)(\bra0-\bra1)}{2} ; \label{POVM2}\\
	  E_c & = \mathbb I-E_a-E_b \ . \label{POVM3}
\end{align}\end{subequations}
This measurement has the benefit of never misidentifying one of the states: the outcome $a$ cannot occur if the state was $\ket0$, so it guarantees that the system was in $(\ket0+\ket1)/\sqrt2$, and vice-versa for outcome $b$. There is, however, a chance of obtaining the inconclusive outcome $c$; this is inevitable, since the two states are not perfectly distinguishable. In any case, employing a POVM is beneficial because it produced reliable results for both states, something orthogonal projectors are unable to achieve.

POVMs, and not orthogonal projectors, are as general as measurements can be according to the postulates of quantum mechanics. But the latter are, in a sense, ``general enough'' for their purpose. Any measurement on a quantum system $S$ can be expressed as the combination of a unitary operation and a von Neumann measurement acting on a larger system composed of $S$ and an ancilla $E$ (which can be thought of as an environment, or as the measuring apparatus). Since the other postulates of quantum mechanics provide the structure necessary for unitary operations and composite systems, every quantum measurement can in principle be understood from the von Neumann perspective\footnote{This seems to justify presenting only this kind of measurement as a postulate, but in reality seldom are actual measurements described with the cumbersome addition of an ancilla. In practice, most measurements are not intended to tell non-orthogonal states apart. Without (need of) such a precise distinction, which is made mostly in quantum information, there is little use for non-orthogonal POVMs.}.  
 In Section~\ref{qFisher}, we will be interested in comparing measurements on a given system $\rho$. We will need to span all possible outcomes and probability distributions, a task for which we must use the fully general POVMs.

We end this Section by mentioning that POVMs allow for an additional way of writing the Bures fidelity, namely
\begin{equation}
F_B (\rho,\sigma) = \min_{\{E_k\}} \left(\sum_k \sqrt{\Tr(\rho E_k)\Tr(\sigma E_k)}\right)^2 = \min_{\{E_k\}} \left( \sum_k \sqrt{P_k(\rho)}\sqrt{P_k(\sigma)} \right)^2 \ ,
\label{BuresPOVM}
\end{equation}
where the minimum is taken over all possible POVMs, and $P_k(\rho)$ is the probability of outcome $k$ occurring with the system in state $\rho$ (respectively for $\sigma$). This can be demonstrated by obtaining a lower bound on $F_B$ using the polar decomposition of $\sqrt\rho\sqrt\sigma$~\cite[p.412]{NielsenChuang}.

\section{Geometric Approach}
\label{GeoApp}

\hspace{5mm} Geometric approaches have greatly contributed to the quantum speed limit, especially by providing a clear interpretation of the bound and a straightforward criterion for its saturation. We first introduce and discuss key aspects of the geometry of Hilbert spaces. This introduction is by no means exhaustive and is rather intended to show the reader knowledgeable in quantum mechanics how a (useful) geometric structure can be built from the space of quantum states. Although no mathematical rigor is lacking in the presented results, demonstrations thereof expressible in familiar quantum-mechanical terms will be favored over more mathematically rigorous versions, which will be referenced along the way. We summarize the main results at the end of the Section.

\subsection{A distance for pure quantum states}\label{distpure}

\hspace{5mm} Any geometric structure of a set is related to the definition of a distance in this set. A distance is a function of two elements of the set, say, $x$ and $y$, and yields a real number $D(x,y)$. The three defining properties of a distance function are that, for any elements $x$, $y$, $z$ of the set,
\begin{subequations}\label{distprop}\begin{align}
	  \mbox{i)}& D(x,y)\geq0 \ \mbox{and} \ D(x,y)=0 \Leftrightarrow x=y \ ; \label{distprop1}\\
	 \mbox{ii)}& D(x,y)=D(y,x) \ \mbox{(symmetric)} \ ; \label{distprop2}\\
	\mbox{iii)}& D(x,z) \leq D(x,y)+D(y,z) \ \mbox{(triangle inequality)} \ . \label{distprop3}
\end{align}\end{subequations}

The first, seemingly innocuous, property has an important consequence to applications to quantum mechanics. Since two vectors of a Hilbert space which are proportional to each other represent the exact same quantum state, we demand a physically meaningful distance function to be null when comparing collinear states, say, $\ket\psi$ and $2e^{i\phi}\ket\psi$. But property i) above requires that any distance rigorously defined in the Hilbert vector space distinguish $2e^{i\phi}\ket\psi$ from $\ket\psi$.

\subsubsection{The projective Hilbert space and its metric}

\hspace{5mm} The best way to circumvent that impediment is to work on the so-called \textit{projective Hilbert space}. A projective space is obtained from a vector space $V$ by identifying vectors which only differ by a nonzero prefactor. The definition of projective space also specifically excludes the null vector. The projective space of three-dimensional flat space 
 can be thought of as the set of straight lines through the origin. Yet another depiction is a unit sphere in which pairs of antipodal points are identified, since each such pair defines the direction of a straight line through the origin. More formally, the projective space is defined as the set of equivalent classes of $V\backslash\{0\}$ by the relation $\sim$, where \mbox{$u\sim v \Leftrightarrow u = \lambda v$} for vectors $u,v\in V$ and $\lambda\neq0$. We note that a projective space is typically not a vector space.

When $V$ is the Hilbert space of quantum states, the projective Hilbert space is obtained. It is a set which does not distinguish $2e^{i\phi}\ket\psi$ from $\ket\psi$ and does not include the null vector in any form.  It is also called the set of ``rays'' of the vector space (in analogy with the straight-line depiction of the projective space), and the common statement that quantum states are given by vectors on the unit sphere coincides with the aforementioned unit-sphere image of the projective space.

Readers familiarized with the Bloch sphere should beware that there is limited analogy between this unit-sphere image of state space and the Bloch sphere, since in the former antipodal points represent the same state and orthogonal states are given by perpendicular rays. Most importantly, vectors within the unit sphere are simply \textit{non-normalized pure states}, not related to mixed states in any way. (Mixed states also obey a form of normalization, $\Tr\rho=1$, and hence would still be on the unit sphere in this unit-sphere representation.) 

The definition of the projective space is extremely suitable to quantum mechanics, which does not physically distinguish collinear vectors and assigns no state to the null vector. This is why it has been pointed out~\cite{Bengtsson} 
 that quantum mechanics is actually described not by the Hilbert space, but by its projective space. Naturally, we still say that quantum mechanics is based on a Hilbert space --- without which we would not be able to talk, e.g., about the dimension of the state space ---, but we recognize that a mathematically more rigorous statement points to the projective Hilbert space as the set of quantum states.


Just as there is a well-known homomorphism between the Hilbert vector space of dimension $n$ and $\mathbb C^n$, the set of $n$-tuples composed of complex entries (usually depicted as column vectors), the projective Hilbert space of quantum states is homomorphic to $P(\mathbb C^n)$, called the \textit{projective complex space}, obtained from $\mathbb C^n$ by the same procedure described above. This can be very helpful because results on our projective space of quantum states can be attained by carrying over results from $P(\mathbb C^n)$, on which the mathematical literature has worked on more profusely~\cite[Vol~II,~pp.133,~159,~273]{KobNomizu}. 

We are interested in Riemannian distances, which, in short, can be obtained from the integration of a differential defined by the Riemannian metric. A metric can be described by a differential form $ds^2$, which indicates how the distance between neighboring points is to be measured (the most usual example of Riemannian metric is that of Euclidean space \mbox{$ds^2=dx^2+dy^2+dz^2$).} Integration of $ds$ along any path will yield a length $\ell=\int_{\rm path}ds$ particular to the chosen path. A Riemannian distance between two elements is the length of the shortest path between them, i.e., the minimum of the length over all paths connecting the elements at hand. This distance is a function only of the two endpoint elements, not of any path between them, and always obeys the properties given in Eq.~\eqref{distprop}.  
We remark that this has been an oversimplified characterization and that there are additional assumptions to the establishment of a Riemannian metric; the interested reader is referred to~\cite[Chap IV]{KobNomizu} for a more complete introduction.

We then need to define a metric valid on the projective Hilbert space. More mathematical treatments achieve this~\cite[Vol~II,~p.159]{KobNomizu} directly on $P(\mathbb C^n)$ by way of general theorems on complex manifolds; we prefer to start from the Hilbert vector space to build a definition befitting the projective space (this is a deliberate choice, there are authors who avoid the projective space almost completely, e.g.~\cite{UhlmannCrell}). In the vector space, a differential can be readily constructed with the usual norm. Let $\ket{d\psi}$ be the infinitesimal variation of state $\ket\psi$, that is, for some parameter $\theta$ on which the state depends, \mbox{$\ket{d\psi}=d\theta (\partial\ket{\psi}/\partial\theta)$}. A simple metric is then given by
\begin{equation}
ds_0^2 = \braket{d\psi|d\psi} \ .
\label{ds2Euclid}
\end{equation}
The integral $\int ds_0$ along a path in the vector space defines a length, from which a distance can be obtained. Nevertheless, $ds_0^2$ is unsuitable for our purposes, since it would assign a nonzero length to paths such as $\ket{\psi(\theta)}=(1+\theta)\ket{\psi_0}$, $\theta\in[0,1]$, and eventually a nonzero distance between parallel states such as $\ket\psi$ and $2\ket\psi$. In other words, $ds_0$ is a metric on the vector space, not on the projective space.

We therefore need a differential form that does not distinguish collinear vectors. It can be constructed by subtracting from $\ket{d\psi}$ its projection on $\ket{\psi}$,
\begin{equation}
\ket{d\psi_\perp} := \ket{d\psi} - \frac{\ket\psi\bra\psi}{\braket{\psi|\psi}}\ket{d\psi} \ ,
\label{dpsiprimario}
\end{equation}
where the fraction above is a normalized projector on $\ket\psi$. Note that $\ket\psi$ is not assumed to be normalized, we instead require that our distance does not distinguish between states differing only by normalization (or phase). It is in fact straightforward to see that collinear variations of $\ket\psi$, of the form \mbox{$\ket{d\psi}=dz\ket{\psi}$} for complex $dz$, yield zero when applied above to any $\ket\psi$. As notation suggests, $\ket{d\psi_\perp}$ is the component of $\ket{d\psi}$ orthogonal to $\ket\psi$. The ``angular variation'' of $\ket{d\psi}$ is defined as
\begin{equation}
\ket{d\psi_{\rm proj}} := \frac{\ket{d\psi_\perp}}{\sqrt{\braket{\psi|\psi}}} = \frac{1}{\sqrt{\braket{\psi|\psi}}}\ket{d\psi} - \frac{\braket{\psi|d\psi}}{\braket{\psi|\psi}^{3/2}}\ket{\psi} \ ,
\label{dpsiortogonal}
\end{equation}
where an overall normalization factor has been included so that $\ket{d\psi_{\rm proj}}$ be a measure of changes on the projective space. The form $\ket{d\psi_\perp}$, orthogonal to $\ket\psi$, can be seen as a circle arc; the extra factor turns it into an angle, in line with the image of the projective space as the set of straight lines (which can be fully characterized by angular components). The need for that extra factor can also be seen by the following example using orthogonal vectors $\ket a$ and $\ket b$. Let us perform the same variation on vectors $0.01\ket a$ and $100\ket a$, which only differ by a prefactor. Say the variation is to add $\ket b$. It is clear that $100\ket a+\ket b$ is quite different from $0.01\ket a+\ket b$, corresponding to different elements of the projective space (equivalently, they correspond to different straight lines). The prefactor added to~\eqref{dpsiortogonal} bears the fact that the greater the modulus of a vector, the smaller its variation by an addition.

The differential form of the distance is given directly by the norm of $\ket{d\psi_{\rm proj}}$,
\begin{equation}
ds_{FS}^2 = \braket{d\psi_{\rm proj}|d\psi_{\rm proj}} = \frac{\braket{d\psi|d\psi}}{\braket{\psi|\psi}} - \frac{|\braket{\psi|d\psi}|^2}{\braket{\psi|\psi}^2} \ ,
\label{ds2FS}
\end{equation}
where the subscript $FS$ denotes that this is the \textit{Fubini-Study metric}, developed independently by Fubini~\cite{Fubini} and Study~\cite{Study}. It can already be seen that this metric is invariant by any unitary $U$ applied to both $\ket\psi$ and $\ket\psi+\ket{d\psi}$. We remark that, although written in terms of vectors, $ds^2_{FS}$ can only be a distance in the projective space, not on the vector space itself.

\subsubsection{Integrating $ds_{FS}$}

\hspace{5mm} We now need to integrate $ds_{FS}$ to obtain the finite distance between two states. This has been done for $P(\mathbb C^n)$ in~\cite{KobNomizu}. Instead of simply carrying over the result, we feel it is instructive to perform the integration and minimization here to explicitly show that, by construction, we do not need to assume any normalization or phase. We do make use of the fact that the geodesic calculated in~\cite[p.277]{KobNomizu} lies entirely in a two-dimensional subspace of the vector space. This bidimensionality can be motivated by the analogy with a unit sphere, whose geodesics --- great circles --- lie in a plane containing the origin~\cite{UhlmannCrell}. Barring the trivial case where initial and final states are collinear, it should be clear that the subspace in question is that spanned by initial and final states.

Let $\ket{\psi_0}$, $\ket{\psi_f}$ denote the initial and final states, respectively, and let $\ket{\psi_1}$ be a state orthogonal to $\ket{\psi_0}$ such that the final state is spanned by $\{\ket{\psi_0},\ket{\psi_1}\}$ (a Gram-Schmidt procedure can always produce a suitable $\ket{\psi_1}$). Since the integration path belongs to the two-dimensional space spanned by $\{\ket{\psi_0},\ket{\psi_f}\}$ or, equivalently, spanned by $\{\ket{\psi_0},\ket{\psi_1}\}$, we can write the state of the system along evolution as
\begin{equation}
\ket{\psi(t)} = f(t) \ket{\psi_0} + g(t) \ket{\psi_1} \ ,
\label{psiintFS}
\end{equation}
where $t$ is a real parameter (not necessarily time), $f$, $g$ are complex functions of $t$, and no normalization is required from $\ket{\psi(t)}$, $\ket{\psi_0}$, $\ket{\psi_f}$ or $\ket{\psi_1}$. We demand $g(0)=0$ and $f(0)\neq0$, such that the initial state of the system be identified with $\ket{\psi_0}$. (Since we only require $\ket{\psi(0)}\sim\ket{\psi_0}$, it is not even necessary that $|f(0)|=1$.) The finite distance $D_{FS}$ is obtained by inserting Eq.~\eqref{psiintFS} into Eq.~\eqref{ds2FS}, integrating $ds_{FS}$ to obtain a length $\ell_{FS}$ and then finding the path (i.e., $f(t)$ and $g(t)$) that minimizes this length. These steps are performed in Appendix~\ref{appFS}, and the result is that the distance between initial state $\ket{\psi_0}$ and final state $\ket{\psi_f}$ measured by the Fubini-Study metric is~\cite{Study}
\begin{equation}
D_{FS}(\ket{\psi_0},\ket{\psi_f}) =\arccos \left( \frac{|\braket{\psi_0|\psi_f}|}{\sqrt{\braket{\psi_0|\psi_0}}\sqrt{\braket{\psi_f|\psi_f}}} \right) \ .
\label{FSarccos}
\end{equation}
We note that $D_{FS}=\pi/2$ if the states are orthogonal, in which case it is maximal. The independence of this distance on the phase and normalization of either state is explicit in the above formula. Wootters~\cite{Wootters} has proposed that this distance also serves as a measure of statistical fluctuations in a sequence of measurements of quantum states. Moreover, $D_{FS}$ is the only Riemannian distance (up to rescaling) invariant by any unitary~\cite{Study,Wootters}.
The argument of the arc cosine above is the square root of the fidelity between the two states, so we can also write
\begin{equation}
D_{FS}(\ket{\psi_0},\ket{\psi_f}) =\arccos\sqrt {F(\ket{\psi_0},\ket{\psi_f})} \ . 
\label{Darccos}
\end{equation}
A consistency check left to the reader is to apply $D_{FS}$ to two neighboring states and obtain\footnote{The relation $\arccos\sqrt{1-x} = x^{1/2}+\mathcal O(x^{3/2})$ may be of help.} $ds_{FS}$.\label{dsD} 

\subsubsection{Geodesics of the Fubini-Study metric}

\hspace{5mm} We can also find the geodesics linking any two given states. Since the geodesics belong to a two-dimensional subspace, we can discuss them using the Bloch sphere\footnote{For readers unfamiliar with the Bloch sphere, it is a standard vector representation of a qubit state very similar to the expectation value $\braket{\vec S}$ of a spin-$1/2$ system, see~\cite[p.168]{Schumacher}.}. From Appendix~\ref{appFS} we know that states of the form Eq.~\eqref{psiintFS} will travel along a geodesic if and only if
\begin{equation}
\frac{g(t)}{f(t)} = \xi (t) \frac{g(t_f)}{f(t_f)} \ ,
\label{fggeod}
\end{equation}
where $\xi(t)$ is any monotonic real function null at $t=0$ and equal to one at $t=t_f$ and $f(t_f)$, $g(t_f)$ are such that the combination in Eq.~\eqref{psiintFS} equals $\ket{\psi_f}$ at $t_f$.

Let us begin by the geodesics linking non-orthogonal states, so that $f(t)\neq0$ for all times and singularities are avoided. Eq.~\eqref{fggeod} then guarantees that the ratio $g(t)/f(t)$ is well-behaved, and we can hence write
\begin{equation}
\ket{\psi(t)} = f(t) \left( \ket{\psi_0} + \frac{g(t)}{f(t)} \ket{\psi_1} \right) \ .
\label{geodqbit}
\end{equation}
We can disregard the overall factor $f(t)$ and substitute Eq.~\eqref{fggeod} in Eq.~\eqref{geodqbit}, obtaining
\begin{equation}
\ket{\psi(t)} =  \ket{\psi_0} + \xi(t) \frac{g(t_f)}{f(t_f)} \ket{\psi_1}  \ ,
\label{geodxi}
\end{equation}
from which one sees that the relative phase between the components is fixed along the path. This necessarily implies that geodesics lie along a great circle of the Bloch sphere. 
 These conclusions can be seen even more clearly if, without any loss of generality, the phase of $\ket{\psi_f}$ is chosen such that $g(t_f)/f(t_f)\in\mathbb R$, making both coefficients of $\ket{\psi(t)}$ real. Bloch vectors with all real components are confined to a plane and, given that these are pure states, to a great circle. 
There are two paths between two non-orthogonal states along a great circle of the Bloch sphere. It is intuitive that the geodesic is given by the shorter of these paths. This is indeed confirmed by Eq.~\eqref{geodxi} in that $\ket{\psi(t)}$ is never orthogonal to $\ket{\psi_0}$.

For orthogonal states, one may use the divisibility of geodesic paths, i.e., the fact that any (compact) section of a geodesic is also a geodesic. Let $\ket{\psi_f}$ be orthogonal to $\ket{\psi_0}$. Because every vector in the vicinity of $\ket{\psi_f}$ is linked to $\ket{\psi_0}$ through a great-circle geodesic, the (infinitely many) great circles linking $\ket{\psi_0}$ to $\ket{\psi_f}$ are the only suitable candidates for the geodesics. It can be easily seen that all such great circles are, in fact, geodesics between the orthogonal states.

We observe that, although the geometry of pure quantum states shown so far is analogous to Euclidean geometry on the Bloch sphere, the analogies break down with the inclusion of mixed states.

\subsubsection{Comments on another distance}\label{another}

\hspace{5mm} We now take a moment to mention another function present in the discussion of distances in quantum state space, the Bures distance\footnote{The nomenclature becomes quite cumbersome here: the \textit{Bures distance} is different from the so-called \textit{Bures angle} (a.k.a. \textit{Bures arc} or even \textit{Bures length}, even though it is a distance, not a length, by geometric definitions), which is a generalization of the Fubini-Study distance; all of them can be written in terms of the \textit{Bures fidelity}.}
\begin{equation}
D_{\rm Bures}(\ket{\psi_0},\ket{\psi_f})  = \sqrt 2 \sqrt{1-\sqrt{F(\ket{\psi_0},\ket{\psi_f})}} \ .
\label{distBures}
\end{equation}
This is a distance in the projective Hilbert space in the sense that it obeys the axioms of Eq.~\eqref{distprop}, but it is not Riemannian, because it cannot be obtained as the shortest path along elements of the \textit{projective} space. It is possible, though, to obtain $D_{\rm Bures}$ as the length of a path along the Hilbert \textit{vector} space. The distinction matters because this path must necessarily go through nonnormalized vectors and this length is only obtained when distinguishing collinear vectors (differing only by normalization). This is shown in Fig.~\ref{grafpqnaoBures}, which represents the vector space as a plane: only through the straight-line (solid) path is $D_{\rm Bures}$ attained. Were the path traveled on the unit circle (dashed line), it would coincide with a special case of the Fubini-Study metric, and $D_{FS}$ would be found\footnote{We warn the reader familiarized with the Bloch sphere that, contrary to first intuition, intermediate states like $\ket{\psi_m}$ in Fig.~\ref{grafpqnaoBures} are pure, nonnormalized states.}.
\begin{figure}[ht]%
\centering
\includegraphics[width=.55\columnwidth]{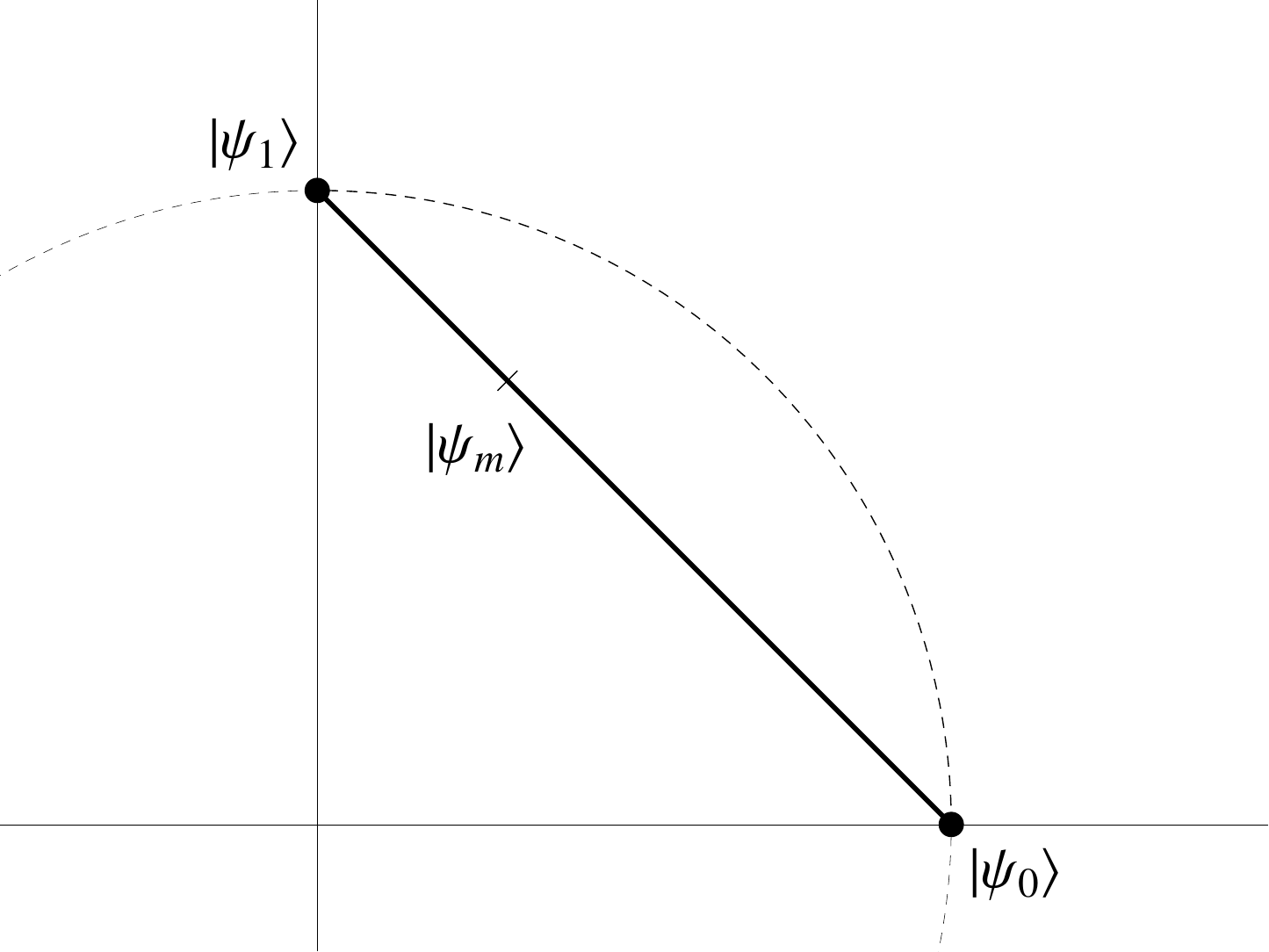}%
\caption{Depiction of a two-dimensional Hilbert space. Integration on the straight line yields $D_{\rm Bures}$, but is unsuited for our purposes, because the distance thus obtained would necessarily distinguish collinear vectors. We note that states inside the unit circle (dashed line), like $\ket{\psi_m}$, are pure, but not normalized.}%
\label{grafpqnaoBures}%
\end{figure}
 This assertion can perhaps be seen best by the fact that the above form is only valid for unit vectors. If we write this distance for vectors of different norm, such as $\ket{\psi_m}$ of Fig.~\ref{grafpqnaoBures}, we get
\begin{equation}
D_{\rm Bures}(\ket{\psi_0},\ket{\psi_m})  = \sqrt{\braket{\psi_0|\psi_0}+\braket{\psi_m|\psi_m}-2|\braket{\psi_0|\psi_m}|} \ .
\label{distBuresnn}
\end{equation}
In other words, the Bures distance is only Riemannian if one distinguishes between collinear vectors. If restricted to unit vectors, it is a valid distance on the projective Hilbert space, but loses its Riemannian character\footnote{For the sake of completeness, we mention that $D_{\rm Bures}$ is the shortest length on the quotient space obtained by identifying vectors which differ only by a phase, as can be seen by the phase-independence of Eq.~\eqref{distBuresnn}.}. Most importantly, it certainly cannot be obtained from $ds_{FS}^2$ (Eq.~\ref{ds2FS}), which is independent of normalization. We will see shortly that the quantities we are interested in must be integrated in the projective space, i.e., without regard to overall normalization, and hence $D_{\rm Bures}$ will not be of much interest to us\footnote{We also note that the Hilbert-Schmidt distance for mixed states, $\sqrt{\Tr[(\rho-\sigma)]}$, reduces to Eq.~\eqref{distBuresnn} for pure states and is subject to the same limitations as the Bures distance.}.


\subsection{Mandelstam-Tamm bound obtained geometrically}

\hspace{5mm} Anandan and Aharonov~\cite{AA,Anandan} have put the knowledge of quantum-state geometry to use in the pursuit of a quantum speed limit. This can be done as follows: let us calculate the differential form $ds_{FS}^2$ of the Fubini-Study metric assuming that the variation of $\ket\psi$ is governed by Schrödinger's equation with Hamiltonian $H$. We have
\begin{equation}
\ket{d\psi(t)} = \ket{\psi(t+dt)} - \ket{\psi(t)} = \frac{H}{i\hbar} \ket{\psi(t)}dt \ ,
\label{dpsischr}
\end{equation}
where now $t$ is, in fact, time. Substitution into Eq.~\eqref{ds2FS} yields
\begin{equation}
ds_{FS}^2 = \frac1{\hbar^2} \left[ \frac{\braket{\psi|H^2|\psi}}{\braket{\psi|\psi}} -  \left( \frac{\braket{\psi|H|\psi}}{\braket{\psi|\psi}} \right)^2 \right] dt^2 = \frac{[\Delta E]^2dt^2}{\hbar^2}\ ,
\label{ds2H}
\end{equation}
where the variance of the Hamiltonian (written for non-normalized $\ket\psi$) can be recognized in the square brackets above. The independence of this distance from global phase and normalization is also evident from this expression. If we integrate this differential along the states through which the system goes during evolution, we obtain the length of the path followed by the state:
\begin{equation}
\ell_{FS} = \int ds_{FS} = \int_0^\tau \frac{\Delta E(t)}{\hbar}dt \ .
\label{comprimento}
\end{equation}

The Mandelstam-Tamm bound, within this framework, amounts to stating that the distance $D_{FS}$ between two states is, by definition, shorter than or equal to the length $\ell_{FS}$ of any path connecting them. In mathematical terms,
\begin{equation}
D_{FS} = \arccos\sqrt{F(\ket{\psi(0)},\ket{\psi(\tau)})} \leq \int_0^\tau \frac{\Delta E(t)}{\hbar}dt \ ,
\label{MTgeom}
\end{equation}
recovering Eq.~\eqref{MTfidelidade}. We note that this demonstration is valid for pure states evolving unitarily.

A slightly different interpretation of the above formula is more useful for the obtention of bounds on evolution time. In a given evolution, the most a final state can be distant from the initial one is the length of the path followed by the system. Accordingly, the minimal time necessary for the distance to reach a chosen value (or, equivalently, for the fidelity to reach a chosen value) is the time at which the path length reaches this value. This can be seen with the help of the graph on Fig.\ref{grafintegral}, where $\Delta E(t)/\hbar$ is plotted against time. Given a certain value $D_1$ of the distance, the minimal time $\tau$ needed to reach it is that which makes the area under the curve $\Delta E(t)/\hbar$ equal $D_1$.
\begin{figure}[ht]
\centering%
\includegraphics[width=.5\columnwidth]{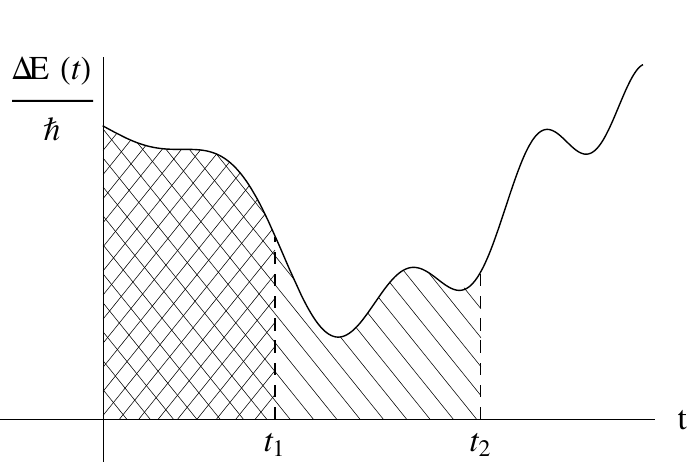}%
\caption{Example of plot of $\Delta E(t)/\hbar$ over time $t$. The bound on the time for a certain distance $D_1$ to be reached is given by the value $t=\tau$ such that the area under the graph equals $D_1$.}%
\label{grafintegral}%
\end{figure}

A breakthrough of the geometric interpretation of the bound was to enable a clear, straightforward criterion for its saturation: it occurs when evolution follows a geodesic. We have seen that geodesics are great circles of the Bloch sphere. On the other hand, any Hamiltonian evolution of a qubit is locally a rotation around some axis of the Bloch sphere. A state evolving along a geodesic must then perform a rotation around a fixed axis and, furthermore, belong to a great circle, which must, then, be the equator relative to this axis.  The only pure states that obey these conditions and hence saturate the Mandelstam-Tamm inequality for unitary evolutions are equiprobable on any pair of energy eigenstates $\ket{E_i}$, $\ket{E_j}$, or
\begin{equation}
\ket{\psi(t)} = \frac{1}{\sqrt 2} \left( \ket{E_i} + e^{-i\phi(t)} \ket{E_j} \right) \ ,
\label{statessaturate}
\end{equation}
where the phase $\phi(t)$ has a monotonic dependence on $t$. Although $H$ must be a rotation around a fixed axis, there can be a time dependence of the form $H(t)=\hbar\omega(t)\sigma$, where $\sigma$ is the Pauli operator relative to the direction of the axis, as long as $\omega(t)$ does not change sign. The phase $\phi(t)$ is related to $H(t)$ by $\phi(t)=\int_0^t \omega(t')dt'$.

\subsubsection{Summary}

\hspace{5mm} The main results of this Section are that Anandan and Aharonov recognized $\Delta E(t)dt/\hbar$ as the differential of a well established metric, the Fubini-Study metric. This allowed the bound to be reinterpreted as stating that the length of any path is greater than or equal to the distance between its endpoints. The (finite) distance measured by this metric is \mbox{$D_{FS}(\ket{\psi_0},\ket{\psi_f}) =\arccos\sqrt {F(\ket{\psi_0},\ket{\psi_f})}$}, whereas the length is the integral of $\Delta E(t)dt/\hbar$, and so the inequality is proved in geometric terms. Saturation occurs on geodesic paths, where the length equals the distance between the endpoints.

We end this Section with a simple analogy. The quantity $\Delta E(t)/\hbar$ is the speed of evolution. It is analogous to what is measured by the speedometer of a car, in that it does not distinguish the direction of the path in any way. Given a traveled course, the information given by the speedometer allows one to calculate the traveled length $\ell$ along the path and, furthermore, the time $\tau$ it takes to travel such length. Since the actual straight-line (geodesic) distance $D$ between starting and finish points may not be greater than the length ($D\leq\ell$), $\tau$ is a lower bound on the time needed to become $D$ apart from the starting point. Only if the path is traveled in a straight line do length and distance coincide.

\section{Fisher information}
\label{Fisher}

\hspace{5mm} We now introduce a quantity which is of great importance to our work, the quantum Fisher information. We begin by a predecessor that naturally appears in estimation theory, the Fisher information (sometimes called  ``\!\textit{classical} Fisher information'' in contrast to the quantum Fisher information, in a bit of a misnomer). 

Historically, the Fisher information arose in the study of the method of maximum likelihood, which is a procedure, started by Fisher in 1912~\cite{Fisher12}, to find the best fit of given data to a theoretical curve with free parameters. This method rivals the minimum variance unbiased estimator, which basically consists in the least squares method constrained to unbiased estimators, as well as the method of moments, which fits $n$ parameters of a theoretical curve so that the $n$ first moments of curve and data coincide (for more on these methods, see~\cite{Kay}). The importance of the Fisher information goes beyond this particular method as it has relevant properties concerning any estimation, as we will see by the Cramér-Rao bound, Eq.~\eqref{CramerRao}. Already suggested in 1922 by Fisher in works such as~\cite{Fisher22}, he explicitly called it an information in 1925 in~\cite{Fisher25}, then later again in 1934 in~\cite{Fisher34}, among others\footnote{Due to the plethora of works by Fisher in those decades, see~\cite[p.564]{Cramer}, it is not easy to pinpoint all the appearances of the Fisher information in the literature of the time.}. A more detailed introduction to both ``classical'' and quantum Fisher information starting from the maximum-likelihood estimation can be found in~\cite{Brunotese} (in Portuguese).

The Fisher information is useful when there is some parameter $x$ that is to be estimated. This parameter has a true value $x_{\rm true}$, a value that one can never exactly obtain, but is considered to exist. The definition further assumes that one performs some sort of procedure related to the parameter, which produces an outcome $k$ out of a given set $\{k\}$. These outcomes can occur with certain probabilities $P_k$, which should depend on the true value $x_{\rm true}$ of $x$; this dependence is denoted by $P_k(x_{\rm true})$. Since $x_{\rm true}$ is not known, one must work, for every possible value of $x$, with $P_k(x)$, the probability distribution as if the parameter had true value $x$. For each $x$, the distribution is normalized as $\sum_{k}P_k(x)=1$. One can naturally repeat the estimation many times to enhance the result. Adaptations for the case of a continuous distribution are straightforward, but do not concern us here.

The estimation procedure itself cannot yield $P_k$ or $x_{\rm true}$. The $P_k(x)$ must be obtained for each $x$, usually by models, and the actual outcomes are then compared to predictions from the models. An estimation can be understood as a way to extract, from the outcome(s) $k$ that has(have) been obtained, a value for the parameter $x$. The simplest form of doing that is perhaps to average the numerical values obtained (a subcase of the method of moments). The maximum-likelihood estimation searches instead for the true value of $x$ which would be most likely to produce the observed outcome(s) by maximizing with respect to $x$ a certain function of the probabilities $P_k(x)$ with $k$ restricted to the set of \textit{observed} outcomes (see~\cite[Sect.2.1.1]{Brunotese} or~\cite[p.498]{Cramer}).

A well-defined estimation procedure then entails a parameter $x$ with an unknown true value $x_{\rm true}$, an outcome set $\{k\}$, and a set of outcome probabilities $\{P_k(x)\}$ for each possible value $x$. The Fisher information with respect to parameter $x$ is defined as
\begin{equation}
F(x) := \sum_{k} P_k(x) \left( \frac{d}{dx} \ln P_k(x) \right)^2 \ .
\label{FisherCl}
\end{equation}
It is intuitive that, for an actual estimation to be relevant, the probability distribution of the outcomes $k$ must in some way depend on $x$. The presence of the derivative of $P_k(x)$ expresses this necessity of an $x$-dependence of $P_k(x)$ for the Fisher information to be nonzero: if the pointer of your meter does not depend in any way of the parameter $x$, it cannot convey any information on $x$. To say that ``even a broken clock is right twice a day'' may be heartwarming or funny, but one knows better than to try to estimate time from it (by ``broken clock'' we consider a clock that has come to a full halt). If a broken clock displays, e.g., the time 9:47, its outcome probabilities are $P_{9:47}=1$ and $P_k=0$ for any other time. The clock has no (Fisher) information on the actual time because these probabilities do not depend on it.

Another useful expression for the Fisher information is
\begin{equation}
F(x)  = \sum_{k} \frac{1}{P_k(x)} \left( \frac{d}{dx} P_k(x) \right)^2 \ ,
\label{Fishersemlog}
\end{equation}
which is directly implied by Eq.~\eqref{FisherCl}.

Let us give a physical example. Estimation of physical parameters is typically made by experiments. The outcomes $k$ can correspond to the sets of raw data obtained after each repetition, but it is more insightful to assign to $k$ the obtained numerical values of the parameter at hand. Let us say a group of undergraduate students is tasked with finding the gravitational acceleration $g$ on the surface of the Earth by an inclined-plane experiment\footnote{Readers familiarized with the undergraduate Physics courses of the University should readily identify the reference to ``FisExp''.}. Imprecision is naturally present due to imperfection on the measurement devices, finite-sized markings for registering position, parallax, among many others\footnote{There may even be a certain contribution due to subideal interest by the students, on rare occasions.}. Considering that uncertainty only allows $g$ to be estimated up to a single decimal place in SI, we expect a higher probability of reaching values around $k=9,8m/s^2$, but the set of possible outcomes $k$ is the set of values spaced by the precision of $0,1m/s^2$,
\begin{equation}
k \in \{0,\!0\tfrac{m}{s^2} ; \ 0,\!1\tfrac{m}{s^2} ;  \ 0,\!2\tfrac{m}{s^2} ; \ 0,\!3\tfrac{m}{s^2} ; ... ; \ 9,\!7\tfrac{m}{s^2} ; \ 9,\!8\tfrac{m}{s^2} ; \ 9,\!9\tfrac{m}{s^2} ; \ 10,\!0\tfrac{m}{s^2} ; \ 10,\!1\tfrac{m}{s^2} ; \ 10,\!2\tfrac{m}{s^2} ; ... \} \ .
\label{cjto}
\end{equation}
Such a coarse graining by precision limitations will always be present in physical experiments and is the reason why we are concerned with discrete outcome sets $\{k\}$.

It is crucial that the probability of obtaining a given outcome depend on the true value $g_{\rm true}$. Suppose the estimation is made in two different places on Earth, the high mountains of Huascarán, in the Peruvian Andes, and the Arctic Sea, such that $g_{\rm true}$ is different in these cases. Furthermore, in addition to the simple didactic laboratory used by the undergraduate students, the measurement of $g$ in each of those places is also done by higher-precision satellites, allowing for an extra decimal place\footnote{For completeness, we mention that actual satellite experiments, as in~\cite{geo}, can be a thousand times more precise; the argument does not depend on such a high precision, though.}. In Fig.\ref{fisexp}, we show a simulation of the distributions of $P_k(g_{\rm true})$ in the four cases, assuming $g_{\rm true}=9,76392m/s^2$ (Huascarán, Andes) and $g_{\rm true}=9,83366m/s^2$ (Arctic Sea); these two values are inspired by measured data taken from~\cite{geo}. The $P_k$ or $g_{\rm true}$ cannot be obtained by measurements; instead, models (of measurement and dynamics) assign $P_k$ for each possible $g$, and the actual measurements are then compared to distributions from the models. 
\begin{figure}[ht]
\centering
\includegraphics[height=.25\columnwidth]{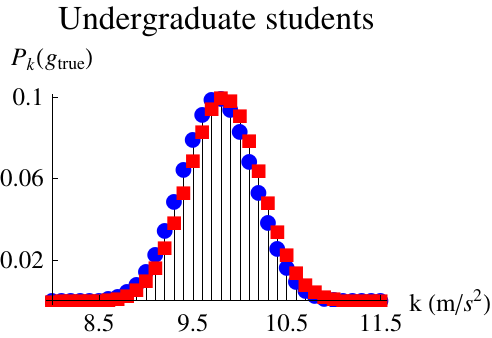} \includegraphics[height=.25\columnwidth]{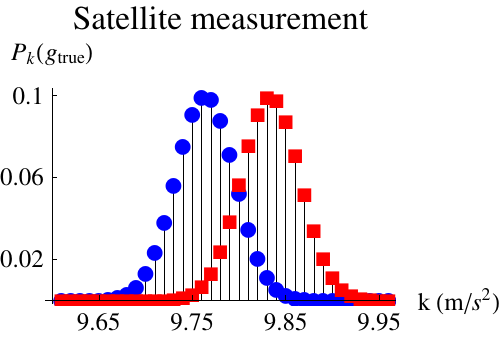}
\caption{Values of $P_k(g_{\rm true})$ for a measurement of $g$ in Huascarán, in the Andes (blue circles) and at the Arctic Sea (red squares). The distributions within each image are different because so is $g_{\rm true}$. Measurement as made in a simple laboratory (left) is compared to that by higher-precision satellites (right).}%
\label{fisexp}%
\end{figure}
The capability of $P_k(g)$ to discern a minute difference in the true value of $g$ is essential for a precise estimation. The higher precision of the satellite experiments is translated in the way it is easier to distinguish the true values of $g$ from the $P_k$ of these measurements.
The Fisher information is a measure of this very property: how much does the outcome distribution change by a change of the underlying true value of the parameter?

This interpretation is helped out by the definition of the Hellinger distance, a distance between probability distributions $\{P_k\}$ for a given set $\{k\}$ of outcomes. If one assumes that two probability distributions are different because they belong to two different values $x$, $x'$ of the parameter, the Hellinger distance $D_H$ can be written as a function of these values~\cite[p.33]{Llorente} (or~\cite[p.16]{Brunotese})
\begin{equation}
D_H(x,x') = \sqrt{\frac 1 2 \sum_{k} \left[\sqrt{P_k(x')}-\sqrt{P_k(x)}\right]^2} \ .
\label{Hellinger}
\end{equation}
We remark nonetheless that $D_H$ is a distance only on the set of probability distributions. It obeys Eq.~\eqref{distprop}, and especially Eq.~\eqref{distprop1}, for the probability distributions $P_k$, $P_k'$, not necessarily for parameter values $x$, $x'$. In other words, $D_H(x,x')$ does not measure variations of the parameter $x$, but instead how distinguishable the distribution $P_k(x)$ becomes if $x$ is varied. If distinct parameter values correspond to the same distribution, as in the broken clock, $D_H$ will be zero. 

These aspects of the Hellinger distance are analogous to some of what has been said about the Fisher information, and we now show how they are mathematically related. The first step is to rewrite the Fisher information as
\begin{equation}
F(x) = 4 \sum_{k} \left( \frac{d}{dx}\sqrt{P_k(x)} \right)^2 \ ,
\label{Fisherraiz}
\end{equation}
and then to write $D_H$ for infinitesimally separated arguments $x$, $x+dx$:
\begin{equation}
D_H^2(x,x+dx) = \frac 1 2 \sum_{k} \left[\sqrt{P_k(x+dx)}-\sqrt{P_k(x)}\right]^2 = \frac 1 2 \sum_{k} \left[\frac{d}{dx}\sqrt{P_k(x)}\right]^2 dx^2 \ .
\label{Hellingerinfinitesimal}
\end{equation}
By comparing the two equations, it is easy to see that, to lowest order in $dx$,
\begin{equation}
D_H^2(x,x+dx) = ds_H^2 = \frac{F(x)}8 \ dx^2 \ ,
\label{Fishercomodistancia}
\end{equation}
i.e., the Fisher information is the core of the infinitesimal form of that distance, measuring how probability distributions grow apart due to an infinitesimal variation of the parameter.
 
Another sign of the importance of the Fisher information is that it bounds the precision of any parameter estimation. This connection is established by the Cramér-Rao bound, shown independently by Rao~\cite{Rao} and Cramér~\cite[p.480]{Cramer}, although some results had been anticipated by Fisher, as recognized in~\cite{Rao47}. The Cramér-Rao relation states that the error $\delta x$ of an estimation is bounded from below:
\begin{equation}
\delta x \geq \frac{1}{\sqrt{F(x)}} .
\label{CramerRao}
\end{equation}
Our interest will be to use the Fisher information not in estimation theory, but as (the square of) a speed related to the geometry of quantum states, so the reader interested in this result (and its proof) is referred to a more detailed discussion to be found in~\cite[Sect.2]{BJP} or in~\cite[Sect.2.1.2 onwards]{Brunotese}, which are works dedicated specifically to parameter estimation.

\subsection{The quantum Fisher information}
\label{qFisher}

\hspace{5mm} The Fisher information can naturally be applied to the estimation of a parameter via a physical measurement, be it on a classical or quantum system. Although no estimation, in either case, can be exact, there is an important difference in the sources of imprecisions in the classical and quantum scenarios. Classically, it is the imperfection of the measurement/estimation process that introduces errors; classical states themselves would allow for perfect estimations if probed by perfect, idealized apparatus. In quantum mechanics, indeterminacy originates additionally from the probabilistic nature of quantum states. Even with ideal measurements, one cannot in general obtain complete information about a parameter on which a quantum state depends, because it is not possible, in general, to distinguish between two non-orthogonal quantum states. The quantum Fisher information aims to quantify how much about a certain parameter can be learned by a quantum state that depends on it, assuming the best measurements possible. The classical counterpart to this quantity would be trivially infinite --- different classical states are always perfectly distinguishable ---; the goal here is to measure how imperfect the information possessed by a quantum state is (alternatively, how finite it is).

An important difference between the original Fisher information and its quantum version is that, whereas the former is defined for each specific estimation procedure, the latter is a function solely of the quantum state and of its dependence on the parameter to be estimated. This is possible because the Fisher information requires a set of outcomes and its occurrence probabilities, while quantum measurement theory, by indicating what the possible measurements are, supplies both of these. The quantum Fisher information is, in fact, the Fisher information of the best measurement available on the quantum state.

Let us now give an introductory example, admittedly oversimplified. Consider that one wishes to measure the coupling constant $\lambda$ of an interaction between a field and a two-level atom\footnote{The interaction in this example is given by the Jaynes-Cummings model on resonance.}. Atom and field will be left to interact for a certain time $T$, and the experiment makes a measurement solely on the atom. The main idea is that the atom starts in a state $\ket a$ and is driven to $\ket b$ by the field. The stronger the coupling constant $\lambda$, the closer to $\ket b$ the final state will be. Let us assume for the sake of simplicity that there is a maximal value to $\lambda$, which is $\lambda=1$, and that the total interaction time $T$ is so chosen that the final state will be exactly $\ket b$ if $\lambda=1$. Other values of $\lambda$ will leave the state after $T$ in a mixture of $\ket a$ and $\ket b$,
\begin{equation}
\rho_T = \left( \begin{array}{cc} \lambda & 0 \\ 0 & 1-\lambda \end{array} \right) \ ,
\label{rhoT}
\end{equation}
with the representation $\ket a= \genfrac(){0pt}{1}{0}{1}$, $\ket b= \genfrac(){0pt}{1}{1}{0}$.

By letting the system interact for a time $T$ and measuring $\rho_T$ in this basis, $\{\ket a,\ket b \}$, one can make a crude estimate of the value of $\lambda$. Since $p_a=1-\lambda$ and $p_b=\lambda$, the Fisher information for this estimation is
\begin{equation}
F_{\{a,b\}}(\lambda) = \frac1{p_a}\left(\frac{\partial p_a}{\partial \lambda} \right)^2 + \frac1{p_b}\left(\frac{\partial p_b}{\partial \lambda} \right)^2 = \frac{(-1)^2}{1-\lambda} + \frac{1^2}{\lambda} = \frac1{\lambda(1-\lambda)} \ ,
\label{exFisher}
\end{equation}
where the subscript $\{a,b\}$ indicates the measurement basis. One can see from this expression that the estimate is worst for values of $\lambda$ close to $1/2$, as expected\footnote{A possibility to reduce error is, of course, to repeat the experiment multiple times. It has been shown~\cite[p.17]{Brunotese} that the Fisher information is \textit{additive}, i.e., in $\nu$ independent runs of an estimation, $F_\nu(x)=\nu F_1(x)$.}. Due to the idealizations in this model, when $\lambda$ equals $0$ or $1$, the measurement outcome is certain. This is reflected by the behavior $F(\lambda)\rightarrow\infty$ on these two values, which corresponds to the possibility of making an error-free measurement. A more realistic description would be to assume some degree of irremovable mixture on the initial and final states, which would produce a well-behaved $F(\lambda)$. We nevertheless keep treating this idealization in order to present some of the important ideas.

As crude as this estimation may be, it is certainly better than to measure in the ``rotated'' basis $\{\ket+,\ket-\}$, with $\ket\pm:=\frac1{\sqrt2}(\ket a\pm\ket b)$. The outcomes on this basis have no dependence on $\lambda$ ($p=1/2$ for both measurements for every $\lambda$) and hence yield no information whatsoever on the parameter, $F_{\{\pm\}}(\lambda)=0$. It can be shown that the original basis $\{\ket a,\ket b\}$ is, in fact, the most capable of distinguishing values of $\lambda$. The quantum Fisher information for the estimation of $\lambda$, denoted $\mathcal F_Q(\lambda)$, of the state $\rho_T$ from Eq.~\eqref{rhoT} is precisely the Fisher information of the best measurement,
\begin{equation}
\mathcal F_Q(\lambda) = F_{\{a,b\}}(\lambda) = \frac1{\lambda(1-\lambda)} \ .
\label{qfilambda}
\end{equation}

Now turning to a more formal discussion, we know from Section~\ref{sectPOVMs} that all possible measurements on a quantum system are given by POVMs.  In the above example, simple orthogonal projectors were used, $E_a = \ket a\bra a$, $E_b = \ket b \bra b$, $E_+ = \ket+\bra+$, $E_-=\ket-\bra-$, and the POVMs employed to obtain the Fisher information were $\{E_a,E_b\}$ for $F_{\{a,b\}}(\lambda)$ and $\{E_+,E_-\}$ for $F_{\{\pm\}}(\lambda)$. In general, the quantum Fisher information for estimation of parameter $x$ is defined as the maximum of the Fisher information over all POVMs, i.e.,
\begin{equation}
\mathcal F_Q(x) := \max_{\{E_k\}} \left\{ \left. F(x) \right|_{\{E_k\}}\right\} = 
\max_{\{E_k\}} \left\{ \sum_k \frac1{\Tr\left(\rho(x)E_k\right)} \left[\frac d{dx} \Tr\left(\rho(x)E_k\right) \right]^2 \right\} \ ,
\label{FisherQ}
\end{equation}
where we have used that $P_k(x)=\Tr(\rho(x) E_k)$.

The Cramér-Rao bound is still valid for $\mathcal F_Q(x)$, but it is now a bound not on the error of a specific estimation procedure, but on the error of any estimation of $x$ from state $\rho(x)$. The quantum Cramér-Rao bound reads
\begin{equation}
\delta x \geq \frac{1}{\sqrt{\mathcal F_Q(x)}} \ .
\label{qCramerRao}
\end{equation}

We are interested, however, in viewing the quantum Fisher information in a different light. We have seen that it is a quantity depending only on the state $\rho$ and on its dependence on the parameter $x$. The example given in Fig.~\ref{fisexp} shows that the ``classical'' Fisher information can measure how much probability distributions grow apart as the parameter varies. Since the quantum version maximizes the Fisher information over all possible measurements, $\mathcal F_Q$ should be sensitive to any variation of $\rho$. In fact, we now show an important result that substantiates this interpretation.

Suppose one is interested in distinguishing two neighboring states, characterized by parameter values $x$ and $x'$, using measurements indicated by the POVM $\{E_k\}$. The Hellinger distance~Eq.\eqref{Hellinger} between the two probability distributions is
\begin{equation}
D_H^2(x,x')=\frac12\sum_k\left(\sqrt{P_k(x)}-\sqrt{P_k(x')}\right)^2 = 1 - \sum_k \sqrt{P_k(x)}\sqrt{P_k(x')} \ .
\label{HellBures1}
\end{equation}
Now let us assume that $x$ and $x'$ are infinitesimally separated, $x'=x+dx$. We know from Eq.~\eqref{Fishercomodistancia} that $D_H^2(x,x+dx)$ is related to the (``classical'') Fisher information $F(x)$, so that
\begin{equation}
1 - \sum_k \sqrt{P_k(x)}\sqrt{P_k(x+dx)} = \left. \frac{F(x)}{8}\right|_{\{E_k\}}dx^2 \ .
\label{preBuresFisher}
\end{equation}
The subscript on the right-hand side serves as a reminder that this equation is valid for any POVM. By choosing the POVM that maximizes this equation,  both left- and right-hand sides at the same time. By writing the above for such a POVM, we obtain, due to the definition of the quantum Fisher information and to Eq.~\eqref{BuresPOVM},
\begin{equation}
1 - \sqrt{F_B[\rho(x),\rho(x+dx)]} = \frac{\mathcal F_Q(x)}8 dx^2 \ ,
\label{preBuresFisher2}
\end{equation}
or, solving for $F_B$,
\begin{equation}
F_B[\rho(x),\rho(x+dx)] = 1 - \frac{\mathcal F_Q(x)}4 dx^2 +\mathcal O(dx^4) \ .
\label{BuresFisher}
\end{equation}
This very important relation shows that, as we have mentioned less formally, the quantum Fisher information is a measure of the rate at which a state becomes distinct following the variation of a parameter. When the parameter in question is time, $\sqrt{\mathcal F_Q(t)}$ can be understood as a speed of the evolving state. This result will be relevant for our geometrically-derived bound in the next Section.

Another important expression for the quantum Fisher information can be obtained by the use of the symmetric logarithmic derivative $L(x)$~\cite{Helstrom,Holevo}. For a given parameter $x$ and a given state $\rho(x)$, $L(x)$ is the Hermitian operator implicitly defined by
\begin{equation}
\frac{d}{dx} \rho (x) = \frac{\rho(x)L(x)+L(x)\rho(x)}{2} \ .
\label{sld}
\end{equation}
This is, in fact, a way to quantize the classical expression $dp/dx = p d(\ln p)/dx$ accounting for the non-commutativity of $\rho$ and $d\rho/dx$. The existence of such an operator is guaranteed by an application of the Riesz-Fréchet theorem to Hilbert spaces~\cite[p.257]{Holevo}.

We reproduce the derivation in~\cite[p.37]{Brunotese} of the relation linking $\mathcal F_Q(x)$ to $L(x)$. It begins with a bound on $\mathcal F_Q(x)$, obtained observing that
\begin{subequations}\label{demorhoL2p1}
\begin{align}
\left[\frac{d}{dx} P_k(x) \right]^2 & = \left[ \frac{d}{dx} \Tr\left( \rho(x) E_k \right) \right]^2 
\label{demorhoL2p1l1}									=  \left[ \Tr \left( \frac{d\rho(x)}{dx}  E_k \right) \right]^2 \\
\label{demorhoL2p1l2}				& = \left[ \Tr \left( \frac{\rho(x)L(x)+L(x)\rho(x)}{2}  E_k \right) \right]^2 \\
\label{demorhoL2p1l3}				& = \left[ \Tr \left( \frac{\rho LE_k+E_kL\rho}{2} \right) \right]^2 = \left[ {\rm Re} \big( \Tr \left( \rho LE_k \right) \big) \right]^2 \\
\label{demorhoL2p1l4}				& \leq \left| \Tr \left( \rho LE_k \right) \right|^2 
																= \left| \Tr \left[ \left(\rho^{1/2}E_k^{1/2}\right)^\dagger\left(\rho^{1/2}LE_k^{1/2}\right) \right] \right|^2 \ . 
\end{align}
But the inner product $(A,B)=\Tr(A^\dagger B)$ obeys the Cauchy-Schwarz inequality, so
\begin{align}
\left[\frac{d}{dx} P_k(x) \right]^2 
									& \leq \Tr \left[ \left(\rho^{1/2}E_k^{1/2}\right)^\dagger\left(\rho^{1/2}E_k^{1/2}\right) \right]  
									    \Tr\left[ \left(\rho^{1/2}LE_k^{1/2}\right)^\dagger\left(\rho^{1/2}LE_k^{1/2}\right) \right] \label{demorhoL2p1l5} \\
									& = P_k(x) \Tr \left[ \rho(x)L(x)E_kL(x) \right] \ . \label{demorhoL2p1l6}
\end{align}
\end{subequations}
The quantum Fisher information is then bounded by
\begin{subequations}\label{demorhoL2p2}
\begin{align}
\mathcal F_Q(x) &   =  \max_{\{E_k\}} \sum_k \left\{\frac{1}{P_k(x)} \left[\frac{d}{dx} P_k(x) \right]^2 \right\} \label{demorhoL2p2l1} \\
								& \leq \max_{\{E_k\}} \sum_k \Tr \left[ \rho(x)L(x)E_kL(x) \right] \label{demorhoL2p2l2} \\
								&   =  \Tr \left[ \rho(x)L^2(x) \right] \ . \label{demorhoL2p2l3}
\end{align}\end{subequations}

To demonstrate the equality, we must show that there is a POVM $\{E_k\}$ such that both inequalities in Eqs.~\eqref{demorhoL2p1} turn into equalities for all $k$. For Eq.~\eqref{demorhoL2p1l4}, equality amounts to
\begin{equation}
{\rm Im} \big( \Tr \left( \rho(x) L(x)E_k \right) \big) = 0 \ ,
\label{demorhoL2Im}
\end{equation}
while equality in the Cauchy-Schwarz relation of Eq.~\eqref{demorhoL2p1l5} is equivalent to 
\begin{equation}
\rho^{1/2}(x)L(x)E_k^{1/2} = \lambda_k(x) \rho^{1/2}(x)E_k^{1/2} \ 
\label{demorhoL2sat}
\end{equation}
for some\footnote{The case $\rho^{1/2}(x)E_k^{1/2}=0$ need not obey Eq.~\eqref{demorhoL2sat}, but attains both equalities in Eqs.~(\ref{demorhoL2p1l4},~\ref{demorhoL2p1l5}) trivially, since $(\rho^{1/2}(x)E_k^{1/2})^\dagger=E_k^{1/2}\rho^{1/2}(x)=0$.} complex $\lambda_k(x)$. We remark that if a POVM obeys Eq.~\eqref{demorhoL2sat} with real $\lambda_k(x)$ for all $k$, both conditions are met. But this last equation can be rewritten as
\begin{equation}
\rho^{1/2}(x)\left( L(x) - \lambda_k(x)\mathbb I \right) E_k^{1/2} = 0 \ .
\label{demorhoauto}
\end{equation}
By taking projector POVMs, such that $E_k^2=E_k=E_k^{1/2}$, we can assure that there is at least one solution to the above for each $k$, because $L(x)$ is Hermitian and hence diagonalizable with real eigenvalues.

We have thus shown that the quantum Fisher information can be written as
\begin{equation}
\mathcal F_Q(x) = \Tr\left( \rho(x) L^2(x) \right) \ .
\label{FisherviaSLD}
\end{equation}
This can be understood as an alternate definition of the quantum Fisher information, somewhat analogous to Eq.~\eqref{FisherCl}. The concision of this expression does not incur in an easier calculation of $\mathcal F_Q(x)$, since the symmetric logarithmic derivative is often cumbersome to obtain. We show in Section~\ref{calculating} a more feasible method of calculating the quantum Fisher information developed by our group~\cite{BRL,BLNR}.

We now turn to the Section where we present our main result.

\section{The most general quantum speed limit}
\label{thebound}
\hspace{5mm}We are now ready to derive our main result, the bound on quantum evolutions be they unitary or not. This is a bound that encompasses any kind of evolution a quantum system can undergo. Because non-unitary evolutions can turn pure states into mixed ones, the first step will be to define a distance valid for mixed as well as pure states. We make extensive use of the purifications presented in Section~\ref{purif}.

\subsection{A distance for all quantum states}\label{distgeneral}

\hspace{5mm}Along the lines of Section~\ref{GeoApp}, we are interested in metrics that can be computed by integrating a differential form. Let us define the differential form of the distance between two neighboring states $\rho$ and $\rho+d\rho$ as the minimal Fubini-Study differential of their respective purifications $\ket\psi$ and $\ket\psi+\ket{d\psi}$,
\begin{equation}
\left.ds_{\alpha}^2\right|_{\rho,\rho+d\rho}:=\left.\min_{\rm purif} \ ds_{FS}^2\right|_{\ket\psi,\ket\psi+\ket{d\psi}} \ ,
\label{ds2misto}
\end{equation}
where the minimum is taken over all purifications of $\rho$ and $\rho+d\rho$, but can equivalently be taken over the purifications of only one of the states. The corresponding length is
\begin{equation}
\ell_{\alpha} := \int_{\rm path} ds_{\alpha} = \int_{\rm path} \min_{\rm purif} \ ds_{FS}  \ .
\label{lmisto}
\end{equation}
We are interested in minimizing not the integrand, but the integrated length. The first condition to do so is to have a non-negative integrand, which $ds_{FS}$ is by definition. A second potential hindrance would be the fact that each state $\rho$ along the path belongs to two differentials on the integration, $(\rho-d\rho,\rho)$ and $(\rho,\rho+d\rho)$, except for the path endpoints. The purification of $\rho$ used to minimize the distance of the first pair could, in principle, not be compatible with minimizing the latter pair. This is however not the case, because we know $ds_{FS}$ can be expressed as a decreasing function of the fidelity $F$ (or overlap) of two pure states, and we have seen in Eq.~\eqref{UhlmTheor} that the maximization of the fidelity of two purifications can be performed varying the purification of only one of the mixed states. We may then write
\begin{equation}
\ell_{\alpha} = \min_{\rm purif}\int_{\rm path} ds_{FS} = \min_{\rm purif} \ \ell_{FS} \ ,
\label{lmisto2}
\end{equation}
where minimization is now performed over all purifications of each state of the path.

A distance can, as before, be defined as the length of the shortest path between two states,
\begin{equation}
D(\rho,\sigma) := \min_{\rm path} \ell_{\alpha} = \min_{\rm path} \left\{ \min_{\rm purif} \ \ell_{FS} \right\}  \ ,  
\label{distrho}
\end{equation}
but the order of the minimizations may be inverted, so that
\begin{equation}
D(\rho,\sigma) = \min_{\rm purif} \left\{ \min_{\rm path} \ \ell_{FS} \right\} = \min_{\rm purif} D_{FS}(\ket\psi,\ket\phi) = \min_{\rm purif} \arccos\sqrt{F(\ket\psi,\ket\phi)}  \ ,
\label{distrhoFS}
\end{equation}
with the minimization being performed over all purifications $\ket\psi$, $\ket\phi$ of $\rho$, $\sigma$. Because $D_{FS}$ is a decreasing function of the fidelity, we can write
\begin{equation}
D(\rho,\sigma) = \arccos\sqrt{ \max_{\rm purif}F(\ket\psi,\ket\phi)} = \arccos\sqrt{F_B(\rho,\sigma)} \ ,
\label{distrhofinal}
\end{equation}
where we have used Uhlmann's theorem, Eq.~\eqref{UhlmTheor}.

This distance is called \textit{Bures angle} (among other names, see footnote on p.~\pageref{another}), and is not only valid as a distance for mixed quantum states, but also possesses the most important of the Riemannian features, that of being able to be obtained as an integral of a differential $ds$. Just as with the Fubini-Study distance on p.~\pageref{dsD}, a consistency check left to reader is to verify that $D(\rho,\sigma)=ds$ for infinitesimally separated $\rho$, $\sigma$.

It is indeed quite intuitive to generalize the Fubini-Study distance to mixed states by replacing the pure-state fidelity for the Bures fidelity, but there are two reasons for carrying the derivation as above. Firstly, it explicitly shows the relation between the finite Bures angle and its differential $ds$. Secondly, there are other functions of two mixed states that recover the pure-state fidelity in the proper limit; this route reinforces the importance of the Bures fidelity by naturally arriving at it via Uhlmann's theorem.

\subsection{The bound}

\hspace{5mm} We now have a proper distance also for mixed states, but what is its relation to the dynamics of a quantum system? This is our main result, obtained in collaboration with B.M. Escher, L. Davidovich, R.L. de Matos Filho and published in~\cite{MMT}.

Let us consider a state $\rho(t)$ which evolves in time from $t=0$ to $t=\tau$ and travels a given path. The length of this path is necessarily greater than (or equal to) the distance between $\rho(0)$ and $\rho(\tau)$. This statement, written in a shorthand notation in which $D(0,\tau)$ is the distance between these two states (and analogously for any $D(t_1,t_2)$ at other times), reads
\begin{equation}
D(0,\tau) \leq \ell_\alpha = \int_{\rm path} ds_\alpha \ .
\label{Dleqlbas}
\end{equation} 
But $ds_\alpha$ can be written in terms of $D$ by noting that
\begin{equation}
D(t,t+t') = \frac{\partial D(t,t+t')}{\partial t'} t' + \mathcal O (t'^2) \ ,
\label{Dexpansao}
\end{equation}
so that, with $\eta=t+t'$ and taking $t'$ to be infinitesimal,
\begin{equation}
ds_\alpha = \left.\frac{\partial D(t,\eta)}{\partial \eta}\right|_{\eta\rightarrow t} dt \ .
\label{dsinf}
\end{equation}
The length is then written as
\begin{equation}
\ell_\alpha = \int_0^\tau \left.\frac{\partial D(t,\eta)}{\partial \eta}\right|_{\eta\rightarrow t} dt \ .
\label{l}
\end{equation}
Since only through the Bures fidelity does the distance depend on the states, and hence on the parameter $\eta$, the integrand can be calculated using the chain rule,
\begin{equation}
\left.\dfrac{\partial D(t,\eta)}{\partial \eta}\right|_{\eta\rightarrow t} = \left[\dfrac{d D(F_B)}{d F_B} \dfrac{\partial F_B(t,\eta)}{\partial \eta}\right]_{\eta\rightarrow t} \ ,
\label{chain}
\end{equation}
where $F_B(t,\eta)=F_B\left(\rho(t),\rho(\eta)\right)$ and $D(F_B)$ is a notation used to stress that the distance $D$ is a function of $F_B$. We know from the relation between these two (Eq.~\ref{distrhofinal}) that $d D(F_B)/d F_B$ tends to infinity in the limit at hand, $F_B\rightarrow1$. Furthermore, Eq.~\eqref{BuresFisher} informs us that the first derivative of $F_B$, $\partial F_B(t,\eta)/\partial \eta$ vanishes for $\eta\rightarrow t$. The indeterminacy can be removed with the aid of l'H\^opital's rule, 
\begin{equation}
\left.\dfrac{\partial D(t,\eta)}{\partial \eta}\right|_{\eta\rightarrow t}= \left.\dfrac{\dfrac{\partial F_B(t,\eta)}{\partial \eta} }{\dfrac{1}{d D(F_B)/dF_B}}\right|_{\eta\rightarrow t}
= \left. \dfrac{ \dfrac{\partial^2 F_B(t,\eta)}{\partial \eta^2} } 
											{\dfrac{d}{dF_B}\left[\dfrac{1}{d D(F_B)/d F_B}\right]\dfrac{\partial F_B(t,\eta)}{\partial \eta}} \right|_{\eta\rightarrow t} .
\label{l'H}
\end{equation}
The numerator is, due to Eq.~\eqref{BuresFisher}, proportional to the quantum Fisher information. The denominator, when multiplied and divided by $dD/dF_B$, yields
\begin{equation}
\left. \left( -\dfrac{1}{\left[d D(F_B)/d F_B\right]^3} \dfrac{d^2 D(F_B)}{d F_B^2}\right) 
																								\left( \dfrac{dD(F_B)}{dF_B} \dfrac{\partial F_B(t,\eta)}{\partial \eta}\right)  \right|_{\eta\rightarrow t} .
\label{denominator}
\end{equation}
The first factor in parentheses can be calculated independently of $t,\eta$ by replacing the limit $\eta\rightarrow t$ by $F_B\rightarrow1$ since the metric $D$ only depends on $t,\eta$ through $F_B(t,\eta)$.
The second factor in parentheses is simply a recurrence of $\left.\partial D(t,\eta)/\partial \eta\right|_{\eta\rightarrow t}$, the term we are calculating. Substituting in \eqref{l'H} and rearranging the terms, one finds
\begin{equation}
\left[\dfrac{\partial D(t,\eta)}{\partial \eta}\right]_{\eta\rightarrow t}^2 = 
\left. \dfrac{\left[\dfrac{d D(F_B)}{d F_B}\right]^3}{\dfrac{d^2 D(F_B)}{d F_B^2}} \right|_{F_B\rightarrow1} \frac{\mathcal F_Q(t)}2  \ .
\label{dDdx'}
\end{equation}
Multiplying and dividing by $2$, the length is given by
\begin{equation}
\ell_\alpha = \left. \sqrt{ \dfrac{2\left[d D(F_B)/d F_B\right]^3}{d^2 D(F_B)/d F_B^2} } \right|_{F_B\rightarrow1} \int_0^\tau\sqrt{\frac{\mathcal F_Q(t)}4} dt \ ,
\label{lengthFisher}
\end{equation}
and substitution in Eq.~\eqref{Dleqlbas} leads to the general bound
\begin{equation}
\left.\sqrt{\frac{d^2 D(F_B)/d F_B^2}{2\left[d D(F_B)/d F_B\right]^3}}\right|_{F_B\rightarrow1} D(0,\tau) \leq \int_0^\tau \sqrt{\frac{\mathcal F_Q(t)}{4}}\,dt \ .
\label{generalbound}
\end{equation}

Let us make three comments about the formula above. First, the integral of the quantum Fisher information is to be interpreted as a path integral parametrized by time $t$. $\mathcal F_Q(t)$ is a local quantity, dependent on the state at $t$ and its neighboring values, and hence dependent on the chosen path. Another relevant remark is that the freedom to regauge the distance by a constant $k$, $D\rightarrow kD$, does not affect the bound, since the $D$-dependent prefactor in the equation cancels such a constant. Lastly, because $d D(F_B)/d F_B$ diverges at $F_B\rightarrow1$, the bound can be rewritten as
\begin{equation}
\left.\sqrt{\frac{|d^2 D(F_B)/d F_B^2|}{2\left\{1+[d D(F_B)/d F_B]^2\right\}^{3/2}}}\right|_{F_B\rightarrow1} D(0,\tau) \leq \int_0^\tau \sqrt{\frac{\mathcal F_Q(t)}{4}}\,dt ,
\label{boundcurv}
\end{equation}
and this shows that this prefactor depends on the curvature of the graph of $D(F_B)$ as a function of $F_B$ at $F_B\rightarrow1$.

We can now replace $D(F_B)$ for $D(F_B)=\arccos\sqrt{F_B}$:
\begin{equation}
\arccos\sqrt{F_B\left[\rho(0),\rho(\tau)\right]} \leq \int_0^\tau \sqrt{\mathcal F_Q(t)/4} \ dt \ ,
\label{boundarccos}
\end{equation}
where the prefactor tends to one. This is the main result of the present thesis. It can be summarized as: the length of the traveled path in state space must be less than (or equal to) the distance between initial and final states. For infinitesimal variations, the length is in fact equal to the distance. This implies that the relation above is an equality in the immediate vicinity of $\tau=0$. Eq.~\eqref{BuresFisher} applied to time serves as a suitable expansion to show that the bound is saturated up to second order in time:
\begin{equation}
F_B[\rho(0),\rho(dt)] = 1 - \frac{\mathcal F_Q(0)}4 dt^2 +\mathcal O(dt^4) \ .
\label{BuresFishert}
\end{equation}

Although originally defined in quantum metrology, the quantum Fisher information for time estimation is closely related to the dynamics of the system. This is seen best by remembering that $\mathcal F_Q(t)=\Tr[\rho(t) L^2(t)]$ and seeing that
\begin{equation}
\dot\rho = \frac12 \left[ \rho(t) L(t) + L(t)\rho(t) \right] \ ,
\label{sldt}
\end{equation}
where the dot over $\rho$ represents the time derivative. In fact, there is a closed form for $L(t)$ for unitary evolutions which clearly reflects the dynamics of the system. Any unitary evolution can be written as
\begin{equation}
\rho(t) = U(t) \rho(0) U^\dagger(t) \ ,
\label{rhoU}
\end{equation}
with $U(t)U^\dagger(t)=\mathbb I=U^\dagger(t)U(t)$, and equivalently as
\begin{equation}
\dot\rho(t) = \frac1{i\hbar}\left[i\hbar \ \dot U(t)U^\dagger(t),\rho(t)\right] = \left[H(t),\rho(t)\right] \ ,
\label{rhopontoH}
\end{equation}
with $H(t)=i\hbar \ \dot U(t)U^\dagger(t)$ being the Hamiltonian governing the evolution, $i\hbar \dot U = H U$.

For a system always in pure states, the relation $\rho^2(t)=\rho(t)$ can be derived with respect to time, so
\begin{align}
& \dot\rho\rho + \rho\dot\rho = \dot\rho \label{Lunit1}\\
& \rho L\rho + L\rho\rho + \rho L\rho +\rho\rho L = \rho L + L\rho  \label{Lunit2} \\
& \rho L\rho = 0 \ ,  \label{Lunit3}
\end{align}
such that
\begin{align}
\rho\dot\rho = \frac12 \left[ \rho\rho L + \rho L\rho \right] = \frac{\rho L}2 \ ; \label{Lunit4} \\
\dot\rho\rho = \frac12 \left[ \rho L\rho +  L\rho\rho \right] = \frac{L \rho}2 \ . \label{Lunit5}
\end{align}
The last two, together with $\rho^2(t)=\rho(t)$, allow us to write
\begin{align}
\mathcal F_Q(t) & = 	\Tr\left(	\rho L^2\right) \label{L2unit1} \\
								& = 	\Tr\left(	\rho L L \rho\right) \label{L2unit2} \\
								& = 4 \Tr\left(	\rho\dot\rho\dot\rho\rho\right) \label{L2unit3} \\
								& = 4 \Tr\left(	\rho\dot\rho^2\right) \label{L2unit4} \\
								& = -\frac{4}{\hbar^2} \Tr\left(	\rho\left[H,\rho \right]^2 \right) \label{L2unit5} \\
								& = \frac4{\hbar^2} \left[\Delta H \right]^2  \ , \label{L2unit6}
\end{align}
where $\left[\Delta H\right]^2$ is the variance of $H(t)$. We thus see that, for unitary pure-state evolutions, the quantum Fisher information for time estimation is simply the energy variance of the system. Substituting this result in the general bound, Eq.~\eqref{boundarccos}, we exactly recover the Mandelstam-Tamm bound for unitary evolutions, Eqs.~\eqref{MTtimedep}, \eqref{MTgeom}. The result of Eq.~\eqref{L2unit6} will be of utmost importance to the method we will present in Section~\ref{calculating} to calculate the quantum Fisher information.

This geometric derivation provides, as before, a clear criterion for saturation. Equality will prevail on Eq.~\eqref{boundarccos} if, and only if, the state travels on a geodesic. We note that pure-state geodesics are still geodesics on the space of density operators, but that there are more geodesic paths, even for a qubit. In Chapter~\ref{cap4}, we apply this bound to different examples, and in some cases we will see saturation, i.e., evolution on geodesics.

A last comment is that related derivations are possible for parameter estimation, with which the precision of estimates of parameter $\theta$ can be bound employing the quantum Fisher information for estimation of $\theta$, $\mathcal F_Q(\theta)$.

\section{Additional results}\label{additional}

\hspace{5mm}We now present additional bounds, which may not have the generality of the main result above, but are steps in the way of establishing the Margolus-Levitin bound in a more structured, geometric framework. We later show that one of these bounds can be saturated in conditions under which neither the Mandelstam-Tamm nor the Margolus-Levitin bounds are. The contents of this Section can be considered as partial results in a work in progress.

This calculation is valid for pure states only, evolving u\-ni\-tar\-i\-ly under a time-in\-de\-pend\-ent Ham\-il\-to\-ni\-an. We base our derivation partially on the calculations by Zwierz~\cite{Zwierz}. If the governing Hamiltonian is denoted $H$, the state $\ket{\psi(t)}$ of the system (initially in $\ket{\psi_0}$), obeys
\begin{equation}
i\hbar \frac d{dt} \ket{\psi(t)} = \left[H-g(t)\right] \ket{\psi(t)} \ ,
\label{eqSchrmed}
\end{equation}
where $g(t)$ is a real function explicitly depicting the freedom to choose the ground-state energy. The distance $D_{FS}$ between initial and final states varies in time according to
\begin{align}
\frac d{dt} D_{FS}(\ket{\psi_0},\ket{\psi(t)}) & = \frac d{dt} \arccos |\braket{\psi_0|\psi(t)}| 																					\label{dDFSdtl1} \\
																								& = -\frac{1}{\sqrt{1-|\braket{\psi_0|\psi(t)}|^2}}\frac d{dt}|\braket{\psi_0|\psi(t)}|   \label{dDFSdtl2} \\
																								& \leq  \frac{1}{\sqrt{1-|\braket{\psi_0|\psi(t)}|^2}} \left| \frac d{dt}|\braket{\psi_0|\psi(t)}| \right| \label{dDFSdtl3} \ ,
\end{align}
where we used the fact that $\frac d{dt} D_{FS} \leq |\frac d{dt} D_{FS}|$. This implies that the bound can only be saturated while the distance decreases monotonically. Furthermore, since $|\braket{\psi_0|\psi(t)}|=\cos D_{FS}$, the square root on the last equation reduces to $\sin D_{FS}$, so that we can write
\begin{equation}
\sin D_{FS}(0,t) \ \frac d{dt} D_{FS}(0,t) \leq  \left| \frac d{dt}|\braket{\psi_0|\psi(t)}| \right| \ ,
\label{sinDdDdt}
\end{equation}
where the argument $\ket{\psi_0},\ket{\psi(t)}$ of $D_{FS}$ has been simplified to $0,t$ for cleanness. The derivative of $|\!\braket{\psi_0|\psi(t)}\!| = \sqrt{\braket{\psi_0|\psi(t)}\braket{\psi(t)|\psi_0}}$ can be bounded by
\begin{align}
\left| \frac d{dt}|\!\braket{\psi_0|\psi(t)}\!| \right| 
& = \frac{|i\!\braket{\psi_0|\psi(t)}\!\braket{\psi(t)|H\!-\!g(t)|\psi_0}-i\!\braket{\psi_0|H\!-\!g(t)|\psi(t)}\!\braket{\psi(t)|\psi_0}\!|}
																																																					{2\hbar|\!\braket{\psi_0|\psi(t)}\!|}				\label{limdomod1} \\
& = \frac{|{\rm Im}\left[\braket{\psi_0|H\!-\!g(t)|\psi(t)}\!\braket{\psi(t)|\psi_0}\right]|}{\hbar|\!\braket{\psi_0|\psi(t)}\!|}													\label{limdomod2} \\
& \leq \frac{\left|\braket{\psi_0|H\!-\!g(t)|\psi(t)}\!\braket{\psi(t)|\psi_0}\right|}{\hbar|\braket{\psi_0|\psi(t)}|} = \frac{\left|\braket{\psi_0|H\!-\!g(t)|\psi(t)}\right| \ |\!\braket{\psi(t)|\psi_0}\!|}{\hbar|\!\braket{\psi_0|\psi(t)}\!|}																	\label{limdomod3} \\
& = \frac{\left|\braket{\psi_0|H\!-\!g(t)|\psi(t)}\right|}{\hbar} \ , \label{limdomod4}
\end{align}
such that
\begin{equation}
\sin D_{FS}(0,t) \ \frac d{dt} D_{FS}(0,t) \leq  \frac{\left|\braket{\psi_0|H\!-\!g(t)|\psi(t)}\right|}{\hbar}
\label{dDFSdtl4} \ .
\end{equation}

This bound is valid for any real function $g(t)$ but, unlike the equation of motion, it is (non-trivially) affected by the choice of $g(t)$. We could simply minimize the right-hand side over $g(t)$ in order to arrive at the tightest bound possible from the above equation, but we can obtain a more insightful bound making a different minimization\footnote{It is left as an exercise to the reader to show that the direct minimization of Eq.~\eqref{dDFSdtl4} yields $g(t)={\rm Re} \frac{\braket{\psi_0|H|\psi(t)}}{\braket{\psi_0|\psi(t)}}$. The impracticality of this expression can be seen from the difficulty to assign a physical meaning to this quantity.}.

Let us expand  $\ket{\psi_0}$ in the energy eigenbasis, $\ket{\psi_0}=\sum_nc_n\ket{E_n}$. We obtain, for the numerator of the right-hand side of Eq.~\eqref{dDFSdtl4},
\begin{equation}
\left|\sum_n |c_n|^2 \left[E_n-g(t)\right] e^{-i\frac{E_n}\hbar t}\right| \leq \sum_n \left||c_n|^2 \left[E_n-g(t)\right] e^{-i\frac{E_n}\hbar t}\right| = \sum_n |c_n|^2 \left| E_n-g(t) \right|  \ ,
\label{decomp}
\end{equation}
where we have used the triangle inequality for complex numbers. The right-hand side of this equation, when substituted in Eq.~\eqref{dDFSdtl4}, still leads to a bound on $dD_{FS}/dt$. We then choose to minimize the quantity on the right-hand side. The value of $g(t)$ yielding this minimum is independent of $t$, and we write it simply as $g$. Our goal is, then, to minimize
\begin{equation}
\sum_n |c_n|^2 \left|E_n-g\right|  = \braket{\psi_0|\big|H-g\big||\psi_0}
\label{somatmod}
\end{equation}
with respect to $g\in\mathbb R$, given $c_n$ and $E_n$, i.e., given a probability distribution for the set of energy values. This minimization is performed in Appendix~\ref{Appmed}, and yields
$g$ equal to the \textit{median} of the energy distribution, $E_\med$.

We can obtain
\begin{equation}
\sin D_{FS}(0,t) \ \frac d{dt} D_{FS}(0,t) \leq  \frac{\braket{\psi_0|\big|H-E_\med\big||\psi_0}}{\hbar}  
\label{medfraco}
\end{equation}
using the fact that $\left|\braket{\psi_0|H\!-\!g(t)|\psi(t)}\right|\leq\braket{\psi_0|\big|H\!-\!g(t)\big||\psi_0}$ granted by Eq.~\eqref{decomp}. This substitution makes the bound less tight, but has the advantage of possessing a simple time-dependence, since the right-hand side of the above is time-independent. Integration with respect to time in this case is straightforward,
\begin{equation}
1-\cos [D_{FS}(0,\tau)] \leq  \frac{\braket{\psi_0|\big|H-E_\med\big||\psi_0}}{\hbar} \tau 
\label{medfracoint}
\end{equation}
and, upon inversion, one finds
\begin{equation}
\tau \geq \frac{1-\cos [D_{FS}(0,\tau)]}{\braket{\psi_0|\big|H-E_\med\big||\psi_0}} \hbar = \frac{1-\sqrt{F_B(0,\tau)}}{\braket{\psi_0|\big|H-E_\med\big||\psi_0}} \hbar \ .
\label{medfracoinv}
\end{equation}

A tighter bound, however, is obtained replacing $g=E_\med$ directly in Eq.~\eqref{dDFSdtl4}. This is licit because this equation is valid for any $g$. We thus arrive at
\begin{equation}
\sin D_{FS}(0,t) \ \frac d{dt} D_{FS}(0,t) \leq  \frac{|\braket{\psi_0|H-E_\med|\psi(t)}|}{\hbar} \ .
\label{boundmediana}
\end{equation}

Integrating with respect to time, we obtain
\begin{equation}
D_{FS}(0,\tau) \leq \arccos\left[1-\int_0^\tau\frac{|\braket{\psi_0|H-E_\med|\psi(t)}|}\hbar dt\right] \ .
\label{boundmedint}
\end{equation}

Both bounds can be thought of as of the Margolus-Levitin kind due to their dependence on the average energy, but are independent from the known Margolus-Levitin bound in the sense that they do not recover it. We show in Section~\ref{applicmedian} that the latter bound, Eq.~\eqref{boundmedint}, can be tight in cases where neither the Mandelstam-Tamm nor the Margolus-Levitin bound are, a fact which vouches for its usefulness.

\subsubsection{Chapter Summary and Next Steps}
\hspace{5mm}We have derived our novel results in this Chapter, preceded by the necessary framework. The most general (and most important) bound is given by Eq.~\eqref{boundarccos} in Section~\ref{thebound}.  We begin the next Chapter by presenting an important method for calculating the quantum Fisher information in Section~\ref{calculating}, instrumental to the application of our main bound to physical examples, which is done later in Section~\ref{applicnonU}. Application of our additional bounds is left to Section~\ref{applicmedian}.

\end{chapter}

\begin{chapter}{Applying the bounds}
\label{cap4}

\hspace{5mm} After having presented our novel results in the last Chapter, we now devote ourselves to the application of those bounds to physical evolutions. We start by introducing in Section~\ref{calculating} an important technique, developed by collaborators B.M. Escher, N. Zagury, R.L. de Matos Filho and L. Davidovich~\cite{BRL,BLNR}, for obtaining the quantum Fisher information, a quantity central to our main result of Eq.~\eqref{boundarccos}. In Section~\ref{applicnonU} we proceed to the actual application of the bound to non-unitary evolutions. Section~\ref{applicmedian} is left to the application of the additional bounds of Eqs.~(\ref{medfracoinv},\ref{boundmedint}).

\section{Calculating the quantum Fisher information}
\label{calculating}

\subsection{A purification-based expression}
\hspace{5mm} An apparent disadvantage of our bound of Eq.~\eqref{boundarccos} is the difficulty to calculate the quantum Fisher information. We here present a method which enables the calculation of this quantity based on purifications. 

The basic idea is that the non-unitary evolution of a given system can be described by the unitary evolution of its purification. We have seen in Eq.~\eqref{L2unit6} that the quantum Fisher information of a unitary evolution is easy to evaluate. We will see how the quantum Fisher information of the original system and that of its purification are related.

Let us consider, as in Section~\ref{purif}, a system $S$ in state $\rho(t)$ evolving non-unitarily. A purification of $\rho(t)$ requires an auxiliary system, denoted $E$ (for ``environment''), and is given by a joint state $\ket{\psi_{S,E}(t)}$ such that
\begin{equation}
\Tr_E\big(\ket{\psi_{S,E}(t)}\bra{\psi_{S,E}(t)}\big) = \rho(t) \ \ \forall \ t \ .
\label{condpurifval}
\end{equation}
Let us consider $\ket{\psi_{S,E}(t)}$ to be an arbitrary purification of $\rho(t)$ out of the infinitely many possible. Because $\ket{\psi_{S,E}(t)}$ contains all information about $\rho(t)$, it is physically reasonable that the quantum Fisher information of the former is at least as great as that of the latter. In fact, the quantum Fisher information of $\ket{\psi_{S,E}(t)}$, denoted $\mathcal C_Q(t)$, acts as an upper bound on the quantum Fisher information of $\rho(t)$. This can be shown by using the definition of quantum Fisher information, Eq.~\eqref{FisherQ}, as well as the fact that a POVM element $E_k^{(S)}$ acting only on $S$ can be understood as $E_k^{(S)}\otimes\mathbb I^{(E)}$, where $\mathbb I^{(E)}$ is the identity on $E$:
\begin{equation}
\Tr_S(\rho(t) E_k^{(S)}) = \Tr_{S,E}( \ket{\psi_{S,E}(t)}\bra{\psi_{S,E}(t)} E_k^{(S)}\otimes\mathbb I^{(E)} ) \ .
\label{IdE}
\end{equation}
We can then write the quantum Fisher information of the state of $S$ as
\begin{equation}
\mathcal F_Q(t) =  \max_{\{E_k^{(S)}\otimes\mathbb I_{}^{(E)}\}} \left\{ \left. F(t) \right|_{\{E_k^{(S)}\otimes\mathbb I_{}^{(E)}\}} \right\} \ ,
\label{FQemSE}
\end{equation}
whereas that of the purification reads
\begin{equation}
\mathcal C_Q(t) =  \max_{\{E_k^{(S,E)}\}}\left\{ \left. F(t) \right|_{\{E_k^{(S,E)}\}} \right\} \ ,
\label{CQemSE}
\end{equation}
where $E_k^{(S,E)}$ is a POVM element acting on the joint system $S+E$. The quantum Fisher information of the purification is then the result of a maximization over all POVMs on $S+E$, a larger family than that of the original system, i.e., $\mathcal F_Q(t)$ is maximized on a subset of that used for $\mathcal C_Q(t)$. Hence $\mathcal F_Q(t)\leq\mathcal C_Q(t)$.

The state $\ket{\psi_{S,E}(t)}$ remains pure along its evolution, which must then be describable by a unitary operator $U_{S,E}(t)$. The Hamiltonian governing the evolution can always be found by $H_{S,E}(t)=i\hbar\dot U_{S,E}(t)U_{S,E}^\dagger(t)$, and we know from Eq.~\eqref{L2unit6} that the quantum Fisher information of a unitary evolution is proportional to the variance of its Hamiltonian, so that 
\begin{equation}
\mathcal C_Q(t) = \frac4{\hbar^2} [\Delta H_{S,E}(t)]^2 \ .
\label{CQvar}
\end{equation}
Bringing these results together, we can find a bound on evolution times written in terms of $\Delta H_{S,E}(t)$,
\begin{equation}
\arccos\sqrt{F_B\left[\rho(0),\rho(\tau)\right]} \leq \int_0^\tau \sqrt{\frac{\mathcal F_Q(t)}4} \ dt \leq \int_0^\tau \sqrt{\frac{\mathcal C_Q(t)}4} \ dt = \frac1\hbar \int_0^\tau \Delta H_{S,E}(t) \ dt \ .
\label{boundCQ}
\end{equation}
Written in this form, in terms of any purification of $\rho(t)$, the bound may or may not be tight. The choice of purification in fact affects the saturation of the speed limit.

We show in Appendix~\ref{Fishertight} that for each evolution there is at least one purification whose quantum Fisher information equals that of the original system, i.e., there is a purification such that $\mathcal C_Q(t)=\mathcal F_Q(t)$. It can be obtained by minimizing $\mathcal C_Q(t)$, or equivalently $\Delta H_{S,E}(t)$, over all purifications such that $E$ has the same dimension as $S$.  This means that all saturation features of our main bound carry over to the above if one minimizes over such purifications. 

In some situations it may, however, be advantageous to forgo the exact obtention of $\mathcal F_Q(t)$ and  minimize $\mathcal C_Q(t)$ over a restricted family of purifications for the sake of practicality. We undertake both full and partial optimization in the examples in the next Section.

Before we move on to the non-unitary channels, we mention that which is perhaps the simplest application of this purification procedure: an initially mixed state of $S$ evolving unitarily. It is possible to purify it such that the Hamiltonian governing the evolution of the purification $\ket{\psi_{S,E}(t)}$ is the original Hamiltonian acting on the mixed state. Applying this to Eq.~\eqref{boundCQ}, one can generalize Eqs.~(\ref{MTfidelidade},\ref{MTgeom}) to
\begin{equation}
D(0,\tau) = \arccos\sqrt{F_B\left[\rho(0),\rho(\tau)\right]} \leq \int_0^\tau \frac{\Delta E(t)}\hbar dt \ ,
\label{unitmisto}
\end{equation}
where the standard deviation $\Delta E(t)$ of the Hamiltonian is calculated in the pure or mixed state of system $S$. This result for unitary evolutions has been shown by Uhlmann in~\cite{Uhlmann} also by geometric means. 
In this case, the purification yielding the minimum is in fact the one governed by the original Hamiltonian, so that the bound above is tight under the same conditions of the general bound.

\section{Application to non-unitary channels}
\label{applicnonU}

\hspace{5mm} We now consider some examples of non-unitary evolutions, or channels, to which we apply our bound. As mentioned previously, our description is made in terms of purifications. Given a state $\rho(t)$ of system $S$ that evolves non-unitarily, let  $\ket{\psi_{S,E}(t)}$ be a purification of it. The evolution of this pure state in time can always be described by a suitable unitary operator, denoted $U_{S,E}(t)$, so that $\ket{\psi_{S,E}(t)}=U_{S,E}(t)\ket{\psi_{S,E}(0)}$.

We know from the previous Section that, in order to obtain the quantum Fisher information (or a useful bound on it), we need the variance of the Hamiltonian $H_{S,E}(t)$ respective to $U_{S,E}(t)$,
\begin{equation}
H_{S,E}(t)=i\hbar\dot U_{S,E}(t)U_{S,E}^\dagger(t) \ .
\label{HdeU}
\end{equation}
This variance is, in principle, to be evaluated at time $t$, i.e., in state $\ket{\psi_{S,E}(t)}$. There is nevertheless a tactic to make this evaluation easier: the expectation value of an operator in state $\ket{\psi_{S,E}(t)}$ can be cast as an expectation value in the initial state $\ket{\psi_{S,E}(0)}$ if the evolution operators are ``absorbed'' by $H_{S,E}(t)$. For an operator that satisfies
\begin{equation}
\mathfrak H_{S,E}(t) = U_{S,E}^\dagger(t) H_{S,E}(t) U_{S,E}(t) \ ,
\label{frakturintro}
\end{equation}
its averaging obeys
\begin{equation}
\braket{\psi_{S,E}(0)|\mathfrak H_{S,E}(t)|\psi_{S,E}(0)} = \braket{\psi_{S,E}(t)|H_{S,E}(t)|\psi_{S,E}(t)}
\label{frakturequiv}
\end{equation}
and the same goes for any polynomial function of $H_{S,E}$. This definition presents a practical advantage: the form of the state in which the expectation values are taken is simplified, because there is no time dependence anymore, but the operator maintains a similar level of complexity when expressed in terms of $U_{S,E}$,
\begin{equation}
\mathfrak H_{S,E}(t):= i \hbar U_{S,E}^\dagger(t) \dot U_{S,E}(t)  
\label{fraksimple}
\end{equation}
(compare Eq.~\ref{HdeU}). This equation can, in fact, be taken as a definition of $\mathfrak H_{S,E}(t)$. This tactic is not restricted to $U_{S,E}$, along this Chapter we define several analogous operators based on different unitaries. The use of fraktur/gothic or other modified fonts always denotes the employment of this tactic for initial-state averaging, see Table~\ref{tabela}. We note that, for self-commuting Hamiltonians, $[H(t),H(t')]=0 \ \ \forall \ t,t'$, this tactic does not change the operator, i.e., $\mathfrak H(t)=H(t)$.

Another important point is that the purification $\ket{\psi_{S,E}(t)}$ of the process characterized by $\rho(t)$ is not unique. The freedom to choose different purifications can be expressed in the following manner: given a first purification $\ket{\psi_{S,E}(t)}$ and its unitary evolution operator $U_{S,E}(t)$, every other purification of the process can be obtained through
\begin{equation}
u_E(t)U_{S,E}(t)\ket{\psi_{S,E}(0)} \ ,
\label{uzinhouzao}
\end{equation}
by varying $u_E(t)$, a unitary operator acting only on system $E$ (note that $u_EU_{S,E}$ is also unitary). When we need to minimize over all purifications, as required by Section~\ref{calculating}, it suffices to minimize over all unitary operators $u_E(t)$ acting on the environment $E$ only.

Moreover, we are interested in minimizing the variance of the Hamiltonian corresponding to the unitary operator $u_E(t)U_{S,E}(t)$. We once again use the tactic for initial-state averaging, so that we are faced with the problem of minimizing the variance of
\begin{equation}
\mathcal H_{S,E}(t) := i\hbar \left[u_E(t)U_{S,E}(t)\right]^\dagger \frac d{dt}\left[u_E(t)U_{S,E}(t)\right] \ .
\label{defmathcalH}
\end{equation}
It is straightforward to show that $\mathcal H_{S,E}$ can be written as 
\begin{equation}
\mathcal H_{S,E}(t) = \mathfrak H_{S,E}(t) + U_{S,E}^\dagger(t)\mathfrak h_E(t) U_{S,E}(t) \ , 
\label{mathcalHsimp}
\end{equation}
where $\mathfrak h_E(t):=i\hbar u_E^\dagger(t) \dot u_E(t)$ is defined from $h_E(t)$ analogously to the way $\mathfrak H_{S,E}(t)$ was from $H_{S,E}(t)$, see Table~\ref{tabela}. The minimization over $u_E(t)$ then amounts to minimizing over all possible Hermitian operators $\mathfrak h_E(t)$.

\begin{table}[ht]
\centering
\begin{tabular}{|c|c|c|}\hline
Evolution operator & Hamiltonian 	& Corrected Hamiltonian  \\ \hline
$U_{S,E}(t)$ 			 & $H_{S,E}(t)$ & $\mathfrak H_{S,E}(t)$ \\ \hline
$u_E(t)$ 		 			 & $h_E(t)$ 		& $\mathfrak h_E(t)$ 	 	 \\ \hline
$u_E(t)U_{S,E}(t)$ &  						& $\mathcal H_{S,E}(t)$  \\ \hline
\end{tabular}
\caption{Notational correspondence between evolution operator and Hamiltonian. The last column indicates the notation for the Hamiltonian corrected for initial-state averaging, as defined by Eq.~\eqref{fraksimple} for the first line and analogously for the rest of the table.}
\label{tabela}
\end{table}

Turning once more to the mixed state evolving unitarily, the suitable first purification is that whose evolution is governed by the original Hamiltonian. It is then straightforward to see that $u_E(t)=\mathbb I$, corresponding to $\mathfrak h_E(t)=0$, yields the minimal variance of $\mathcal H_{S,E}(t)$ from Eq.~\eqref{mathcalHsimp}.


\subsection{Amplitude-damping channel}\label{ampdamp}

\hspace{5 mm} The first example to which we apply our bound is the so-called amplitude-damping channel. This channel is useful for describing energy loss: a qubit (system $S$) initially in the excited state $\ket e$ that decays onto the ground state $\ket g$. We suppose there is a probability $P(t)$ of $\ket e$ not decaying to $\ket g$ in the interval $[0,t]$ (conversely, a probability $1-P(t)$ of a $\ket e\rightarrow\ket g$-decay up to time $t$; it should be clear that $P(0)=1$). This evolution of $S$ is described by
\begin{equation}
\rho(0)= \left( \begin{array}{cc} \rho_{gg} & \rho_{ge} \\ \rho_{ge}^* & \rho_{ee} \end{array} \right) \rightarrow \rho(t) = \left( \begin{array}{rc} \rho_{gg} + [1-P(t)]\rho_{ee} & \sqrt{P(t)}\rho_{ge} \\ \sqrt{P(t)}\rho_{ge}^* & P(t)\rho_{ee} \end{array} \right) \ ,
\label{ampdamprho}
\end{equation}
with the representation $\ket g=\genfrac(){0pt}{1}{1}{0}$, $\ket e= \genfrac(){0pt}{1}{0}{1}$. The most prevalent case of the amplitude-damping channel is when $\rho_{ge}=0=\rho_{gg}$ and the system is initially in the excited state $\ket e$. In this scenario, the qubit evolves along mixtures of $\ket e$ and $\ket g$ with varying weights. In the Bloch sphere, this corresponds to a path along the axis connecting $\ket e$ to $\ket g$, see Fig.~\ref{ampdampBloch}.
\begin{figure}[ht]
\centering
\includegraphics[height=.5\columnwidth]{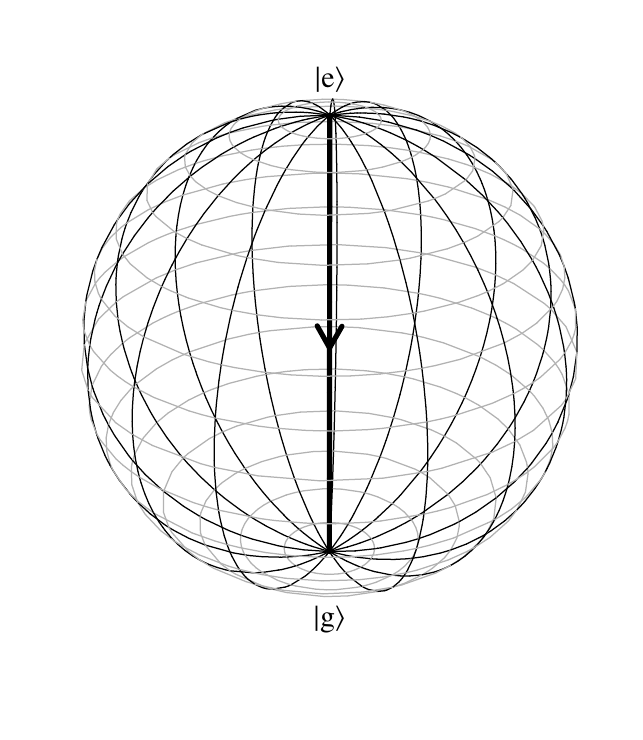}
\caption{Depiction of the Bloch sphere showing the path followed by an initially excited state ($\ket e$) in the amplitude-damping channel. This path is a geodesic between the orthogonal states $\ket e$ and $\ket g$, in addition to the great semi-circles also shown.}%
\label{ampdampBloch}%
\end{figure}

For the purposes of calculating the quantum Fisher information, we now need to write the evolution of $S$ in the form of a purification, which explicitly includes an auxiliary system $E$. System $E$ only needs to be a qubit, so whenever the initial state of $S$ is pure the channel can be described by
\begin{subequations}\label{amplitudechannel}\begin{align}
\ket g\ket0_E &\rightarrow \ket g\ket0_E \label{amp1} \ ,\\
\ket e\ket0_E &\rightarrow  \sqrt{P(t)} \ket e\ket0_E +\sqrt{1-P(t)} \ket g\ket1_E \ , \label{amp2}
\end{align}
\end{subequations}
where $\ket0_E$ is chosen as the initial state of $E$ and $\ket1_E$ is orthogonal to it\footnote{It is left as an exercise to the reader to show that Eq.~\eqref{ampdamprho} is recovered by taking the partial trace (over $E$) of Eq.~\eqref{amplitudechannel}}. This is actually the most common way of describing the amplitude-damping channel. We presented the $S$-contained form of Eq.~\eqref{ampdamprho} to emphasize that we are interested in treating the evolution of $S$, not of $S+E$, and that the auxiliary system $E$ is, as far as this treatment is concerned, a tool which we take full liberty to explore.

The evolution of this purification is given by the following unitary operator on $S+E$
\begin{equation}
U_{S,E}(t)=\exp[-i \left(\arccos\sqrt{P(t)}\right) (\sigma_+\sigma_-^{(E)}+\sigma_-\sigma_+^{(E)})] \ ,
\label{Uamplitude}
\end{equation}
where $\sigma_\pm$ are raising and lowering operators of qubit $S$ ($\sigma_+\ket g=\ket e$, $\sigma_-\ket e=\ket g$, $\sigma_\pm^2=0$), and analogously for $\sigma_\pm^{(E)}$ on $E$. The Hamiltonian for this evolution (corrected for initial-state averaging) is
\begin{equation}
\mathfrak H_{S,E}(t) := i\hbar U_{S,E}^\dagger(t) \dot U_{S,E}(t) = \hbar \frac{d\arccos\sqrt{P(t)}}{dt} \left(\sigma_+\sigma_-^{(E)}+\sigma_-\sigma_+^{(E)} \right) \ , 
\label{Hfrakampdamp}
\end{equation}
from which we obtain the bound $\mathcal C_Q(t)$ on the quantum Fisher information:
\begin{equation}
\frac{\mathcal C_Q(t)}4 = \frac1{\hbar^2}\left[\Delta \mathfrak H_{S,E}(t)\right]^2 = \braket{\sigma_+\sigma_-} \left(\frac{d\arccos\sqrt{P(t)}}{dt}\right)^2 \ ,
\label{CQampdamp}
\end{equation}
where the variance is calculated in the initial state of $S+E$ (hence the expectation value $\braket{\sigma_+\sigma_-}$ is to be calculated in the initial state of $S$). The corresponding bound on evolution time is given by
\begin{equation}
D(0,\tau) \leq \sqrt{\braket{\sigma_+\sigma_-}} \int_0^\tau \left|\frac{d\arccos\sqrt{P(t)}}{dt}\right| dt \ .
\label{boundampdamp}
\end{equation}
A first result is that if the initial state of $S$ is $\ket g$, the bound shows exactly that the distance $D(0,\tau)$ remains zero, as it must according to Eq.~\eqref{amp1}. Furthermore, if $P$ is a monotonic (decreasing) function of $t$, integration can be performed at once and yields
\begin{equation}
D(0,\tau) = \arccos\sqrt{F_B(0,\tau)} \leq \sqrt{\braket{\sigma_+\sigma_-}} \arccos\sqrt{P(\tau)} \ .
\label{boundampdamp2}
\end{equation}

An example is the exponential damping $P(t)=e^{-\gamma t}$, for which 
\begin{equation}
D(0,\tau)\leq\sqrt{\braket{\sigma_+\sigma_-}}\arccos{e^{-\gamma\tau/2}} \ ,
\label{boundampdampex}
\end{equation}
or
\begin{equation}
\tau \geq \frac2\gamma \ln \left( \sec \frac{D}{\sqrt{\braket{\sigma_+\sigma_-}}} \right) \ .
\label{boundampdampexpl}
\end{equation}

If system $S$ is initially in state $\ket e$, the expectation value $\braket{\sigma_+\sigma_-}=1$ and the bound is saturated, since for this case $F_B(0,\tau)=P(\tau)$, see Eq.~\eqref{boundampdamp2}.
We can draw a couple of conclusions from this fact. Firstly, saturation implies that $\mathcal C_Q(t)$ above is the minimum over different purifications for initial states $\ket e$ or $\ket g$, i.e. $\mathcal C_Q(t)=\mathcal F_Q(t)$. This is, in fact, an exception: we see in the next Subsections that the first, most intuitive purification does not usually yield the minimum, and an optimization must be performed.

Secondly, saturation implies that the path from $\ket e$ to $\ket g$ through mixed states of Fig.~\ref{ampdampBloch} is a geodesic. We have seen in Section~\ref{GeoApp} that great circles of the Bloch sphere are the geodesics between two orthogonal states; the inclusion of mixed states adds this path (among others) to the set. We should also note that the geometry of the Bures angle is radically different from the usual, Euclidean geometry on the Bloch sphere, since both a diameter and a great semi-circle have here the same length (as shown in Fig.~\ref{ampdampBloch}). 

Another feature of the result is that the bound saturates for any monotonic $P(t)$. This is a manifestation of the geometric nature of the bound, since varying $P(t)$ among monotonic functions can be understood as a rescaling of the time parameter, still on the route of Fig.~\ref{ampdampBloch}. A non-monotonic $P(t)$ would imply going back and forth along the path, which cannot constitute a geodesic, and the bound is no longer saturated.

A special case of this channel is a two-level atom interacting with an external field modeled by the resonant Jaynes-Cummings Hamiltonian when there is only one quantum of energy in the compound system. If we take the atom to be system $S$ and the external field as $E$, the interaction is given by $H_{S,E}(t)=\hbar g (\sigma_+\sigma_-^{(E)}+\sigma_-\sigma_+^{(E)})=\mathfrak H_{S,E}(t)$, $g$ being a coupling constant. Eq.~\eqref{amplitudechannel} is obeyed with $P(t)=\cos^2gt$ up to the first root of $\cos gt$, which is the relevant short-time regime for speed limits. The bound is then also saturated, since it predicts
\begin{equation}
F_B(0,\tau)\geq \cos^2gt \ .
\label{JaynesCummings}
\end{equation}

\subsection{Single-qubit dephasing}\label{1qubit}

\hspace{5 mm} We now present an interaction typically used to describe the generation of decoherence, the dephasing. In this Markov evolution, a qubit (system $S$) loses information on the relative phase between its $\ket0$ and $\ket1$ components. This channel can arise physically when a system interacts with many external degrees of freedom without energy loss. An example is a particle scattering off of many different atoms in a medium, with each scattering being elastic. Energy is a constant of motion, but the relative phase accumulated between the energy eigenstates depends on the time during which scattering took place. This time is usually not known (partly because it is a function of the precise path in physical space traveled by the particle, which is also unknown), hence there is a gradual loss of phase information.

The system Hamiltonian is $\hbar\omega_0Z/2$, with $Z$ being the Pauli operator on $S$ ($Z\ket0=+\ket0$, $Z\ket1=-\ket1$) and analogously for $X$, $Y$. The evolution can be written as 
\begin{equation}
\frac{d\rho}{dt} = -i \frac{\omega_0}2[Z,\rho] - \frac\gamma2 \left( \rho - Z\rho Z \right) \ ,
\label{dephasmaster}
\end{equation}
where, in addition to the Hamiltonian term ($\omega_0$), there is a term proportional to $\gamma$ modeling the loss of phase information. In terms of the Bloch sphere, the term in $\omega_0$ is responsible for a rotation around the $z$-axis, whereas the term in $\gamma$ reduces the $x$,$y$ components of the vector, or the distance from the vector to the $z$-axis. The trajectory is contained in a plane perpendicular to the $z$-axis, and the expectation value $\braket Z$ is constant throughout the motion, see Fig.~\ref{dephBloch}. For sufficiently long times, the state is an incoherent mixture of $\ket0$, $\ket1$.
It is customary to treat the dephasing channel in the interaction picture so that the Hamiltonian part of the evolution (rotation around the $z$-axis) does not manifest itself in the state of the system, but it is not licit to do so in our problem for the reasons pointed out on p.~\pageref{Schrpicture}.

\begin{figure}[ht]
\centering
\includegraphics[height=.5\columnwidth]{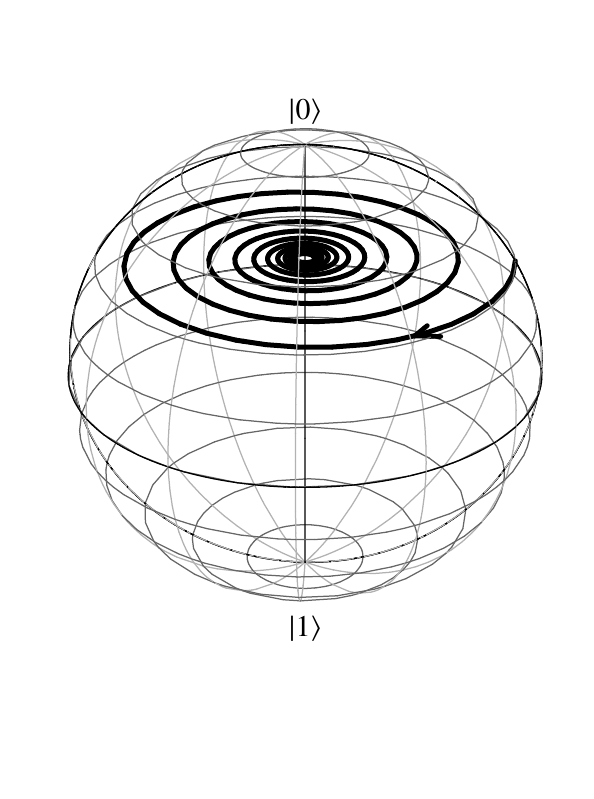}
\caption{Depiction of the Bloch sphere showing the effect of the dephasing channel (thick line) on a superposition of $\ket0$ and $\ket1$. The rotation around $z$ occurs with (angular) frequency $\omega_0$, the radius shrinkage has characteristic time $1/\gamma$.}%
\label{dephBloch}%
\end{figure}

Let us now introduce a purification describing the dephasing channel. The auxiliary system $E$ once again only needs to be a qubit. With the initial state of $S$ assumed pure, $S+E$ can be initially separable, and hence the initial state of $E$ can be chosen to be $\ket0_E$:
\begin{subequations}\label{dephasingchannel}\begin{align}
\ket0\ket0_E &\rightarrow e^{- i\omega_0t/2} \left( \sqrt{P(t)} \ket0\ket0_E + \sqrt{1-P(t)} \ket0\ket1_E \right) \ , \label{deph1} \\
\ket1\ket0_E &\rightarrow e^{+ i\omega_0t/2} \left( \sqrt{P(t)} \ket1\ket0_E - \sqrt{1-P(t)} \ket1\ket1_E \right) \ , \label{deph2}
\end{align}
\end{subequations}
with\footnote{It is left once again as an exercise to the reader to show that Eq.~\eqref{dephasmaster} is recovered by taking the partial trace (over $E$) of Eq.~\eqref{dephasingchannel} and deriving the resulting $\rho$ with respect to time.} $P(t):=(1+e^{-\gamma t})/2$. The purity of the initial state of $S$ poses no restriction, since, in this channel, every evolution starting in a mixed state of $S$ can be seen as part of a longer evolution (starting earlier in time) from a pure state.

The matching evolution operator on $S+E$ is
\begin{equation}
 U_{S,E}(t) = e^{-i\omega_0 t  Z/2} e^{-i\arccos\sqrt{P(t)} \ Z \ Y^{(E)}} ,
\label{evol}
\end{equation}
where $Y^{(E)}$ is a Pauli operator acting on qubit $E$. The corresponding Hamiltonian corrected for initial-state averaging is
\begin{equation}
\mathfrak H_{S,E}(t) = \frac{\hbar \omega_0}2 Z + \frac{\hbar\gamma/2}{\sqrt{e^{2\gamma t}-1}}Z Y^{(E)} \ . 
\label{Hdephas1}
\end{equation}
The bound $\mathcal C_Q(t)$ on the quantum Fisher information obtained from this $U_{S,E}(t)$ is
\begin{equation}
\frac{\mathcal C_Q(t)}4 = \frac1{\hbar^2}\left[\Delta \mathfrak H_{S,E}(t)\right]^2 = \omega_0^2 \frac{[\Delta Z]^2}4 + \frac{\gamma^2/4}{e^{2\gamma t}-1} \ ,
\label{CQdeph1}
\end{equation}
where $[\Delta Z]^2=\left(1-\braket{Z}^2\right)$ is the variance of $Z$, again to be evaluated in the initial state of $S$. A quantum speed limit can already be obtained from it,
but this is just one of the many possible purifications of this channel; any additional unitary operator $u_E(t)$ applied to the right-hand side of Eq.~\eqref{dephasingchannel} maintains its partial trace unaltered. To minimize $\mathcal C_Q(t)$ over different purifications, we have to minimize the variance of $\mathcal H_{S,E}(t)$ from Eq.~\eqref{mathcalHsimp} over all possible operators $\mathfrak h_E(t)$. Since $E$ is a qubit, we only need to minimize over all Hermitian $2\times2$ operators, which can in general be written in terms of the Pauli operators and the identity. Furthermore, the addition of an identity does not affect the variance of an operator, so that, for our purposes, the most general form of $\mathfrak h_E(t)$ is
\begin{equation}
\mathfrak h_E(t) = \alpha(t) X^{(E)} + \beta(t) Y^{(E)} + \delta(t)Z^{(E)} \ ,
\label{frakhdeph1}
\end{equation}
where $\alpha(t)$, $\beta(t)$, $\delta(t)$ are real parameters to minimize over, and $X^{(E)}$, $Y^{(E)}$, $Z^{(E)}$ are Pauli operators acting on qubit $E$. This minimization is performed in Appendix~\ref{dephchannelderivation}, and the result is
\begin{equation}
\min\frac{\mathcal C_Q(t)}4 = \frac{\mathcal{F}_Q(t)}4 = \frac{[\Delta Z]^2}4 \left( \omega_0^2e^{-2\gamma t} + \frac{\gamma^2}{e^{2\gamma t}-1} \right) ,
\label{deltaHmin1}
\end{equation}
dependent on the initial-state variance $[\Delta Z]^2$ of $Z$. 
This leads to the following implicit bound on $\tau$
\begin{equation}
D(0,\tau) \leq \frac{\Delta Z}2 \int_0^\tau \sqrt{\omega_0^2e^{-2\gamma t} + \frac{\gamma^2}{e^{2\gamma t}-1}} dt \ ,
\label{deph1int}
\end{equation}
where $\Delta Z=\sqrt{[\Delta Z]^2}$ is the standard deviation of $Z$.

A first feature of the bound is that it correctly portrays the non-evolution of eigenstates of $Z$. Moreover, in the two extreme cases $\gamma\rightarrow0$ or $\omega_0\rightarrow0$, Eq.~\eqref{deph1int} reduces to simple analytical expressions. For the former, the evolution becomes once again unitary, and the Mandelstam-Tamm bound is recovered [compare Eq.~\eqref{MTfidelidade}]
\begin{equation}
D(0,\tau) \leq \omega_0\frac{\Delta Z}2\tau = \frac{\Delta E}{\hbar}\tau \ .
\label{dephunit}
\end{equation} 
For the latter, the bound reduces to
\begin{equation}
D(0,\tau) \leq \frac{\Delta Z}2\arccos\left(e^{-\gamma\tau}\right) \ ,
\label{dephomeganuloD}
\end{equation}
from which one sees that the distance from initial to final states cannot exceed $D=\pi/4$. Equivalently, we can write
\begin{equation}
\tau \geq \frac1\gamma \ln \left(\sec\frac{2D}{\Delta Z}\right) \ .
\label{dephomeganulotau}
\end{equation}
The bound can be saturated in this ``pure-dephasing'' ($\omega_0=0$) case: this happens when the initial state of the system is on the equator of the Bloch sphere ($\Delta Z=1$), for which the bound yields
\begin{equation}
F_B(0,\tau) \geq \frac12\left(1+e^{-\gamma\tau}\right) = P(\tau)
\label{dephsatur}
\end{equation}
and the actual value of $F_B$ is indeed $F_B(0,\tau) = P(\tau)$.

We can also analyze the asymptotic behavior of the bound for $\tau\rightarrow\infty$. Integration of Eq.~\eqref{deph1int} over the region $(0,\infty)$ arrives at the elliptic integral of the second kind\footnote{We adopt the notation where the second argument $m$ is the so-called \textit{parameter} of the elliptic integral in all occurrences of $E(y,m)$.} $E(y,m)$,
\begin{equation}
D(0,\infty) \leq \frac{\Delta Z}{2} \sqrt{r^2+1} E\left(\frac\pi2,\frac{r^2}{r^2+1}\right)  \ ,
\label{dephinfty}
\end{equation}
with $r:=\omega_0/\gamma$. Given the way the bound has been constructed, a bound for $D(0,\infty)$ serves as bound for the distance at any other time $\tau$. In Fig.~\ref{exclwindow} we have plotted the bound for $D(0,\infty)$ for $\Delta Z=1$, an expression bounding the distance at any time and for any initial state. From this graph, we see that, analogously to Eq.~\eqref{dephomeganuloD}, the bound guarantees that orthogonal states are not reachable in finite time for certain values of $r$, namely $r<r_{\rm crit}\approx2.60058$ (compare the amplitude-damping channel, where $\ket e$ evolves to $\ket g$).
\begin{figure}[ht]
\centering
\includegraphics[height=.35\columnwidth]{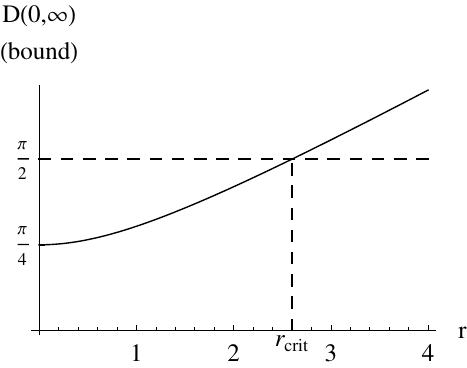}
\caption{Single-qubit dephasing: bound on the distance $D(0,\infty)$ at time $\tau\rightarrow\infty$ as a function of parameter $r=\omega_0/\gamma$, Eq.~\eqref{dephinfty}. It is readily seen that, for $r<r_{\rm crit}\approx2.60058$, no initial state can become orthogonal in finite time.}%
\label{exclwindow}%
\end{figure}

The integral in the general result on Eq.~\eqref{deph1int} is in fact solvable in terms of elliptic integrals of the second kind $E(y,m)$,
\begin{equation}
D(0,\tau) \leq \frac{\Delta Z}{2} \sqrt{r^2+1} \left[E\left(\frac\pi2,\frac{r^2}{r^2+1}\right) - E\left(\arcsin \ e^{-\gamma\tau},\frac{r^2}{r^2+1}\right) \right] \ .
\label{deph1ellipt}
\end{equation}
In Fig.~\ref{dephgenplot}, we plot the relative discrepancy between the bound and the distance as calculated exactly from its evolution. The exact distance in the dephasing channel obeys
\begin{equation}
D_{\rm exact}(0,\tau) = \frac12 \left[ 1+\braket Z^2 + [\Delta Z]^2 e^{-\gamma\tau}\cos \omega_0\tau \right] 
\label{deph1Dexact}
\end{equation}
and we see that the bound remains close to the exact result up to the first minimum of the latter, which is the region of interest for quantum speed limits. This is a display of the usefulness of the bound even when it is not strictly tight.

\begin{figure}[ht]
\centering
\includegraphics[height=.35\columnwidth]{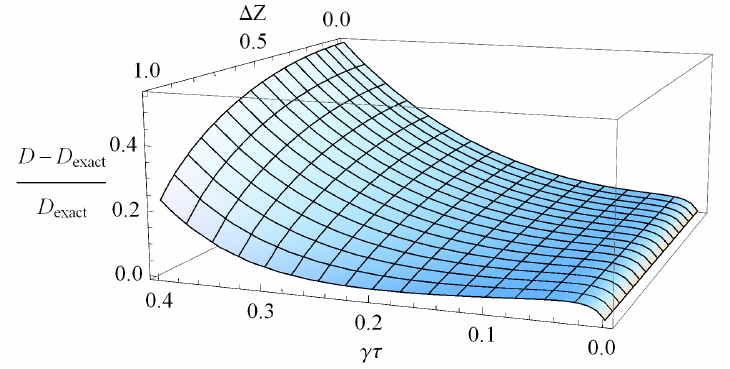}
\caption{Plot of the relative discrepancy from the bound (Eq.~\ref{deph1ellipt}) to exact calculations (Eq.~\ref{deph1Dexact}) in the single-qubit dephasing as a function of dimensionless time $\gamma\tau$, with $r=8$. The various $\Delta Z$ (longitudinal axis) correspond to different initial states.}
\label{dephgenplot}%
\end{figure}

\subsection{Dephasing and entanglement}\label{Nqubit}

\hspace{5 mm} Another display of the usefulness of the bound lies in the discussion of the relation between entanglement and quantum evolution time. This issue concerns quantum systems composed of many ($N$) subsystems and consists in understanding the interplay between correlations among the subsystems (especially entanglement) and the speed of evolution of the system as a whole.

\subsubsection{Established results for unitary evolutions}

\hspace{5mm}Several works~\cite{GiovannettiPRA,GiovannettiEPL,Batle,Borras2006,Zander,Frowis} have shown that in unitary evolutions entanglement is a resource that allows for a speed-up of quantum evolution. Let us then compare the speed of separable and entangled states. We assume that the subsystems do not interact among themselves in order to guarantee that no entanglement is created in the initially separable states.

We begin by applying the quantum speed limit to a separable pure state in a unitary evolution, as done in~\cite{GiovannettiPRA}. Let us assume the bound saturates for some time $\tau$ for the whole state,
\begin{equation}
\tau = \frac\hbar{\Delta E} \arccos \sqrt{F_B(0,\tau)} \ .
\label{MTdenovo}
\end{equation}
Each subsystem $j$ has fidelity $F_{B,(j)}$ compared to its initial state and energy spread $\Delta E_{(j)}$, with $F_B=\prod\nolimits_jF_{B,(j)}$ and $[\Delta E]^2=\sum_j[\Delta E_{(j)}]^2$. Now, the quantum speed limit can be applied to each subsystem, such that
\begin{equation}
\tau \geq \frac\hbar{\Delta E_{(j)}} \arccos \sqrt{F_{B,(j)}(0,\tau)} 
\label{MTsubj}
\end{equation}
and, inserting Eq.~\eqref{MTdenovo} into Eq.~\eqref{MTsubj},
\begin{equation}
\frac\hbar{\Delta E} \arccos \sqrt{F_B(0,\tau)} \geq \frac\hbar{\Delta E_{(j)}} \arccos \sqrt{F_{B,(j)}(0,\tau)} \ .
\label{sepcomp}
\end{equation}
Moving $\Delta E_{(j)}$ to the left-hand side, squaring and summing on $j$, one finds
\begin{equation}
\arccos^2 \sqrt{\prod\limits_jF_{B,(j)}(0,\tau)} \geq \sum_j \arccos^2 \sqrt{F_{B,(j)}(0,\tau)} \ ,
\label{sepcompsimp}
\end{equation}
where $F_B=\prod\nolimits_jF_{B,(j)}$ was also used. But a property of the function $\arccos^2$ is that it is subadditive in the sense that the left-hand side of the above must be less than or equal to the right-hand side\footnote{The relation $\arccos^2\left(\prod\nolimits_jx_j\right) \leq \sum_j\arccos^2(x_j)$ for $0\leq x_j\leq1$ can be proven first for two variables and then by induction. For two variables, one can use that, for any given value of $x_2$, equality holds for $x_1=1$ and that the derivative of the left-hand side is greater than that of the right-hand side for $0\leq x_1<1$.}. The only solution for Eq.~\eqref{sepcompsimp}, then, is for equality to hold. But the condition for equality is that all arguments $F_{B,(j)}=1$ except for a single $j'$, for which $F_{B,(j')}=F_B$. This means that a separable state cannot reach the quantum speed limit unless only a single subsystem evolves and the remaining are stationary. For mixed separable states, $\rho = \sum_n p_n \ket{\phi_n}\bra{\phi_n}$, the same is true of each term $\ket{\phi_n}$ of the decomposition~\cite{GiovannettiPRA}.

In contrast, an $N$-party Greenberger-Horne-Zeilinger (GHZ) state of the form $(\ket{000...}+\ket{111...})/\sqrt2$ --- an example of a fully entangled state --- can reach the quan\-tum speed limit. Saturation occurs on an evolution under a Hamiltonian $H=\sum_jH_j$ with $\ket0_j$ and $\ket1_j$ being the eigenstates of $H_j$. This can be seen by taking $H_j=\hbar\omega_0Z_j/2$, where $Z_j$ is the $Z$ Pauli operator acting on qubit $j$. One obtains $\sqrt{F_B(0,\tau)}=\cos(N\omega_0\tau/2)$, with $\Delta E=N\hbar\omega_0/2$.

A measure of the speed-up due to entanglement is obtained by comparing, for $N\gg1$, the $N$-dependence of the evolution time of states symmetric on all subsystems. For our purposes, we assume each subsystem has energy spread $\Delta E_{\rm sub}$ (unaffected by variation of $N$) and the global Hamiltonian is of the form $H=\sum_jH_j$. For separable states, the bound on each subsystem is tighter than that for the state as a whole, so relation~\eqref{MTdenovo} is eclipsed by the more relevant relation
\begin{equation}
\tau \geq \frac\hbar{\Delta E_{\rm sub}}\arccos\sqrt{F_B^{1/N}(0,\tau)} = \frac{\hbar}{\Delta E_{\rm sub}}\sqrt{\ln\left[\frac1{F_B(0,\tau)}\right]} \frac1{\sqrt N} + \mathcal O\left(\frac1{N^{3/2}}\right)\ ,
\label{sepsymmunit}
\end{equation}
where it was used that the fidelity for each subsystem relative to its initial state equals $F_B^{1/N}(0,\tau)$. We then say that there is a $\tau\sim1/\sqrt N$ scaling of the time $\tau$ with the number of subsystems $N$, where $\tau$ is the time necessary to reach a certain distance (or fidelity). Every separable state is limited by this scaling, there are states that reach this bound, e.g., an initial state $[\ket0+\ket1]^{\otimes N}/\sqrt2^N$ under $H_j=\hbar\omega_0Z_j/2$. Note that the scaling does not depend on the value of $F_B$.

On the other hand, this scaling does not hold for entangled systems, the GHZ state serves as a counterexample. It is symmetric in all subsystems and, assuming the same evolution under $H_j=\hbar\omega_0Z_j/2$, it obeys
\begin{equation}
\tau = \frac{\arccos\sqrt{F_B(0,\tau)}}{\Delta E} = \frac1N \frac{\arccos\sqrt{F_B(0,\tau)}}{\Delta E_{\rm sub}} \ ,
\label{tauGHZunit}
\end{equation}
with a clear $\tau\sim1/N$ scaling. The change in scaling from $1/\sqrt{N}$ to $1/N$ in going from separable to entangled states expresses the evolution speed-up due to the presence of entanglement.

It can then be said that separable states are ``slow'' as compared to the speed that entangled ones can reach\footnote{It should be noted that entangled states are not \textit{necessarily} fast, as can be seen by the evolution of a fully entangled initial state $\ket{010101...}+\ket{101010...}$ under the same $H=\sum_jH_j$ as before. Assuming an even number of subsystems, this state does not evolve at all.}. Fröwis~\cite{Frowis} has shown this shift in behavior in a multi-qubit system by showing that the quantum Fisher information of a unitary evolution of $N$ qubits is upper-bounded by $N$ for separable states and by $N^2$ in general\footnote{The authors of~\cite{Frowis} and references therein assume dimensionless time units to arrive at a dimensionless quantum Fisher information. The employment of $t$ with dimension of time does not affect the scaling in any way, though.}. Inserting those upper bounds in our main quantum speed limit of Eq.~\eqref{boundarccos} demonstrates the change in scaling.  This has also been discussed in~\cite{GiovannettiPRA,GiovannettiEPL}, examples have been given in \cite{Batle,Borras2006,Zander}. Let us see how these scalings behave in a non-unitary evolution.

\subsubsection{Non-unitary case: $N$-qubit dephasing}

\hspace{5mm} We study the dephasing of $N$ qubits, each of them undergoing the dephasing channel of Eq.~\eqref{dephasmaster} or Eq.~\eqref{dephasingchannel}. The auxiliary system is also a set of $N$ qubits, and the only interaction between $S$ and $E$ occurs between the $j$-th qubit of $S$ and the $j$-th of $E$.

The unitary operator describing our first purification is 
\begin{equation}
 U_{S,E}(t) =  \prod_{j=1}^N e^{-i\omega_0 t  Z_j/2} e^{ - i\arccos\sqrt{P(t)}  Z_j Y^{(E)}_j } ,
\label{evolN}
\end{equation}
where $Z_j$ is a Pauli operator acting on the $j$-th qubit of $S$, and $Y^{(E)}_j$, on the $j$-th of $E$. The corresponding corrected Hamiltonian is
\begin{equation}
\mathfrak H_{S,E}(t) = \sum_j \frac{\hbar \omega_0}2 Z_j + \frac{\hbar\gamma/2}{\sqrt{e^{2\gamma t}-1}}Z_j Y_j^{(E)}   
\label{HdephasN}
\end{equation}
and we must minimize the variance of
\begin{equation}
\mathcal H_{S,E}(t) = \mathfrak H_{S,E}(t) + U_{S,E}^\dagger(t)\mathfrak h_E(t) U_{S,E}(t) 
\label{mathcalHsimp2}
\end{equation}
with respect to different $\mathfrak h_E(t)$. Minimizing over the complete set of $\mathfrak h_E(t)$, composed of all Hermitian $2^N\times2^N$ matrices, is impractical. We optimize instead over a three-parameter family of operators, the choice of which hinges on the symmetry of the system:
\begin{equation}
\mathfrak h_E(t) = \alpha(t) \sum_jX_j^{(E)} + \beta(t) \sum_jY_j^{(E)} + \delta(t) \sum_jZ_j^{(E)} \ .
\label{frakhdephN}
\end{equation}

This minimization is performed in Appendix~\ref{dephchannelderivation}. The result is expressed in terms of the variance $[\Delta\mathcal Z]^2$ and average $\braket{\mathcal Z}$ of $\mathcal Z:=\sum_jZ_j/N$. We also define a parameter $q:=[\Delta\mathcal Z]^2/\left(1-\braket{\mathcal Z}^2\right)$. The minimum is
\begin{equation}
\mathcal C_Q^{\rm opt}(t) = [\Delta\mathcal Z]^2\left[\frac{\omega_0^2N^2}{Nq(e^{2\gamma t}-1)+1} + \frac{\gamma^2N/q}{e^{2\gamma t}-1}\right] \ ,
\label{deltaHminNq}
\end{equation}
where expectation values are to be taken on initial states.

The parameter $q$ is restricted to $0\leq q\leq1$ and acts as an indicator of the degree of the correlations among the qubits of $S$. A GHZ state (full correlation between qubits) obeys $q=1$; for any separable state, $q\leq1/N$, with equality occurring for symmetric states. The value $q=0$ can be attained by anticorrelated states of the form $\ket{010101...}$ or $[\ket{010101...}+\ket{101010...}]/\sqrt 2$.

Let us then evaluate the $\tau\times N$ behavior of the bound. We begin by taking $q=1$, which corresponds in general to an initial state of the form $\sqrt{p}\ket{000...}+e^{i\phi}\sqrt{1-p}\ket{111...}$. The bound then reads
\begin{equation}
2D(0,\tau) \leq \sqrt{N} \Delta\mathcal Z \int_0^{\gamma\tau} \sqrt{ r^2 \dfrac{N}{N(e^{2u}-1) + 1} +  \frac{1}{e^{2u}-1} } du \ ,
\label{boundGHZ}
\end{equation}
with $r:=\omega_0/\gamma$. An equally general upper bound for $D(0,\tau)$ leads to an analytically expressible bound on $\tau$. It is found by considering $N(e^{2\gamma\tau}-1) \gg 1$:
\begin{equation}
2D(0,\tau) \leq \sqrt{N} \Delta\mathcal Z \int_0^{\gamma\tau} \sqrt{ \frac{r^2+1}{e^{2u}-1} } du = \sqrt{N} \Delta\mathcal Z \sqrt{r^2 + 1} \arctan \sqrt{ e^{2\gamma\tau}-1 } \ .
\label{boundGHZmed}
\end{equation}
Notice that the bound maintains its generality by making this substitution, but becomes less tight. Solving for $\tau$, we find
\begin{equation}
\tau \geq \frac1\gamma \ln \left[\sec \frac{2D}{\sqrt{N}\Delta\mathcal Z\sqrt{r^2+1}} \right] \ , 
\label{GHZsolvtau}
\end{equation}
which, for $N\gg1$, leads to
\begin{equation}
\tau \geq \frac1N \frac{4D^2}{\gamma(r^2+1)} \ ,
\label{GHZeqtauN}
\end{equation} 
where the $\tau\sim1/N$ behavior is explicit. There is an alternative estimate yielding a better approximation of the integral in Eq.~\eqref{boundGHZ}, obtained in the $r\gg1$ limit:
\begin{align}
2D(0,\tau) \leq {} & \sqrt{N} r \Delta\mathcal Z \int_0^{\gamma\tau}\frac1{\sqrt{e^{2u}-(1-\frac1N)}}du \label{rgg1p1}\\
	& = \sqrt{N} r \Delta\mathcal Z \sqrt{\frac{N}{N-1}}\left[\arctan\sqrt{\frac{e^{2\gamma\tau}-(1-\frac1N)}{1-\frac1N}} -\arctan\frac1{\sqrt{N-1}}\right] \ .
\label{rgg1p2}
\end{align}
Expanding for high values of $N$, one obtains
\begin{equation}
\tau \geq \frac1N\frac{2D}{\omega_0\Delta\mathcal Z}\left(1+\frac{D}{r\Delta\mathcal Z}\right) \ ,
\label{boundGHZcoeff}
\end{equation}
which not only shows the $\tau\sim1/N$ dependence, but also produces a very good fit of the numerically calculated bound, see Fig.~\ref{tauNGHZ}.
\begin{figure}[tb]\centering
\includegraphics[width=.3\columnwidth]{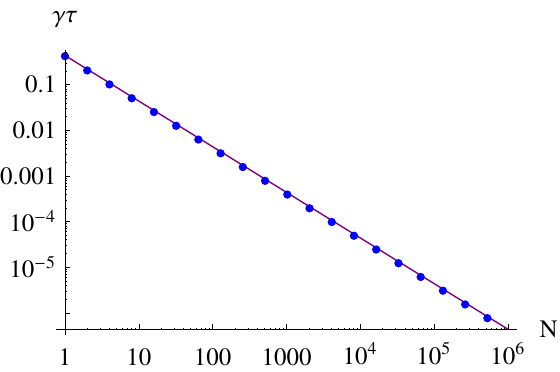} 
\includegraphics[width=.3\columnwidth]{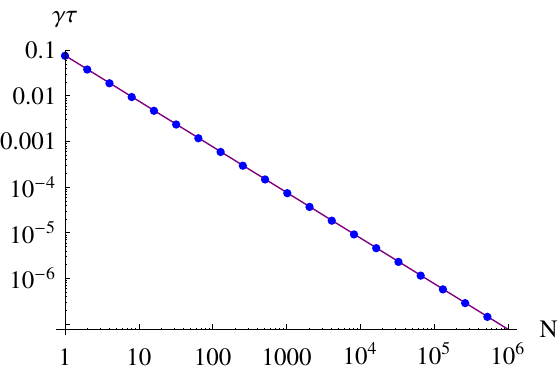} 
\includegraphics[width=.3\columnwidth]{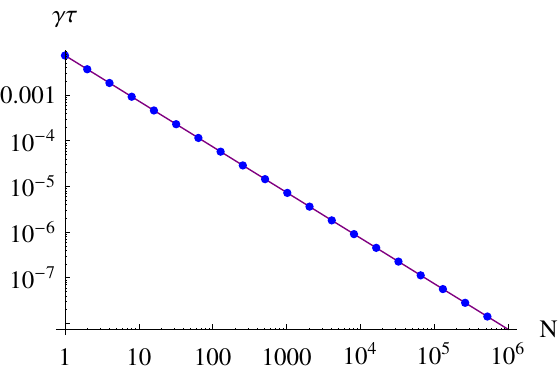}
\caption{Lower bound on time $\gamma\tau$ for fully entangled GHZ state ($\Delta\mathcal Z=1$) to reach $D=94\%$ of the maximal distance (${F_B}=1\%$), calculated numerically from \eqref{boundGHZ}, as a function of $N$, with $r=8$, $40$ and $400$, respectively (error bars smaller than symbols). The straight line, proportional to  $1/N$, obeys \eqref{boundGHZcoeff}.}%
\label{tauNGHZ}
\end{figure}

The $\tau\sim1/N$ behavior for $q=1$ can also be corroborated by an exact calculation with an initial GHZ state $[\ket{000...}+\ket{111...}]/\sqrt 2$. This yields
\begin{equation}
\cos^2 D = F_B(0,\tau) = \frac{1+e^{-N\gamma\tau}\cos N\omega_0\tau}{2} ,
\label{FDiretoGHZ}
\end{equation}
where the $\tau\sim1/N$ scaling is clear from the joint dependence on the product $(N\tau)$, even if solving for $\tau$ is not trivial.

In summary, the fully entangled states found by setting $q=1$ present, on this non-unitary channel, a behavior analogous to that of unitary evolutions, namely, it is capable of a ``fast'' evolution $\tau\sim1/N$. Let us now turn to the evolution of separable states under this same channel.

Any separable state has $q\leq1/N$, where equality occurs for separable states symmetric on all qubits (more generally, for states such that $\braket{Z_j}=\braket{\mathcal Z}$ for every $j$). We take $q=1/N$ as our paradigm for separable states. The bound in this case reads
\begin{equation}
2 D(0,\tau) \leq  \sqrt{N} \sqrt{1 - \braket{\mathcal Z}^2} \sqrt{r^2+1} 
	  \left[ E\left(\frac{\pi}{2},\frac{r^2}{r^2+1}\right) 	- E\left(\arcsin e^{-\gamma\tau}, \frac{r^2}{r^2+1}\right) \right] \ .
\label{boundSep}
\end{equation}
Notice that we write the initial-state dependence as $1 - \braket{\mathcal Z}^2$ ($\in [0,1]$) because $\Delta\mathcal Z$ is now bounded by $1/\sqrt{N}$ and we are interested in making the $N$-dependence explicit. The result is quite similar to that of a single qubit, Eq.~\eqref{deph1ellipt}, except for a factor $\sqrt{N}$ (notice that for a single qubit $\Delta Z=\sqrt{1-\braket Z^2}$).

We plot in Fig.~\ref{tauNsepbound} the bound on the time needed to arrive at a $1\%$ fidelity relative to the initial state (equivalently, $D(0,\tau)\approx94\% \pi/2$) as a function of $N$. We observe a striking result: the $\tau\times N$ dependence is not the same for all values of $N$. By the change of slope on the logarithmic graph, it is clear that there are two different power laws and a transition between them.

\begin{figure}[ht]
\centering
\includegraphics[width=.5\columnwidth]{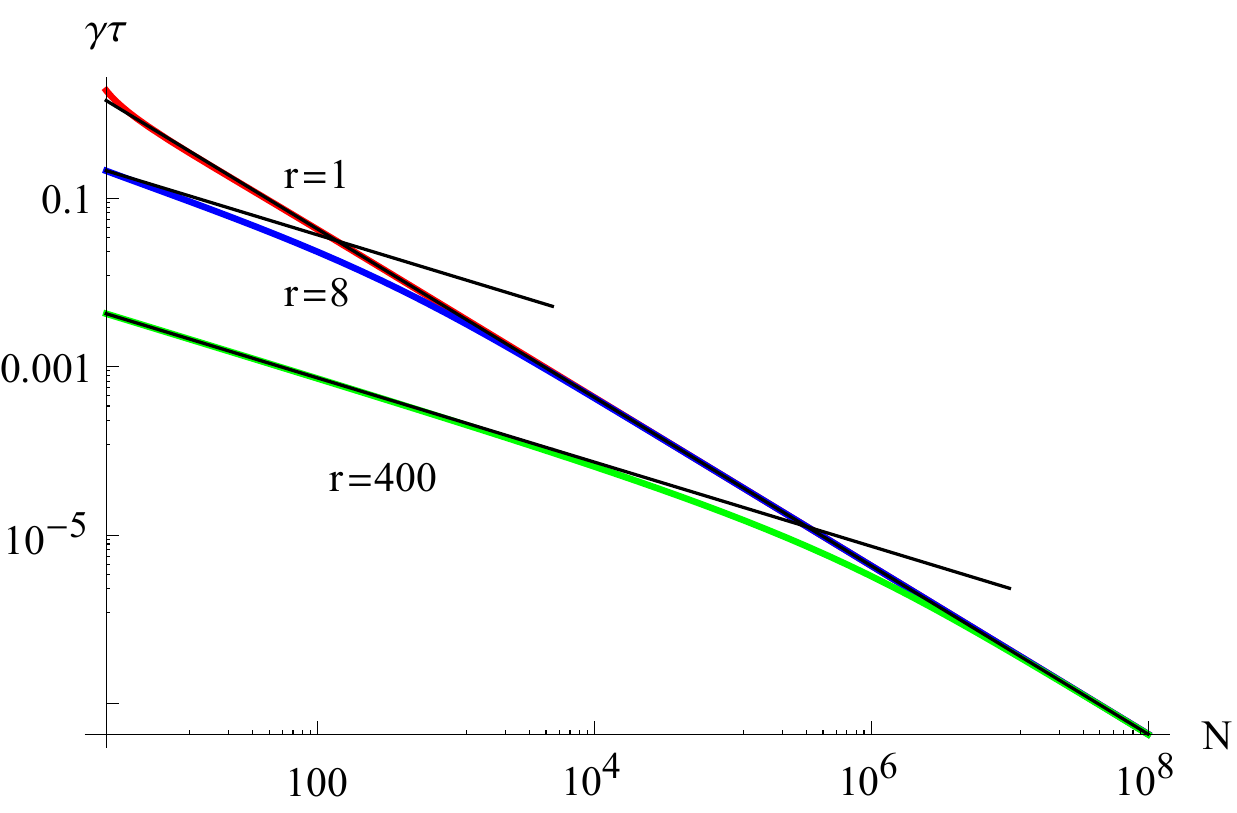} 
\caption{$N$-qubit dephasing: lower bound on the time necessary for separable, symmetric state undergoing to reach $D=94\%$ of the maximal distance (${F_B}=1\%$ relative to its initial state) as a function of the number of qubits $N$, Eq.~\eqref{boundSep}. Two different asymptotic behaviors can be seen, $\tau\sim1/\sqrt N$ and $\tau\sim1/N$, obeying Eqs.~(\ref{boundSep-rtauN},~\ref{boundSep-u}).}
\label{tauNsepbound}%
\end{figure}

We can find asymptotic expressions to better understand this behavior. The steeper slope can be fitted by expanding Eq.~\eqref{boundSep} to lowest order in $\gamma\tau$ and solving for $\tau$ (equivalent to the $N\gg1$ limit, since $\tau\times N$ is unequivocally decreasing),
\begin{equation}
 \tau \geq \frac{1}{N} \frac{ 2 D^2 } {\gamma  (1 - \braket{\mathcal Z}^2)} \ ,
\label{boundSep-u}
\end{equation}
which shows the $\tau\sim1/N$ dependence for $N\gg1$. As in Fig.~\ref{tauNsepbound}, the asymptotes are the same for every $r$. To obtain the other asymptotic limit, one should notice that for a less slanted region to appear $r$ must be sufficiently high. Taking the $r\gg1$ limit in Eq.~\eqref{boundSep}, one finds
\begin{equation}
2 D(0,\tau) \leq \sqrt{1 - \braket{\mathcal Z}^2} \sqrt{N} r (1-e^{-\gamma\tau}) \ .
\label{boundSep-r}
\end{equation}
We can still use $\gamma\tau<1$, which then yields
\begin{equation}
\tau \geq \frac{1}{\sqrt N} \frac{2 D}{\omega_0\sqrt{1 - \braket{\mathcal Z}^2}} \ ,
\label{boundSep-rtauN}
\end{equation}
exhibiting a $\tau\sim1/\sqrt N$ behavior. We can also estimate when the transition between the two behaviors occurs as when the two expansions of Eqs.~(\ref{boundSep-u},~\ref{boundSep-rtauN}) coincide. This corresponds to
\begin{equation}
N_{\rm tr} \simeq r^2 \frac{D^2}{1 - \braket{\mathcal Z}^2} \ ,
\label{Ntr}
\end{equation}
which amounts to a $N_{\rm tr} \sim r^2$ scaling. Eqs.~(\ref{boundSep-u},~\ref{boundSep-rtauN}) are the asymptotes shown in Fig.~\ref{tauNsepbound}.

We have thus shown that for separable states the relation $\tau\times N$ given by the bound presents two different behaviors. For $\gamma^2 N \ll \omega_0^2$, the bound exhibits a ``slow'' $\tau\sim1/\sqrt N$ relation, but as $N$ increases there is a transition to a ``fast'' $\tau\sim1/N$ dependence, achieved when $\gamma^2 N \gg \omega_0^2$. This is remarkably different from the unitary case, where no separable state presents such a ``fast'' evolution. An important advantage of the bound is that it is able to present such result in a more general manner, independent of initial state. Such ``fast'' evolutions are of interest and could have potential applications in fields such as computation. A non-unitary reset operation that is fast independently of initial-state entanglement can be of great use in this area.

To corroborate even further the result, we have selected a specific initial state to calculate exact results for $D(0,\tau)$ and compare to our bound, namely, the state $(\ket0+\ket1)^{\otimes N}/\sqrt2^N$. Fig.~\ref{tauNsep} displays the bound for separable states together with exact calculations. It shows that our bound is capable of depicting important features of the quantum systems at hand even without being saturated. We note that the bound correctly predicts not only the twofold behavior of $\tau\times N$, but the scaling of $N_{\rm tr}$ at which the transition between the different slopes takes place.

\begin{figure}[ht]
\centering
\includegraphics[width=.5\columnwidth]{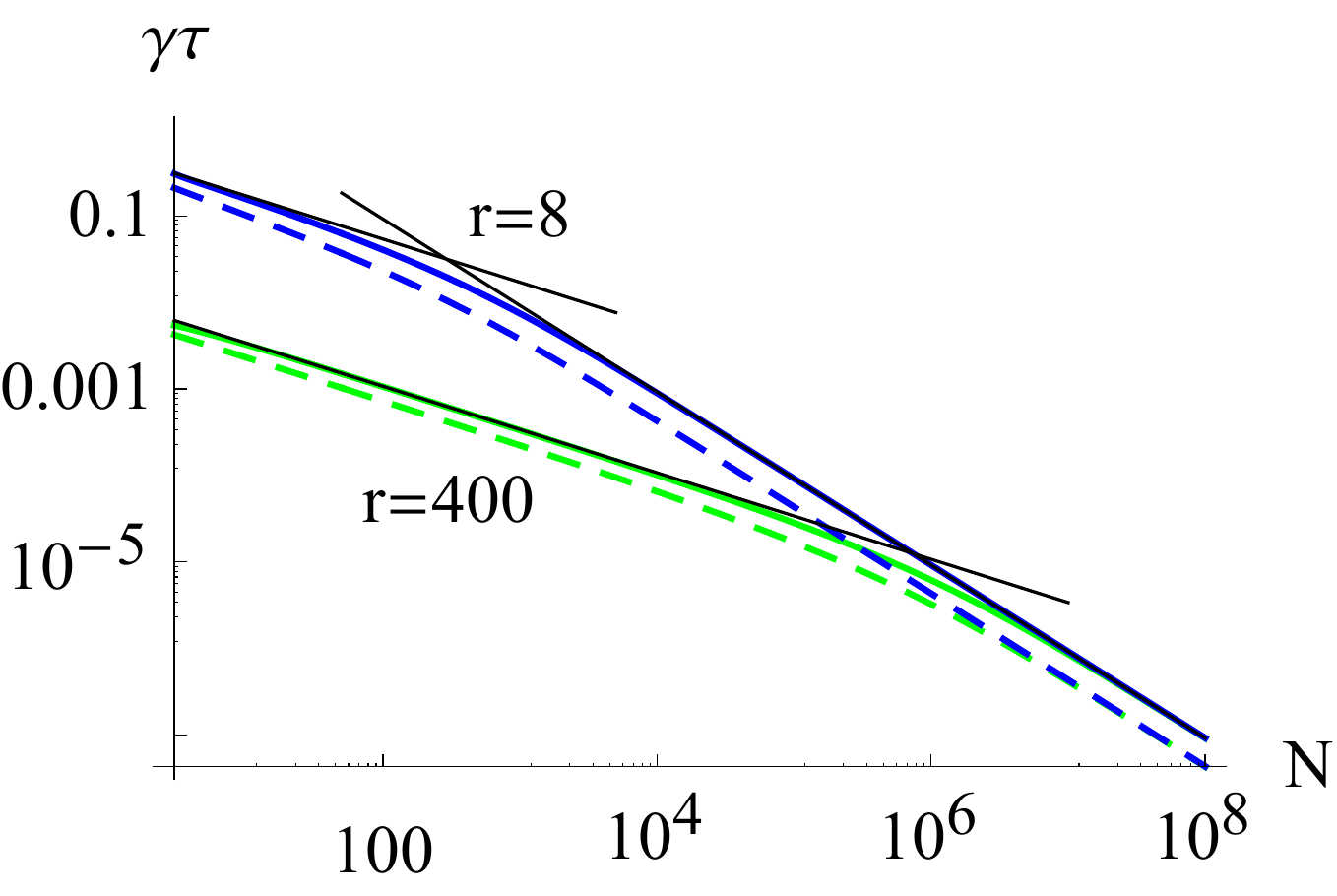} 
\caption{$N$-qubit dephasing: comparison between lower bound (dashed line) and exact calculation (solid line) for the time necessary for separable, symmetric state to reach $D=94\%$ of the maximal distance (${F_B}=1\%$) as a function of the number of qubits $N$, Eqs.~(\ref{boundSep},~\ref{FDiretoSep}), resp.  The asymptotes, proportional to $1/\sqrt N$, $1/N$, are taken from Eq.~\eqref{exactexpans}.}
\label{tauNsep}%
\end{figure}

The exact calculations for $(\ket0+\ket1)^{\otimes N}/\sqrt2^N$, $1-\braket{\mathcal Z}=1$, yield
\begin{equation}
\cos^2 D = F_B(0,\tau) =  \frac{1}{2^N}\left(1+e^{-\gamma\tau}\cos\omega_0\tau \right)^N ,
\label{FDiretoSep}
\end{equation}
which can be rewritten and expanded as
\begin{align}
&\gamma\tau + \frac{r^2-1}{2} \gamma^2\tau^2 +\mathcal{O}(\tau^3) =  4\frac{\ln\sec D}{N} + \mathcal{O}(\frac1{N^2}) \ . \label{exactexpans}
\end{align}
The two regimes can be seen in the above equation: for $N\gg1$ and $\gamma\tau\ll1$, the first term of the left-hand side is dominant, yielding a $\tau\sim1/N$ dependence; for smaller values of $N$, higher values of $\tau$ (still obeying $N>1$ and $\gamma\tau<1$) and $r\gg1$, the second term is dominant, and $\tau\sim1/\sqrt{N}$. The asymptotes of Fig.~\ref{tauNsep} are the $\tau\times N$ relations obtained taking only the first or the second term of the left-hand side of Eq.~\eqref{exactexpans} above. We can estimate when the transition occurs by finding the value of $\tau$ that leads to equal contributions from both terms, which is $\gamma\tau_{\rm tr}=2/(r^2-1)$. The corresponding value of $N$ is
\begin{equation}
N_{\rm tr} =  (r^2 - 1) \ln \sec D ,
\label{Nr2}
\end{equation}
confirming the $N_{\rm tr} \sim r^2$ scaling.

In summary, we have applied our main bound, Eq.~\eqref{boundarccos}, to three different quantum non-unitary evolutions: the amplitude-damping channel, the single-qubit dephasing and the multi-qubit dephasing. The first of these showed an example of an evolution on a mixed-state geodesic connecting pure states, illustrating nontrivial aspects of mixed-state geometry. As a path through a geodesic, it also exemplifies the saturation of the bound. Next, the single-qubit dephasing was able to yield a relation, the bound on $D(0,\infty)$, which restricted the possibility of a system becoming orthogonal in general grounds, independent of initial state and valid for any finite time. It also served as a case to demonstrate how the optimization procedure is performed. Lastly, the $N$-qubit dephasing allowed for the discussion of the interplay between entanglement and evolution speed, produced nontrivial results and showed that non-unitary evolution can be radically different from the usual unitary intuition.


\section{Application of the additional bound}\label{applicmedian}

\hspace{5mm} We end this Chapter presenting an application of our additional, median-based bound of Section~\ref{additional}, Eqs.~(\ref{medfracoinv},~\ref{boundmedint}). This bound, valid for pure states evolving unitarily, is applied to a three-level state, initially in
\begin{equation}
\ket{\psi_0} = c_0\ket0 + c_1\ket1 + c_2\ket2 \ ,
\label{medinit}
\end{equation}
with the governing Hamiltonian $H\ket n=n\hbar\omega\ket n$. If the median $E_\med=0\hbar\omega$, the result reduces to a Margolus-Levitin bound, in the sense that it depends on the average energy respective to the ground state. If $E_\med=2\hbar\omega$, the average energy is taken respective to the highest energy level, but the resulting expression is still analogous to a Margolus-Levitin bound. More interesting is the intermediate case $E_\med=1\hbar\omega$. We set $|c_1|^2=1/2$ in order to guarantee this value of the median. There is then only one relevant free parameter, the value of $p_2=|c_2|^2$ (since $|c_0|^2 = 1/2-p_2$). 

Let us first examine the version of the bound dependent on $\braket{\psi_0|\big|H-E_\med\big||\psi_0}$, Eq.~\eqref{medfracoint}. In this case, $\braket{\psi_0|\big|H-E_\med\big||\psi_0}=\hbar\omega/2$, and 
\begin{equation}
\cos [D_{FS}(0,\tau)] = \sqrt{F_B(0,\tau)} \geq 1 - \frac{\omega\tau}2 \ .
\label{exmedfraco}
\end{equation}
This bound is always surpassed by the Mandelstam-Tamm bound for the same evolution,
\begin{equation}
\sqrt{F_B(0,\tau)} \geq \cos\left[\sqrt{\frac14+2|c_2|^2(1-2|c_2|^2)} \ \omega\tau\right] \ ,
\label{exmedMTcomp}
\end{equation}
but can often beat the Margolus-Levitin bound on the time to become orthogonal, $\tau_{ML}\geq \pi/[\omega(1+4p_2)]$. The three are plotted in Fig.~\ref{medpre} for $p_2=1/4$, which is the value for which Eq.~\eqref{exmedfraco} presents the best results. Also plotted is the fidelity as by exact calculations, which yield
\begin{equation}
\cos [D_{FS}(0,\tau)] = \sqrt{F_B(0,\tau)} = \frac{1+\cos\omega\tau}2 - 2p_2(1-2p_2)\sin^2\omega\tau \ .
\label{medianaFexato}
\end{equation}

\begin{figure}[htb]
\centering
\includegraphics[width=.4\columnwidth]{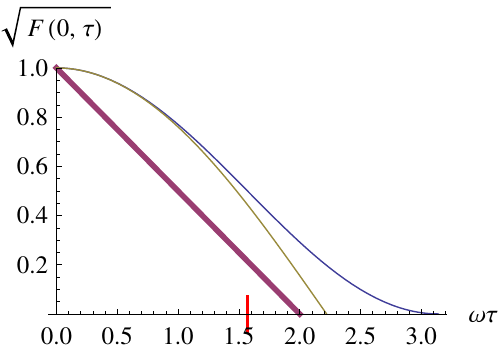}%
\caption{Plot of the first, weaker version of the median-based bound, dependent on $\braket{\psi_0|\big|H-E_\med\big||\psi_0}$, applied to the three-level system for $p_2=0.25$. The thick (purple) line represents the bound as in Eq.~\eqref{exmedfraco}, actual evolution is given by the uppermost (blue) curve, the Mandelstam-Tamm bound, plotted for comparison, is the intermediate (tan-colored) curve (Eq.~\ref{exmedMTcomp}). The Margolus-Levitin bound on orthogonality time $\tau_{ML}$ is shown as a red mark on the $\omega\tau$ axis.}%
\label{medpre}%
\end{figure}

A tighter result is attained with the second version of the median-based bound, dependent on $\left|\braket{\psi_0|H-E_\med|\psi(t)}\right|$. We have, for this three-state evolution,
\begin{equation}
\left|\braket{\psi_0|H-E_\med|\psi(t)}\right| = \hbar\omega\sqrt{\frac14-2p_2(1-2p_2)\cos^2\omega t} \ ,
\label{absmediana}
\end{equation}
and the bound of Eq.~\eqref{boundmedint}, obtained by integration, is expressed in terms of elliptic integrals of the second kind $E(y,m)$:
\begin{equation}
\cos [D_{FS}(0,\tau)] = \sqrt{F_B(0,\tau)} \geq 1 - 2\left|\frac14-p_2\right| E \left(\omega\tau , -\frac{p_2(\frac12-p_2)}{(\frac14-p_2)^2}\right) \ .
\label{medellipt}
\end{equation}
This version is actually competitive with both the Mandelstam-Tamm and Margolus-Levitin bounds: for certain values of $p_2$, it is the best of the three, as shown in Fig.~\ref{medbom}. This happens in the vicinity of $p_2=1/4$, value at which the bound is saturated throughout the evolution. We note that for $p_2=1/4$ Eq.~\eqref{medellipt} is indefinite, but the bound is well-defined and reads
\begin{equation}
\cos [D_{FS}(0,\tau)] = \sqrt{F_B(0,\tau)} \geq \frac{1+\cos\omega\tau}2 \ .
\label{bmedp2umquarto}
\end{equation}

\begin{figure}[htb]\centering%
\includegraphics[width=.32\columnwidth]{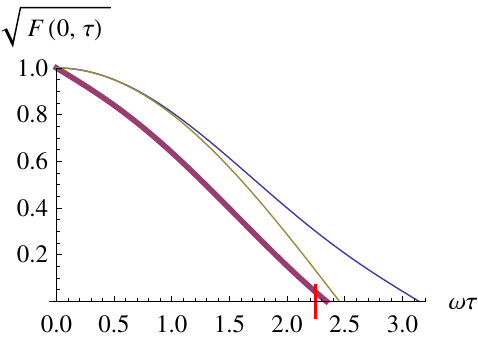}
\includegraphics[width=.32\columnwidth]{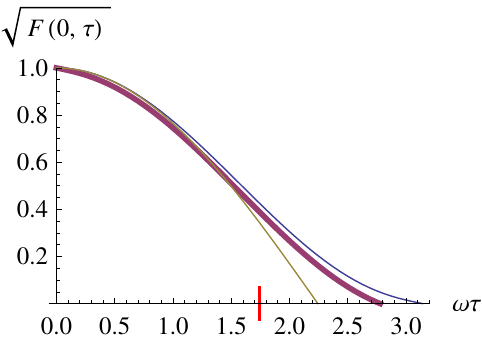}
\includegraphics[width=.32\columnwidth]{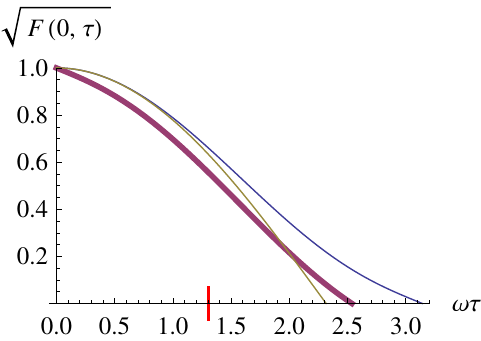}
\caption{Plot of the stronger version of the median-based bound, dependent on $\left|\braket{\psi_0|H-E_\med|\psi(t)}\right|$, applied to a three-level system for $p_2=0.1$, $0.2$ and $0.35$, resp. --- different $p_2$ correspond to different initial states. The thick (purple) line represents the bound as in Eq.~\eqref{medellipt}, actual evolution is given by the uppermost (blue) curve, the Mandelstam-Tamm bound, plotted for comparison, is the remaining (tan-colored) curve (Eq.~\ref{exmedMTcomp}). The Margolus-Levitin bound on orthogonality time $\tau_{ML}$ is shown as a red mark on the $\omega\tau$ axis.}%
\label{medbom}%
\end{figure}

This application shows that the additional result of Section~\ref{additional} has the potential to lead to useful bounds on evolution times. Although still restricted to unitary evolutions, it has been able to overcome the previously known bounds on certain occasions. The analytically demonstrated saturation at $p_2=1/4$ above for all times up to becoming orthogonal is a feat that has not been achieved with Margolus-Levitin-like bounds.  This median-based results may be a path for a more structured derivation, possibly geometric, of a general bound encompassing that of Margolus and Levitin.

\end{chapter}

\begin{chapter}{Final comments and perspectives}
\label{final}

\hspace{5 mm} We have in this thesis developed different quantum speed limits, which are bounds on how fast a quantum system can evolve. 

In the Introduction, besides presenting motivations for this study, we took the time to discern bounds on evolution time from energy-time uncertainty relations, which share similar mathematical expressions, but relate to conceptually different assertions.

We performed in Chapter~\ref{cap2} a review of the literature on the subject prior to our contributions. We presented the two paradigmatic quantum speed limits, the Mandelstam-Tamm bound and and the Margolus-Levitin bound, the most easily noticeable difference between them being that the former relies on the energy variance of the state and the latter, on the average energy. Another important distinction is that the former is established in a more complete quantum-mechanical framework than the latter.

Chapter~\ref{cap3} is the central chapter of this work, in which we derived our novel results. We first introduced some of the standard quantum-informational tools in Section~\ref{open}. In Section~\ref{GeoApp} we defined and developed a geometric structure for pure quantum states, culminating in a distance measure for this set and in a reinterpretation of the Mandelstam-Tamm bound on geometric foundations, with practical gains especially with regard to the issue of saturation of the bound. Section~\ref{Fisher} was dedicated to presenting the Fisher information, both ``classical'' and quantum. The core of this thesis is found in Section~\ref{thebound}, where we derived the main result, the general bound on evolution times of Eq.~\eqref{boundarccos}, valid for non-unitary as well as unitary evolutions. The greatest novelty of this finding is to extend the Mandelstam-Tamm bound to non-unitary evolutions, a feat deemed impossible on at least one occasion~\cite{JonesKok}. We were also able to preserve important features such as clear conditions for saturation. Additional bounds, based on the median of the energy distribution, were introduced in Section~\ref{additional}. They are still restricted to unitary evolutions and can be considered as a work in progress, but are relevant candidates for future developments.

Applications of the bounds to physical situations are the subject of Chapter~\ref{cap4}. It nevertheless starts with a development for calculating the quantum Fisher information based on purifications, central to the application of our main bound. We saw examples displaying saturation along with useful bounds even when not saturated. We were also able to make use of the bound to discuss the interplay between entanglement of a multipartite system and the speed of its evolution, achieving nontrivial results for non-unitary channels fundamentally distinct from their unitary counterparts.

There have been other results on the quantum speed limit for non-unitary processes after our work had been developed, but prior to the writing of this thesis. We discuss their properties and shortcomings in Appendix~\ref{later}.

We mention two main perspectives for future work. We are of the opinion that the Margolus-Levitin bound is founded, at this moment, on less than ideal grounds. To be able to derive it as a consequence of a more structured framework, possibly entailing a clearer interpretation of its meaning, would be a considerable achievement. We believe that the median-based bound presented in Section~\ref{additional} is a potential candidate for arriving at such a derivation, since it reduces to the Margolus-Levitin bound in some scenarios and stems from geometric considerations.

\label{brach}
The other perspective is the quantum brachistochrone problem, which consists in finding the fastest route from a quantum state to another given a set of allowed paths, or equivalently, given a set of restrictions. This problem is of direct practical importance and is closely related to quantum speed limits. One of the basic results of this topic is that if restriction is placed only on the available energy, the solution saturates the bound on evolution times~\cite{CarliniBr6}. We feel that the experience with the geometry of quantum states and optimization procedures gained along the present Ph.D. course will be of great help towards this goal.

\end{chapter}



\newpage
\phantomsection



\appendix
\begin{chapter}{Integration of the Fubini-Study metric}
\label{appFS}

\hspace{5 mm} We here perform the integration in two-dimensional Hilbert space of the differential form of the Fubini-Study metric from Eq.~\eqref{ds2FS}, 
\begin{equation}
ds_{FS}^2 = \braket{d\psi_{\rm proj}|d\psi_{\rm proj}} = \frac{\braket{d\psi|d\psi}\braket{\psi|\psi} - |\braket{\psi|d\psi}|^2}{\braket{\psi|\psi}^2} \ , 
\label{ds2FS-A}
\end{equation}
to find the finite Fubini-Study distance between two states, as mentioned on p.~\pageref{psiintFS}. Just as in Chapter~\ref{cap3}, we denote by $\ket{\psi_0}$ and $\ket{\psi_f}$ the initial and final states, respectively, and $\ket{\psi_1}$ will be a state orthogonal to $\ket{\psi_0}$ so that $\ket{\psi_f}$ belongs to the span of $\{\ket{\psi_0},\ket{\psi_1}\}$. Any evolution between these states can be parametrized as in Eq.~\eqref{psiintFS} by
\begin{equation}
\ket{\psi(t)} = f(t) \ket{\psi_0} + g(t) \ket{\psi_1} \ ,
\label{psiintFS-A}
\end{equation}
where $t\in[0,t_f]$ is a parameter and $f$ and $g$ are complex functions of it, with $g(0)=0$. None of the states mentioned need to be normalized, so there is no constraint on $f(0)$ other than $f(0)\neq0$ (in other words, we only demand $\ket{\psi(0)}\sim\ket{\psi_0}$, not equality).

The variation $\ket{d\psi}$ is given by
\begin{equation}
\ket{d\psi} = \ket{\psi(t+dt)}-\ket{\psi(t)} = \left( \dot f(t) \ket{\psi_0} + \dot g(t) \ket{\psi_1} \right)dt \ ,
\label{dpsi}
\end{equation}
while the product $\braket{\psi|d\psi}$ is
\begin{equation}
\braket{\psi|d\psi} = \braket{\psi(t)|d\psi}  = \left( f^*(t)\dot f(t) \braket{\psi_0|\psi_0} + g^*(t)\dot g(t) \braket{\psi_1|\psi_1} \right)dt \ ,
\label{psidpsi}
\end{equation}
where the overdot represents derivation with respect to $t$. Substituting in the expression for $ds_{FS}^2$, making the change of variables \mbox{$\tilde f(t):=f(t)\braket{\psi_0|\psi_0}$}, \mbox{$\tilde g(t):=g(t) \braket{\psi_1|\psi_1}$} and taking the square root, we obtain
\begin{equation}
ds_{FS} = \frac{\left| \tilde f(t) \dot{\tilde g}(t) - \dot{\tilde f}(t) \tilde g(t) \right|}{\left|\tilde f(t)\right|^2 + \left|\tilde g(t)\right|^2 }dt \ .
\label{eqdspeq}
\end{equation}
The numerator can be written as the derivative of the quotient $\tilde g(t) / \tilde f(t)$,
\begin{equation}
ds_{FS} = \frac{\left| d\left(\tilde g(t)/\tilde f(t)\right)/dt \right|}{1 + \left|\tilde g(t)/\tilde f(t)\right|^2 }dt = \frac{\left| d\left(\tilde g(t)/\tilde f(t)\right)\right|}{1 + \left|\tilde g(t)/\tilde f(t)\right|^2} \ .
\label{eqdspinteg}
\end{equation}
With the change of variables $z(t):=\tilde g(t) / \tilde f(t)$ (note that $z(0)=0$), the length of the path in the projective space according to the Fubini-Study metric is
\begin{equation}
\ell_{FS} = \int ds_{FS} = \int_0^{z_f} \frac{|dz|}{1+|z|^2} \ ,
\label{intz}
\end{equation}
where the path of integration is specified by $z(t)$, and $z_f :=\tilde g(t_f)/\tilde f(t_f)$ corresponds to state $\ket{\psi_f}$. Setting \mbox{$x={\rm Re}(z)$}, \mbox{$y={\rm Im}(z)$}, we see that this is in fact the line integral of $(1+|z|^2)^{-1}$ on the complex plane, since \mbox{$|dz|=\sqrt{dx^2+dy^2}$}; see Fig.~\ref{grafintFS}. In other words, the integration of the Fubini-Study metric has been mapped onto a line integral on the complex plane equipped with a Euclidean metric \mbox{$|dz|^2=dx^2+dy^2$}.
\begin{figure}[ht]%
\centering
\includegraphics[width=.5\columnwidth]{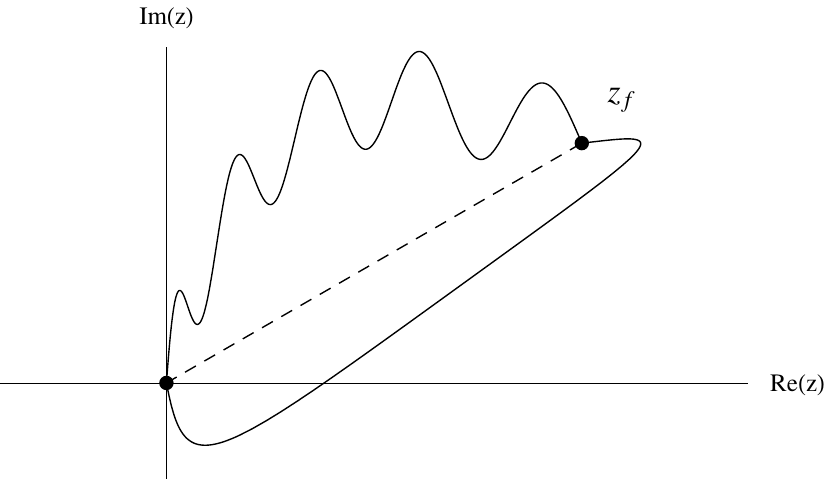}%
\caption{The integration of the Fubini-Study metric is mapped onto the line integral of a radial function on the complex plane (with usual Euclidean metric) from the origin to the final point $z_f$. The straight line is, in fact, the minimizing path.}%
\label{grafintFS}%
\end{figure}

In order to find the distance, we need to find the path between $z=0$ and $z_f$ yielding the minimal length $\ell_{FS}$. Due to the radial symmetry of the problem, it is intuitive to consider the straight line in the complex plane as yielding the minimum. This intuition can be confirmed by the fact that the radial straight line is the shortest path between the endpoints of the complex plane (in Euclidean metric!) and at the same time the path of steepest descent of the integrand throughout -- the second assertion is due to the radial gradient thereof.

Alternatively, this minimization can be performed by standard calculus of variations. Let $\left(x(\eta),y(\eta)\right)$ be a parametrization of the path for some parameter $\eta$. The length we wish to minimize is then written as
\begin{equation}
D_{FS} = \min_{\rm path} \int_{\rm path} \frac{\sqrt{\dot x^2 + \dot y^2}}{1+x^2(\eta) + y^2(\eta)} d\eta \ ,
\label{inteta}
\end{equation}
for fixed initial and final points, where the overdot now represents derivation with respect to $\eta$. The Euler equation for this minimization problem reads
\begin{equation}
\frac{\ddot x\dot y - \dot x\ddot y}{\dot x^2+\dot y^2} + 2 \frac{x\dot y - \dot xy}{1+x^2+y^2} = 0 \ ,
\label{Euler}
\end{equation}
and it is straightforward to see that the straight line through the origin, \mbox{$y(\eta)=Cx(\eta)$}, \mbox{$\dot y=C\dot x$} and \mbox{$\ddot y=C\ddot x$}, where $C$ is a constant, is a solution to the above.

For the outward straight line we then have $|dz|=d|z|$, and
\begin{equation}
D_{FS}(\ket{\psi_0},\ket{\psi_f}) = \int_0^{|z_f|} \frac{d|z|}{1+|z|^2} = \arctan |z_f| = \arctan \left| \frac {g(t_f) \braket{\psi_1|\psi_1} }{ f(t_f)\braket{\psi_0|\psi_0}}  \right| \ ,
\label{sFSchegando}
\end{equation}
which can be rewritten as
\begin{equation}
D_{FS}(\ket{\psi_0},\ket{\psi_f}) = \arccos \left( \frac{|f(t_f)\sqrt{\braket{\psi_0|\psi_0}}|}{\sqrt{|f(t_f)|^2\braket{\psi_0|\psi_0}+|g(t_f)|^2\braket{\psi_1|\psi_1}}} \right) \ .
\label{sFSarccosquase}
\end{equation}
But the (normalized) inner product between initial and final states reads
\begin{equation}
\frac{\braket{\psi_0|\psi_f}}{\sqrt{\braket{\psi_0|\psi_0}}\sqrt{\braket{\psi_f|\psi_f}}} 
                                                                      =  \frac{f(t_f)\sqrt{\braket{\psi_0|\psi_0}}}{\sqrt{|f(t_f)|^2\braket{\psi_0|\psi_0}+|g(t_f)|^2\braket{\psi_1|\psi_1}}}
\label{prodintapend}
\end{equation}
and we conclude that
\begin{equation}
D_{FS}(\ket{\psi_0},\ket{\psi_f}) = \arccos \left( \frac{\left| \braket{\psi_0|\psi_f} \right|}{\sqrt{\braket{\psi_0|\psi_0}}\sqrt{\braket{\psi_f|\psi_f}}} \right) \ , 
\label{FSarccos-A}
\end{equation}
as displayed in Eq.~\eqref{FSarccos} of the main text.

Moreover, we now know that the geodesic is the straight line from the origin to $z_f$. Expressed in terms of $z(t)$, it reads $z(t)= \xi(t) z_f$, where $\xi(t)$ is any nondecreasing real function of $t$ such that $\xi(0)=0$ and $\xi(t_f)=1$. Any condition imposed on $z$ will ignore, as it should, the overall phase and normalization of the state. When written in terms of $f$, $g$, the condition for the geodesic is
\begin{equation}
\frac{g(t)}{f(t)} = \xi (t) \frac{g(t_f)}{f(t_f)} \ ,
\label{fggeod-A}
\end{equation}
from which one sees that on the geodesic the relative phase between the components must be constant and the ratio of their moduli must not decrease during evolution.

\end{chapter}

\begin{chapter}{Minimization yielding the median}
\label{Appmed}

\hspace{5 mm} We have the task of minimizing the function of Eq.~\eqref{somatmod},
\begin{equation}
\sum_n |c_n|^2 \left|E_n-g\right|  = \braket{\psi_0|\big|H-g\big||\psi_0} \ ,
\label{repsomatmod}
\end{equation}
with respect to the real parameter $g$, given the set of energies $E_n$ and the probability distribution $|c_n|^2$.

The first step is to notice that this function tends to infinity if $g\rightarrow\pm\infty$, but is bounded from below because it is non-negative, so the minimum exists. 
This expression is, as a function of $g$, sectionally linear. This can be seen best by taking the derivative with respect to $g$, which gives
\begin{equation}
\sum_n |c_n|^2 \ {\rm sign} \left(g-E_n\right) = \sum_{n \ | \ E_n<g} |c_n|^2 - \sum_{n \ | \ E_n>g} |c_n|^2 \ .
\label{derg}
\end{equation}
The derivative is ill-defined when $g=E_n$ for some value of $n$. A graph of the function is shown in Fig.~\ref{figminmed}. It is linear in each section in which $g$ is between two consecutive values of $E_n$. Notice that the slope of the function always increases from one section to the other, since the increase of $g$ can only remove a term from the second summation in Eq.~\eqref{derg} and add it to the first.
\begin{figure}[ht]
\centering
\includegraphics[width=.5\columnwidth]{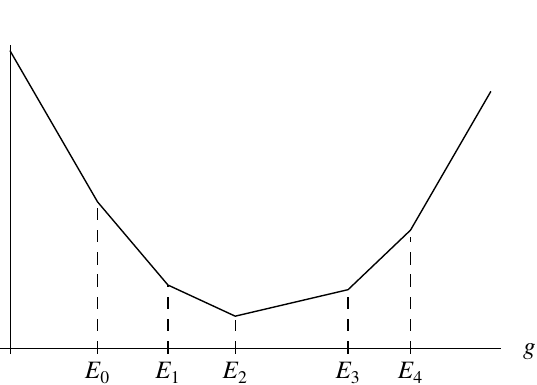}%
\caption{An example of the graph of the function we are minimizing as a function of $g$. The slope changes whenever $g=E_n$ for some $n$.}%
\label{figminmed}%
\end{figure}
The point where the slope goes from negative to positive is where the minimum of this function is achieved. This corresponds to the energy $E_{n*}$ such that the probability of all energies below $E_{n*}$ is less than one half, $\sum_{n<n*}|c_n|^2<\frac12$ and, at the same time, the probability of all energies above $E_{n*}$ is also less than one half, $\sum_{n>n*}|c_n|^2<\frac12$. This value $E_{n*}$ is, by definition, the \textit{median} of the energy distribution, $E_\med$.

However, there may not be a single point where the function goes from decreasing to increasing. This happens when there is a value of $g$ such that the probability of all energies below $g$ exactly equals one half as well as that of all energies above $g$,
\begin{equation}
\sum_{n \ | \ E_n<g} |c_n|^2 = \frac12 = \sum_{n \ | \ E_n>g} |c_n|^2 \ .
\label{condslopezero}
\end{equation}
 The corresponding graph has a section of slope zero, as in Fig.~\ref{figminmedzero}.
\begin{figure}[ht]
\centering
\includegraphics[width=.5\columnwidth]{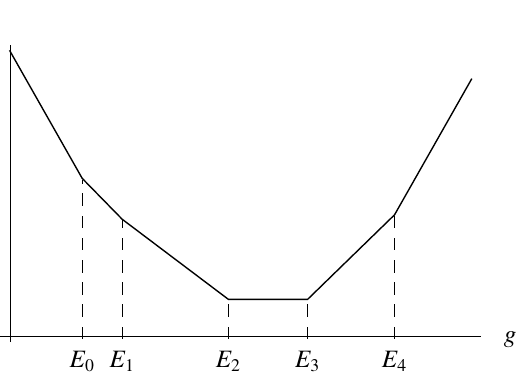}%
\caption{Graph of the function we are minimizing as a function of $g$ with a zero-slope region.}%
\label{figminmedzero}%
\end{figure}
The usual definition of the median in this case is the arithmetic average of the two values closest to the null slope. Since our function is constant in this interval, this value is one of those which yield the minimum of our function. The minimum is achieved by $E_\med$ also in this case.

We can then state with no loss of generality that the minimum of Eq.~\eqref{somatmod}, or~\eqref{repsomatmod}, is found when $g=E_\med$, and equals
\begin{equation}
\sum_n |c_n|^2 \left|E_n-E_\med\right|  = \braket{\psi_0|\big|H-E_\med\big||\psi_0} \ .
\label{mindosomat}
\end{equation}

\end{chapter}

\begin{chapter}{Tightness of the bound on the quantum Fisher information}
\label{Fishertight}

\hspace{5 mm} We have shown in Section~\ref{calculating} that there is an upper bound on the quantum Fisher information $\mathcal F_Q(t)$ of a system $S$ in state $\rho(t)$ based on the purification of this state
\begin{equation}
\mathcal F_Q(t)\leq\mathcal C_Q(t) \ .
\label{FleqC}
\end{equation}
In this Appendix, we show that this bound, $\mathcal C_Q(t)$, can be saturated. $\mathcal C_Q(t)$ is nothing more than the quantum Fisher information of a purification $\ket{\psi_{S,E}(t)}$ of $\rho(t)$ into a larger system $S+E$. We must then show that there is a purification of $\rho(t)$ such that $\mathcal C_Q(t)=\mathcal F_Q(t)$.

We start by considering the Bures fidelity of two states $\rho(t)$ and $\rho(t')$. We know from Uhlmann's theorem, Eq.~\eqref{UhlmTheor}, that for any purification $\ket{\psi_{S,E}(t)}$ of the former there is a purification $\ket{\phi_{S,E}^{\max}(t')}$ of the latter such that 
\begin{equation}
F_B[\rho(t),\rho(t')] = \left|\braket{\psi_{S,E}(t)|\phi_{S,E}^{\max}(t')}\right|^2 \ .
\label{Buresmax}
\end{equation}
Let us expand both sides of this equation on $t'$ around $t'=t$. The left-hand side has been expanded in Eq.~\eqref{BuresFisher} of the main text,
\begin{equation}
F_B[\rho(t),\rho(t')] = 1 - \frac{\mathcal F_Q(t)}4 dt^2 +\mathcal O(dt^3) \ , 
\label{repBuresFisher}
\end{equation}

with $dt=t'-t$, while the expansion of the right-hand side yields
\begin{equation}
\left|\braket{\psi(t)|\phi^{\max}(t')}\right|^2 = \left.I^{(0)}(t')\right|_{t'=t}  + \left.I^{(1)}(t')\right|_{t'=t} dt + \left.\frac{I^{(2)}(t')}2\right|_{t'=t} dt^2 + \mathcal O(dt^3) \ ,
\label{expprod}
\end{equation}
where we have omitted the subscript $_{S,E}$ for cleanness. The zero-order coefficient is simply the inner product squared, $I^{(0)}(t)=\left|\braket{\psi(t)|\phi^{\max}(t)}\right|^2$, and it must equal one, since this is the value of the Bures fidelity between state $\rho(t)$ and itself. In fact, the unit value of $F_B$ in this case allows us to write
\begin{equation}
\ket{\phi^{\max}(t)}\bra{\phi^{\max}(t)} = \ket{\psi(t)}\bra{\psi(t)} \ .
\label{igualdadeppurif}
\end{equation}
The higher-order terms are
\begin{align}
\left.I^{(1)}(t')\right|_{t'=t} = & \left[ \frac{d\bra{\phi^{\max}(t')}}{dt'} \ket{\psi(t)}\!\braket{\psi(t)|\phi^{\max}(t')} + \braket{\phi^{\max}(t')|\psi(t)}\!\bra{\psi(t)}\frac{d\ket{\phi^{\max}(t')}}{dt'} \right]_{t'=t} \ , \label{firstorder}\\
\begin{split}\left.I^{(2)}(t')\right|_{t'=t} = & \left[ \frac{d^2\bra{\phi^{\max}(t')}}{dt'^2}  \ket{\psi(t)}\!\braket{\psi(t)|\phi^{\max}(t')} + \braket{\phi^{\max}(t')|\psi(t)}\!\bra{\psi(t)}\frac{d^2\ket{\phi^{\max}(t')}}{dt'^2} + \right. \\ 
& \ \ \left. + 2  \frac{d\bra{\phi^{\max}(t')}}{dt'}\ket{\psi(t)}\!\bra{\psi(t)}\frac{d\ket{\phi^{\max}(t')}}{dt'} \right]_{t'=t} \ , \end{split}  \label{secondorder}
\end{align}
and, by using Eq.~\eqref{igualdadeppurif}, these simplify to
\begin{align}
\left.I^{(1)}(t')\right|_{t'=t} = & \left. \frac d{dt'}\left(\braket{\phi^{\max}(t')|\phi^{\max}(t')}\right) \right|_{t'=t} \ , \label{firstordersimp}\\
\begin{split}\left.I^{(2)}(t')\right|_{t'=t} = & \left. \frac{d^2}{dt'^2}\left(\braket{\phi^{\max}(t')|\phi^{\max}(t')}\right)\right|_{t'=t} + \\ 
& \ \ - 2 \left[\!\! \left(\frac{d\bra{\phi^{\max}(t')}}{dt'}\right)\!\!\left(\frac{d\ket{\phi^{\max}(t')}}{dt'}\right)  -  \left|\frac{d\bra{\phi^{\max}(t')}}{dt'}\ket{\phi^{\max}(t)}\right|^2 \right]_{t'=t} \ . \end{split}  \label{secondordersimp}
\end{align}
The normalization of $\ket{\phi^{\max}}$ implies that the first and second derivatives of $\braket{\phi^{\max}|\phi^{\max}}$ are zero, so that the first-order term of Eq.~\eqref{firstordersimp} is null and substitution of Eq.~\eqref{secondordersimp} into Eq.~\eqref{expprod} gives, up to second order,
\begin{equation}
\left|\braket{\psi(t)|\phi^{\max}(t')}\right|^2 = 1 - \left[\!\! \left(\frac{d\bra{\phi^{\max}(t')}}{dt'}\right)\!\!\left(\frac{d\ket{\phi^{\max}(t')}}{dt'}\right)  -  \left|\frac{d\bra{\phi^{\max}(t')}}{dt'}\ket{\phi^{\max}(t)}\right|^2 \right]_{t'=t} dt^2  \ .
\label{expquase}
\end{equation}
The evolution of $\ket{\phi_{S,E}^{\max}(t')}$ (reinserting the subscript!) is unitary, since this state is always pure. Considering that its derivative can be written in terms of the Hamiltonian $H_{S,E}(t')$ as $i\hbar \frac d{dt'}\ket{\phi_{S,E}^{\max}(t')} = H_{S,E}(t') \ket{\phi_{S,E}^{\max}(t')}$, the expansion up to second order becomes
\begin{equation}
\left|\braket{\psi(t)|\phi_{S,E}^{\max}(t')}\right|^2 = 
 1 - \frac{\left[\Delta H_{S,E}(t)\right]^2}{\hbar^2} dt^2 \ ,
\label{expfinal}
\end{equation}
where $\left[\Delta H_{S,E}(t)\right]^2$ is the variance of $H_{S,E}(t)$. From the comparison of the two expansions, Eqs.~(\ref{repBuresFisher}, \ref{expfinal}), we can ascertain that
\begin{equation}
\mathcal F_Q(t) = \left.\left. \frac{4}{\hbar^2} \left[\Delta H_{S,E}(t)\right]^2 \right|_{\ket{\phi_{S,E}^{\max}}} = \mathcal C_Q(t)\right|_{\ket{\phi_{S,E}^{\max}}} \ ,
\label{boundtight}
\end{equation}
i.e., there is a purification $\ket{\phi_{S,E}^{\max}(t)}$ of $\rho(t)$ such that the quantum Fisher information of the two ($\mathcal C_Q(t)$ and $\mathcal F_Q(t)$, respectively) coincide. Since the former is an upper bound to the latter, $\mathcal F_Q(t)$ is the minimum over purifications:
\begin{equation}
\mathcal F_Q(t) = \min_{\rm purif} \mathcal C_Q(t) = \min_{\rm purif} \frac{4}{\hbar^2} \left[\Delta H_{S,E}(t)\right]^2 \ .
\label{boundmin}
\end{equation}

Since the equality on Uhlmann's theorem requires the auxiliary system to have the same dimension as the original, that is the requirement for equality to hold for this bound. 
\end{chapter}

\begin{chapter}{Optimization of the bound for the dephasing channel}
\label{dephchannelderivation}

\hspace{5 mm} In this Appendix, we minimize the bound $\mathcal C_Q(t)$ for the dephasing channel. This is done directly in the $N$-qubit case. The single-qubit case is recovered by taking $N=1$, an independent derivation thereof can also serve as an exercise to the reader.

We are tasked with minimizing the initial-state variance of 
\begin{equation}
\mathcal H_{S,E}(t) = \mathfrak H_{S,E}(t) + U_{S,E}^\dagger(t)\mathfrak h_E(t) U_{S,E}(t)  
\label{mathcalHsimp3}
\end{equation}
over the three-parameter family of Eq.~\eqref{frakhdephN}, where the quantities have been defined along Chapter~\ref{cap4} in Eqs.~(\ref{fraksimple},\ref{defmathcalH},\ref{mathcalHsimp}), see also the table on p.~\pageref{tabela}. For $N=1$, this represents a minimization over all possible purifications. It is straightforward to show that this variance can be cast in the form
\begin{equation}
[\Delta\mathcal H_{S,E}(t)]^2 = [\Delta\mathfrak H_{S,E}(t)]^2 + [\Delta\tilde{\mathfrak h}_{E}(t)]^2 + 2 {\rm Re} \left[\braket{\mathfrak H_{S,E}(t)\tilde{\mathfrak h}_{E}(t)} - \braket{\mathfrak H_{S,E}(t)}\braket{\tilde{\mathfrak h}_{E}(t)} \right] \ ,
\label{DeltamathcalH}
\end{equation}
with $\tilde{\mathfrak h}_{E}(t) := U_{S,E}^\dagger(t)\mathfrak h_E(t) U_{S,E}(t)$ (for any operator $A$, $[\Delta A]^2$ denotes its variance).

Let $|\psi_0\rangle|0\rangle_E$ be the initial state of $S+E$. From $\mathfrak H_{S,E}(t)$ of Eq.~\eqref{HdephasN},
\begin{equation}
\braket{\mathfrak H_{S,E}(t)} = N \frac{\hbar\omega_0}2 \braket{\mathcal Z} \ ,
\label{HdephasNmedia}
\end{equation}
with $\mathcal Z:=\sum_jZ_j/N$, and
\begin{equation}
[\Delta\mathfrak H_{S,E}(t)]^2 = N^2 \frac{\hbar^2\omega_0^2}{4} [\Delta\mathcal{Z}]^2 + N \frac{\hbar^2\gamma^2/4}{e^{2\gamma t}-1} .
\label{medDeltaH2}
\end{equation}

From Eqs.~(\ref{frakhdephN},\ref{evolN}), one obtains
\begin{equation}
\braket{\tilde{\mathfrak h}_{E}(t)} = -\alpha(t) N \braket{\mathcal Z} 2\sqrt{P(t)}\sqrt{1-P(t)} + \delta(t)N[2P(t)-1] \ ,
\label{htilmedio}
\end{equation}
where $P(t):=(1+e^{-\gamma t})/2$, and
\begin{equation}\begin{split}
[\Delta\tilde{\mathfrak h}_{E}(t)]^2 = & N \left[\alpha^2(t)+\beta^2(t)+\delta^2(t)\right]
																	+ \alpha^2(t) \left( N^2[\Delta\mathcal{Z}]^2 - N \right) 4P(t)[1-P(t)]  \\
 &- N \delta^2(t) [2P(t)-1]^2 + 2 \alpha(t) \delta(t) N \braket{\mathcal Z} 2\sqrt{P(t)}\sqrt{1-P(t)}[2P(t)-1] \ .
\end{split}\label{medDeltah2}
\end{equation}

From the previous equations,
\begin{equation}\begin{split}
2 {\rm Re} \left[\langle\mathfrak H_{S,E}(t)\right.&\left.\tilde{\mathfrak h}_{E}(t)\rangle - \braket{\mathfrak H_{S,E}(t)}\braket{\tilde{\mathfrak h}_{E}(t)} \right] = \\
& = - 2\alpha(t) N^2 \hbar\omega_0 [\Delta\mathcal Z]^2 \sqrt{P(t)}\sqrt{1-P(t)} + \beta(t)\frac{N\hbar\gamma}{\sqrt{e^{2\gamma t}-1}} \braket{\mathcal Z} \ , 
\end{split}\label{corrhH}
\end{equation}
and $[\Delta\mathcal H_{S,E}(t)]^2$, according to \eqref{DeltamathcalH}, is the sum of \eqref{medDeltaH2}, \eqref{medDeltah2}, and \eqref{corrhH},
\begin{equation}
\begin{split}
[\Delta\mathcal H_{S,E}(t)]^2 = & N^2 \frac{\hbar^2\omega_0^2}{4} [\Delta\mathcal{Z}]^2 + N \frac{\hbar^2\gamma^2/4}{e^{2\gamma t}-1} 
							+ N \left[\alpha^2(t)+\beta^2(t)+\delta^2(t)\right] + \beta(t)\frac{N\hbar\gamma}{\sqrt{e^{2\gamma t}-1}} \braket{\mathcal Z} + \\
					&			+ \alpha^2(t) \left( N^2[\Delta\mathcal{Z}]^2 - N \right) 4P(t)[1-P(t)] - N \delta^2(t) [2P(t)-1]^2 + \\ 
					& + 2 \alpha(t) N \sqrt{P(t)}\sqrt{1-P(t)} \left\{ \delta(t) \braket{\mathcal Z} 2[2P(t)-1] - N \hbar\omega_0 [\Delta\mathcal Z]^2 \right\} \ .
\end{split}
\label{megazord}
\end{equation}

One now minimizes over $\alpha(t)$, $\beta(t)$, $\delta(t)$ for each $t$. From $\partial [\Delta\mathcal H_{S,E}(t)]^2 / \partial \beta = 0$, one finds
\begin{equation}
\beta(t)= - \frac{\hbar\gamma}{2\sqrt{e^{2\gamma t}-1}} \langle \hat{\mathcal Z}\rangle .
\label{beta}
\end{equation}
Conditions $\partial [\Delta\mathcal H_{S,E}(t)]^2 / \partial \alpha = 0$ and $\partial [\Delta\mathcal H_{S,E}(t)]^2 / \partial \delta = 0$ lead to a linear system of equations, with solutions
\begin{align}
\alpha(t)= {} & + \frac{\hbar\omega_0}2 \frac{e^{\gamma t} \sqrt{e^{2\gamma t}-1} N q }	{1 +(e^{2\gamma t}-1)N q } ,
\label{alpha} \\
\delta(t)= {} & - \frac{\hbar\omega_0}2 \frac{e^{2\gamma t} N q }	{ 1 + (e^{2\gamma t}-1)N q }  \braket{\mathcal Z} ,
\label{delta}
\end{align}
where $q:=[\Delta\mathcal Z]^2/\left(1-\braket{\mathcal Z}^2\right)$. By replacing Eqs.~(\ref{beta},\ref{alpha},\ref{delta}) into Eq.~\eqref{megazord},  one finds
\begin{equation}
\mathcal C_Q^{\rm opt}(t) = \frac4{\hbar^2}[\Delta\mathcal H_{S,E}(t)]^2 =   
  [\Delta\mathcal Z]^2\left[\frac{\omega_0^2N^2}{Nq(e^{2\gamma t}-1)+1} + \frac{\gamma^2N/q}{e^{2\gamma t}-1}\right] \ ,
\label{resultN}
\end{equation}
as in Eq.~\eqref{deltaHminNq}. The single-qubit result of Eq.~\eqref{deltaHmin1} is recovered by taking $N=1$ (notice that $q=1$ in this case).
\end{chapter}

\begin{chapter}{Later results in the literature}
\label{later}

\hspace{5mm} After the completion and submission to publication of our work, other authors have also derived bounds on non-unitary evolution. We now comment on these results.

In~\cite{rival}, a work subsequent to ours both in publication as preprint and submission to a refereed journal, del Campo and collaborators have also obtained bounds on non-unitary evolutions. Their bounds, albeit technically correct, lack some of the interesting features of our results.

Firstly, although they recognize that the well-established Bures fidelity $F_B$ would be the natural choice~\cite[p.1]{rival}, del Campo et alli base their comparison of initial and final states --- $\rho_0$ and $\rho_t$, respectively ---  on a figure of merit
\begin{equation}
f(t)=\frac{\Tr\left(\rho_0\rho_t\right)}{\Tr (\rho_0^2)}
\label{tipofidelidade}
\end{equation}
without justification other than its past employment on pure states and/or unitary evolutions. We note that for pure states evolving unitarily both $F_B$ and $f$ above --- as well as many other possible functions --- recover the pure-state fidelity $F(\ket\psi,\ket\phi)=|\braket{\psi|\phi}|^2$. The figure of merit $f$ above does yield $f=0$ for orthogonal states and $f=1$ for coinciding ones, but a meaning for its intermediate values has not been offered\footnote{It would seem particularly telling of the arbitrariness of this figure of merit that on a different section of the same paper~\cite[Eq.~(10)]{rival}, a different function $f(t)=\Tr\left(\rho_0\rho_t\right)$ is used for the same purpose. Comparing the main text of~\cite{rival} with its supplemental material, though, we are inclined to pardon this different version of $f$ as a typo.}. 

Another shortcoming of their result is that neither of their two bounds can be saturated in a non-unitary evolution, because their derivation either makes use of $\Tr(\rho_t^2)\leq1$ or of a Cauchy-Schwarz inequality whose saturation occurs only if $\rho_t\propto\rho_0$, and saturation of either bound at time $\tau$ would require either $\Tr(\rho_t^2)=1$ or $\rho_t\propto\rho_0$ for all $t\in(0,\tau)$. Considering that the novelty of general bounds is their application to non-unitary processes, such a limitation severely restricts the potential usefulness of the result in~\cite{rival}.

On the other hand, even when confined to unitary evolutions and pure states, the bound does not fare much better. The authors agree that a bound should reduce to the Mandelstam-Tamm bound in this case; their results fail to do so. Their bounds, applied to a unitary evolution of pure states, yield orthogonalization\footnote{For non-orthogonal final states, the bounds in~\cite{rival} fare even \textit{worse}.} times $\tau\geq\hbar/(\sqrt2\Delta E)$ and $\tau\geq\hbar/(2\Delta E)$, more than a factor of $2$ worse than the tight Mandelstam-Tamm bound, $\tau\leq\pi\hbar/(2\Delta E)$. In light of this reasoning, the claim of the authors of~\cite{rival} of generalizing the Mandelstam-Tamm bound seems inappropriate.

A last comment on this paper is that a confusing notation is used. Their bounds are implicit on time $\tau$, just as Eq.~\eqref{MTtimedep} or Eq.~\eqref{boundarccos}. Each of their results can be schematically cast into the form
\begin{equation}
\int_0^\tau \Big( t,{\rm evolution} \Big) dt \geq  {\rm function}[f(\tau)]  \ ,
\label{maq1}
\end{equation}
where the integrand is determined by the evolution and is time-dependent, and the right-hand side is some function of the figure of merit $f$ at final time $\tau$. The bounds are implicit because the integrand is time-dependent and integration can only be performed for each specific evolution. The authors nevertheless choose to multiply by $\tau/\tau$ in order to express this in terms of a time average $\frac1\tau\int_0^\tau\Big( t,{\rm evolution} \Big) dt=\overline{\Big( t,{\rm evolution} \Big)}$:
\begin{equation}
\tau \geq \frac{{\rm function}[f(\tau)]}{\frac1\tau\int_0^\tau\Big( t,{\rm evolution} \Big) dt} = \frac{{\rm function}[f(\tau)]}{\ \overline{\Big( t,{\rm evolution} \Big) } \ }  \ .
\label{maq2}
\end{equation}
This apparent explicitation of $\tau$ on the left-hand side and its omission on the denominator on the right-hand side have proven to be quite confusing. It can lead to the interpretation that the right-hand side be a bound on time $\tau$, whereas it actually depends on $\tau$.  We have come across researchers in the very field of Quantum Information who have been mislead by this notation, and we believe that it should be avoided completely.

Another article published after our results was that of Deffner and Lutz in~\cite{DeffLutzbound}. Although it has advantages over~\cite{rival}, it shares some of the criticism above. The authors here make use of the actual fidelity between initial and final states, restricting themselves to pure initial states. The alleged generalization of both Margolus-Levitin (ML) and Mandelstam-Tamm (MT) bounds is not achieved. As for the ML bound, there is a technical flaw in the relation that links the average energy to one of the obtained bounds. Applying the general bound to a unitary evolution under Hamiltonian $H_t$, the relation also depends on the state $\rho_t$ at time $t$ and  reads
\begin{equation}
\Tr\{|H_t\rho_t|\} = \Tr\{H_t\rho_t\} = \braket{H_t} \ , \ \ {\rm (incorrect!)}
\label{DLcgd}
\end{equation}
claimed to be valid for $H_t$ having non-negative spectrum, see~\cite[Eq.~(10)]{DeffLutzbound}. The mistake is due to the fact that the product $H_t\rho_t$ is assumed Hermitian, which need not be the case. A simple counterexample of a qubit in state $\rho=(\mathbb I+X)/2$ under a Hamiltonian $H=\mathbb I+Z$ disproves the relation. As such, the results of~\cite{DeffLutzbound} have not been related to the average energy, and no relation to the ML bound have been shown. As for the claim of recovering the MT bound, the corresponding result, when applied to a unitary evolution under a constant Hamiltonian, reads
\begin{equation}
\int_0^\tau \Delta E(t) dt \geq \frac\hbar{\sqrt2}(1-F) \ ,
\label{MTfalseta}
\end{equation}
where $F$ is the fidelity between initial and final states. It recovers one of the two bounds present in~\cite{rival}. In spite of the functional similarity to Eq.~\eqref{MTtimedep}, the different prefactor and functional dependence of Eq.~\eqref{MTfalseta} on $F$ imply a less tight bound, which cannot be saturated in any case. The most central result of~\cite{DeffLutzbound} is a bound that is expressible in terms of the operator norm (even when applied to unitary evolutions, it is expressible neither in terms of average energy nor of energy variance, hence is fundamentally different from ML or MT bounds). A purported advantage of this result is that it would be sharper than the MT bound~\cite[p.1]{DeffLutzbound}, but the comparison is made not with the actual MT bound, but with the weaker Eq.~\eqref{MTfalseta} and thus does not suffice to corroborate the relevance of the result.

Saturation is observed numerically in some instances for this bound, but it is shown to occur only for specific instants, not along an evolution. Clear criteria for saturation are also lacking, as well as an interpretation of the bound.

Deffner and Lutz also make use of the confusing notation illustrated in Eq.~\eqref{maq2}; they go so far as to use a specific value of $\tau$ to calculate the average on the right-hand side and treat the resulting fraction as a bound on $\tau$! (See~\cite[Fig.~1]{DeffLutzbound}, which is supposed to be a plot of the bound on $\tau$ obtained by taking $\tau=1$.)  This impedes a proper interpretation of their results. Only saturation can be (cumbersomely) evaluated from their plots and it is, as mentioned, only seen numerically for a specific value of $\tau$. This serves as an example of the damage done by the confusing choice of an apparently explicit $\tau$ as in Eq.~\eqref{maq2}.

\end{chapter}


\label{ofim}

\end{document}